\documentclass[pra,twocolumn,amsfonts,amssymb,amsmath,floatfix,tightenlines,superscriptaddress,nofootinbib]{revtex4-2}
\usepackage{bm}
\def\pg{\textrm{pg}}
\def\sc{\textrm{sc}}

\def\pair{\textrm{pair}}
\def\BKT{\textrm{BKT}}
\def\coh{\textrm{coh}}
\def\dia{\textrm{dia}}

\def\cond{\textrm{cond}}
\def\excited{\textrm{excited}}

\newcommand{\STO}{SrTiO$_3$}
\newcommand{\LSCO}{La$_{2-x}$Sr$_x$CuO$_4$}
\newcommand{\BSCCO}{Bi$_2$Sr$_2$CaCu$_2$O$_{8+\delta}$}
\newcommand{\YBCO}{Y$_{0.8}$Ca$_{0.2}$Ba$_2$Cu$_3$O$_{7-\delta}$}

\newcommand{\vect}[1] {\mathbf{#1}}

\newcommand{\xik}{\xi_{\mathbf{k}}}
\newcommand{\ek}{\epsilon_{\mathbf{k}}}

\newcommand{\Ek}{E_{\mathbf{k}}}

\newcommand{\mb}[1]{{\mathbf{#1}}}
\newcommand{\uk}{u_{\mathbf{k}}}
\newcommand{\vk}{v_{\mathbf{k}}}

\newcommand{\veck}{\vect{k}}
\newcommand{\vecq}{\vect{q}}

\newcommand{\PsiBCS}{ \Psi^{\textrm{BCS}}}

\usepackage{amsfonts,amssymb}
\usepackage[subnum]{cases}
\usepackage{mathrsfs}
\usepackage[tone]{tipa}
\usepackage{amsmath}
\usepackage{graphicx}
\usepackage{footmisc}
\usepackage[colorlinks=true,citecolor=blue,
linkcolor=blue,urlcolor=blue]{hyperref}
\usepackage{bm}          
\usepackage{soul} 
\usepackage{color}
\usepackage{longtable}
\usepackage{ textcomp }
\usepackage{verbatim}
\usepackage[english]{babel}
\usepackage{times}
\usepackage[title,titletoc]{appendix}

\newcommand{\phik}{\varphi_{\mathbf{k}}}

\begin{document}

\title{\textbf{When Superconductivity Crosses Over: From BCS to BEC}}

\author{Qijin Chen}
\affiliation{Hefei National Research Center for Physical Sciences at the Microscale and School of Physical Sciences, University of Science and Technology of China,  Hefei, Anhui 230026, China}
\affiliation{Shanghai Research Center for Quantum Science and CAS Center for Excellence in Quantum Information and Quantum Physics, University of  Science and Technology of China, Shanghai 201315, China}
\affiliation{Hefei National Laboratory, University of  Science and Technology of China, Hefei 230088, China}
\author{Zhiqiang Wang}
\affiliation{Department of Physics and James Franck Institute, University of Chicago, Chicago, Illinois 60637, USA}
\affiliation{Hefei National Research Center for Physical Sciences at the Microscale and School of Physical Sciences, University of Science and Technology of China,  Hefei, Anhui 230026, China}
\affiliation{Shanghai Research Center for Quantum Science and CAS Center for Excellence in Quantum Information and Quantum Physics, University of  Science and Technology of China, Shanghai 201315, China}
\affiliation{Hefei National Laboratory, University of  Science and Technology of China, Hefei 230088, China}
\author{Rufus Boyack}
\affiliation{Department of Physics and Astronomy, Dartmouth College, Hanover, New Hampshire 03755, USA}
\author{Shuolong Yang}
\affiliation{Pritzker School of Molecular Engineering, University of Chicago, Chicago, Illinois 60637, USA}
\author{K. Levin}
\affiliation{Department of Physics and James Franck Institute, University of Chicago, Chicago, Illinois 60637, USA}

\date{\today}

\begin{abstract}

  New developments in superconductivity, particularly through
  unexpected and often astonishing forms of superconducting materials,
  continue to excite the community and stimulate theory. It is now
  becoming clear that there are two distinct platforms for
  superconductivity: natural and synthetic materials. The study of these artificial materials
  has greatly expanded in the last decade or so,
  with the discoveries of new forms of superfluidity in artificial
  heterostructures and the exploitation of proximitization. 
  Natural superconductors continue to surprise through the Fe-based pnictides and
  chalcogenides, and nickelates as well as others. It is the goal of
  this review to present this two-pronged investigation into
  superconductors, with a focus on those that we have come to
  understand belong somewhere between the Bardeen-Cooper-Schrieffer (BCS) and Bose-Einstein
  condensation (BEC) regimes. We characterize in detail the nature of
  this ``crossover" superconductivity, which is to be distinguished
  from crossover superfluidity in atomic Fermi gases.  In the process, we
  address the multiple ways of promoting a system out of the BCS and
  into the BCS-BEC crossover regime within the context of concrete
  experimental realizations. These involve natural materials, such as
  organic conductors, as well as artificial, mostly two-dimensional
  materials, such as magic-angle twisted bilayer and trilayer
  graphene, or gate-controlled devices, as well as one-layer and
  interfacial superconducting films. This work should be viewed as a
  celebration of BCS theory by showing that even though this theory
  was initially implemented with the special case of weak correlations
  in mind, it can in a very natural way be extended to treat the case
  of these more exotic strongly correlated superconductors.
\end{abstract}

\maketitle

\tableofcontents

\section{Introduction: Background and History}

There has been a recent explosion of papers addressing a new form
of superconductivity. Superconductivity as traditionally addressed
within the famous Bardeen-Cooper-Schrieffer (BCS) theory~\citep{Bardeen1957} arises in
metals when an attractive interaction is present.
We often refer to this attractive interaction as the pairing ``glue".
This attraction causes fermions to form pairs, called Cooper pairs, which
are in some sense ``bosonic"~\citep{SchriefferBook}. Because of this connection to bosonic statistics,
the ground state of the pairs
can effectively counteract the Pauli exclusion principle.  Thus, as in a Bose system, 
the ground state of fermion pairs can now be macroscopically
occupied and the system, thereby, condenses into its ground state.
This BCS form of condensation, however, is not the same as the phenomenon of Bose-Einstein
condensation (BEC), appropriate to Bose systems, in which all fermionic degrees of freedom
have disappeared.

But something different from the BCS picture is found in a new generation of superconductors in which it
appears that there is
an anomalously strong pairing glue (of unspecified origin).
We refer to these systems as strongly correlated superconductors, and we characterize their form of superconductivity
as being described by a machinery which is neither the more familiar BCS theory
nor does it correspond to BEC; 
here the fermionic degrees of freedom are not completely absent.
These superconductors are said to be described by ``BCS-BEC crossover theory".
This new type of condensation phenomenon appears to be also present
in ultracold atomic Fermi gases, where it has been widely studied
~\cite{Chen2005,Giorgini2008,Randeria2014}.

An exciting fact is that there is now a very large class of
recently discovered superconductors that appear to
exhibit BCS-BEC crossover-like characteristics. These include:
iron-based superconductors, organic superconductors, magic-angle
twisted bilayer (MATBG) and trilayer graphene (MATTG), gate-controlled
two-dimensional devices, interfacial superconductivity, and
magnetoexcitonic condensates in graphene heterostructures.
One might also contemplate that the high-transition-temperature cuprates are also included in this class.

It is useful to note here an important characteristic of BCS-BEC
crossover theory, beyond the ground state discussed above.
In this theory, because the pairing interaction 
is stronger than in conventional
materials, it follows that fermion pairs form before they Bose condense
at the superfluid transition temperature, $T_\text{c}$. 
We will find that this property leads to a variety of experimental implications.
It is in striking contrast to the
well-established theory of BCS, 
where (because the attractive interaction is extremely weak)
pairing and condensation occur at exactly the same
temperature. There is no hint that a given BCS superconductor
will undergo the phase transition at any temperature above $T_\text{c}$.

This Review article is written to address these issues in considerable depth
and to describe what has been observed in
these two-dimensional (2D) and three-dimensional (3D)
superconductors that appear to be somewhere between BCS and BEC. We will
show how their various experimentally measured characteristics
relate to BCS-BEC crossover, paying special attention to 2D materials, where there seems to be a
surprisingly large number of examples. In the process, we present a
theoretical understanding of the crossover formalism at general temperatures.

Lest there be any confusion at the start, throughout this Review what
we mean by ``BCS-BEC crossover" is \textbf{not} the onset or proximity
to the BEC regime as defined by some researchers, but an intermediate regime between BCS
and BEC, where a significant departure from strict BCS theory is
apparent.  It should also be emphasized that what is being discussed
here pertains to the theoretical ``machinery" of superconductivity
rather than the microscopic pairing mechanism.

\begin{figure*}
\centering
\includegraphics[width=6.0in]{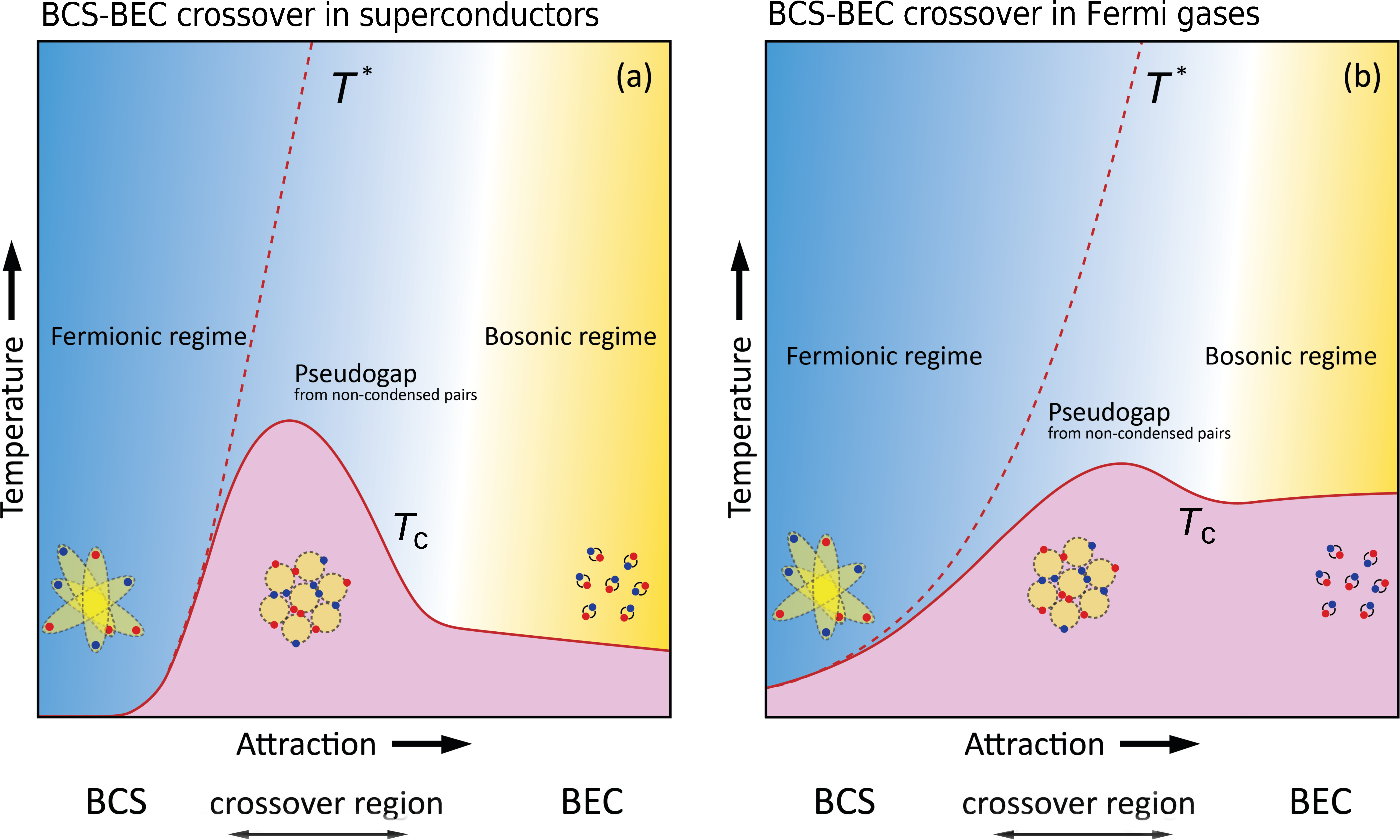}
\caption{Contrasting behavior of the 3D $s$-wave BCS-BEC crossover
  phase diagram for (a) superconductors, as in the attractive Hubbard
  model, and (b) Fermi gases with contact interactions and a free-particle dispersion. Note the contrasting behavior in the BEC regime
  where $T_\text{c}$ approaches either zero (a) or a finite number (b).  Also
  important is the ubiquitous dome shape in the solid-state
  system. The minimum or shoulder in both $T_\text{c}$ curves marks a
  transition to a different physical regime, as it corresponds to the
  onset of a bosonic superfluid, with $\mu = 0$. It is important to emphasize
  here that \textbf{the crossover regime begins at the point where
    the two temperature scales $T^*$ (corresponding to the opening of
    a pairing gap) and $T_\text{c}$ become distinct}.  Microscopic units for
  the superconducting case are provided in Fig.~\ref{fig:10}(a) in a
  later section of the Review.  }
\label{fig:1}
\end{figure*}

We will begin the discussion of BCS-BEC crossover by following the
original discovery papers~\cite{Leggett1980,Eagles1969}, which focus on
a particular choice of ground state, namely that having the form
originally introduced in BCS theory.  While there is a body of
literature on alternative approaches to BCS-BEC crossover in the solid
state, (some of which is reviewed here), we will focus mainly on this
so called ``BCS-Leggett" ground state and its finite-temperature
implications~\cite{Kadanoff1961} rather than on variants that have
ground states that are incompletely characterized and less well understood.

The appreciation of this broader applicability of BCS theory and its
straightforward extension to a form of Bose condensation underlines
how remarkable the original contribution of Bardeen, Cooper, and
Schrieffer was. It should be noted that their discovery has provided
support and a crucial framework for multiple Nobel prizes (of the
order of 10 or so) besides their
own, including nuclear and particle physics. In this way, the
recognition of its even greater generality is particularly significant.

This recognition can be credited to two physicists: A. J. Leggett~\cite{Leggett1980} and D. M. Eagles~\cite{Eagles1969}. Leggett's contribution was motivated by the discovery of a BCS-like triplet-pairing state in the neutral superfluid helium-3. He emphasized that this form of fermionic superfluidity has features that are clearly distinct from conventional superconductors; here the Cooper pairs have complex degrees of freedom. Moreover, the underlying attraction that leads to superconductivity in this neutral system must derive from a distinct pairing mechanism~\cite{Levin1983}.

In making his claims, Leggett pointed to the sweeping generality of the BCS ground state:
\begin{equation}
 \PsiBCS=\Pi_{\vect{k}}
  \left(\uk+\vk a_{\vect{k},\uparrow}^{\dagger} a_{-\vect{k},\downarrow}^{\dagger}\right)|0\rangle,
\label{eq:1}
\end{equation}
where $a_{\vect{k}, \uparrow}^{\dagger} a_{-\vect{k}, \downarrow}^{\dagger}$
creates a pair of fermions with opposite spins and opposite momenta,
${\bf k}$ and $-{\bf k}$, from the vacuum ($|0\rangle$).

The broader applicability of this wavefunction is accessed by self consistently adjusting
the variational parameters $\uk$ and $\vk$ as one varies the strength
of the attractive interaction. This accommodates a continuous
evolution from weak to strong pairing.  One can replace $\uk$ and
$\vk$ by more experimentally relevant parameters: the fermionic
chemical potential $\mu$ and the zero-temperature fermionic
excitation gap parameter $$\Delta_0 \equiv \Delta(T=0).$$
These are two important parameters that we will refer to throughout this Review.  
Notably, the wave function $\PsiBCS$ supports a smooth
transition between a BCS and a BEC-like phase. The former is characterized by a large pair size, a small
$\Delta_0$, and a chemical potential equal to the non-interacting
Fermi energy ($E_\text{F}$). In
this latter case the pair size is small, $\Delta_0$ is large (comparable to or even larger than 
$E_\text{F}$), and $\mu$ is negative.

It should be emphasized that this BEC phase is specific to the
ground-state fermionic wave function and need not represent that of a
true weakly interacting Bose gas.  Importantly, within a generalized
BCS framework it is relatively straightforward to address finite
temperatures above and below $T_\text{c}$ ~\cite{Kadanoff1961}; this is, in
part, a consequence of the fact that the pairing formalism is closely
related to an exactly solvable many-body problem~\cite{Richardson1963}.

In a related way, Eagles~\cite{Eagles1969} also made ground-breaking
observations. He should be credited with emphasizing the concept of
``pairing without superconductivity''. This preformed-pair
normal-state scenario is at the heart of BCS-BEC crossover theory,
once the attraction strength is beyond the BCS regime. He should also be credited
with drawing attention to the possibility that superconductivity in
lightly doped semiconductors can be described by a
form of BCS-BEC crossover. Indeed, we will see in this Review that
there is currently renewed interest in these superconductors with low carrier density.

\subsection{Early theoretical work: Extending BCS-BEC crossover theory to finite temperatures}

In 1985, Nozi\'eres and Schmitt-Rink (NSR) began to think about going
beyond the ground state and including the effects of finite
temperature.  They wrote a famous paper~\cite{Nozieres1985} that
brought attention back to the earlier work by Eagles and Leggett and
presented an in-depth discussion of the ground state given in
Eq.~\eqref{eq:1}. Moreover, they suggested an approach for computing
the transition temperature $T_\text{c}$.
It should be noted, however, that the extrapolated ground state
associated with NSR's finite-temperature theory is
different~\cite{Diener2008} from the expression $\Psi^\text{BCS}$ in
Eq.~\eqref{eq:1}.  Importantly, the NSR paper was the first to
emphasize that BCS-BEC crossover theory in a solid-state lattice
system assumes a character in the strong-coupling BEC regime quite
different from that of a Fermi gas.

The schematic plot in Fig.~\ref{fig:1} relates to this observation. It compares the 
phase diagram for BCS-BEC crossover in (a) a lattice as contrasted with (b) a Fermi gas. A central
difference arises from the kinetic energy degrees of freedom associated with the motion of fermions in solids having a periodic lattice as distinct from their motion in free space.
The most striking consequence is that in a solid, 
$T_\text{c}$  in the BEC regime can become arbitrarily small as the pairing strength increases.
Indeed, we emphasize this distinction in the present Review, as it bears on the relevance (or lack thereof) of the ultracold atomic Fermi gas superfluids to the solid-state superconductors we discuss here.  

Related work in the form of a review was written by Micnas and
co-workers in 1990 ~\cite{Micnas1990} addressing superconductors in
the BEC-like or strong-attraction limit.  In their approach, a local
pairing scenario was adopted, rather like treating a hard core Bose
gas on a lattice. The emphasis was on clarifying the various
alternative phases that compete with superconductivity.
Subsequently, the finite-temperature theory of the NSR paper was
followed by work from S\'a de Melo, Randeria, and
Engelbrecht~\cite{SadeMelo1993}, which provided a
functional-integral reformulation.

Around the same time, and in collaboration with Trivedi and others~\cite{Trivedi1995}, these researchers presented a series of papers using Quantum Monte Carlo (QMC) simulation techniques 
to address normal-state features of the attractive 2D Hubbard model. This was thought to be relevant to high-temperature superconductivity and its anomalous ``pseudogap'' phase. 
This phase corresponds to a ``normal" state above $T_\text{c}$ in which there is a gap for fermionic excitations. 
In their work, it was presumed that
the pseudogap is associated with pairing in the absence of condensation~\footnote{They noted their particular numerics supported the interpretation of the pseudogap (or equivalently a normal-state excitation gap) as a ``spin gap" in which the charge degrees of freedom did not equally participate.}.

The onset temperature for such a normal-state gap is called
$T^*$. Although there are a number of competing explanations,
understanding the origin of this pseudogap, which shows up in
thermodynamics and transport~\cite{Timusk1999}, has been a central
focus in the cuprate field.  We emphasize that the pseudogap, as well
as the distinct temperature scales $T^* \neq T_\text{c}$, play an important
role in general BCS-BEC crossover physics and will be discussed in
more detail throughout this Review article. They are also depicted in the
schematic comparison plot in Fig.~\ref{fig:1}.

\subsection{BCS-BEC crossover in cold-atom experimental research}

Because the cold-atom systems constitute ideal laboratories
for investigating the phenomena of BCS-BEC crossover
(albeit in a Fermi gas),
it is useful next to summarize the groundbreaking achievements beginning around 2003, when 
Fermi condensates in trapped atoms were first reported. 
Condensation was initially observed ~\cite{Greiner2003,Jochim2003} at strong coupling in the BEC regime (where $\mu < 0$) and shortly thereafter~\cite{Regal2004,Zwierlein2004} at intermediate coupling (in a ``unitary'' gas, where the chemical potential was positive). These experiments should be recognized by the solid-state physics community as a true ``tour de force''. Researchers managed to surmount multiple challenges stemming from the fact that the atomic gases are charge neutral, they are confined to inaccessible traps, and moreover, there is no direct way of measuring their temperature.

As a result, in the first few generations of experiments, ``proof'' of superfluidity was established indirectly through magnetic field sweeps. 
These sweeps make use of a Feshbach resonance to take a gas in the more fermionic regime and quickly change the magnetic field thereby projecting the system onto the strong-pairing regime. In this limit, a bimodality in the density profiles of the fermion pairs, 
with a narrow central peak on top of a broad distribution,
reveals the presence of a condensate along with thermally excited pairs. 
Over the next year or two, subsequent experiments made claims for superfluidity through measurements of the specific heat~\cite{Kinast2005} and later it was quite spectacularly established through direct observation of quantized vortices~\cite{Zwierlein2005}.

With increased understanding of these Fermi gas superfluids, the community then focused on additional probes such as transport~\cite{Sommer2011,Joseph2015} and additional complexities associated with spin-imbalanced or polarized gases~\cite{Partridge2006,Zwierlein2006} (very much like superconductors in magnetic fields) as well as in optical lattices~\cite{Chin2006}. Along these lines, there were interesting accompanying theoretical contributions~\cite{Radzihovsky2010,Chien2006} as well as those which contemplated even more exotic (e.g., spin-orbit coupled and topological) phases~\cite{Wu2015,Anderson2015,He2013,Zhang2014}. Also notable were the contrasts with solid-state superconductors centered around low viscosity or ``perfect fluids''~\cite{Kovtun2005,Guo2011a} in the Fermi gases and ``bad metals''~\cite{Gunnarsson2003,Guo2011} associated with highly resistive transport as in the cuprate superconductors.

The collective contribution of the dedicated experimental groups who met the challenge of finding and characterizing these Fermi condensates deserves enormous respect. Among the groups were those of Jin~\cite{Greiner2003,Regal2004,Stewart2008}, Ketterle~\cite{Zwierlein2004,Zwierlein2003}, Grimm~\cite{Jochim2003,Bartenstein2004}, Thomas~\cite{Ohara2002,Kinast2004}, Hulet~\cite{Zhang1999,Strecker2003}, and Salomon~\cite{Bourdel2004}.

Among the first theorists to apply BCS-BEC crossover theory to the cold gases were Y. Ohashi and A. Griffin~\cite{Ohashi2002}
who implemented the theory of Nozieres and Schmitt-Rink ~\cite{Nozieres1985}. This was followed by work from our group~\cite{Stajic2004} which, shortly before the 2003 discovery, called attention to the expected importance of a pseudogap in these cold gases. This, in turn, helped motivate experimental efforts beginning with early observations of possible pseudogap signatures~\cite{Jochim2003} using radio frequency (RF) spectroscopy~\cite{Chin2004}.
Later research by Jin and her colleagues~\cite{Stewart2008} introduced a rather ingenious analogue of angle resolved photoemission spectroscopy (ARPES) to investigate the pseudogap in more detail. These experiments have been revisited more recently by removing some of the trap complications,
using a so-called ``box" trap, where pseudogap effects appear more prominent~\cite{Yao2023}.

In addition to this focus on the pseudogap, substantial effort was devoted to the unitary gas, intermediate between BCS and BEC, where the scattering length becomes infinite. Here precise numbers for thermodynamic features, variables in the equation of state, and special inter-relationships~\cite{Nascimbene2010,Tan2008,Ku2012} provided a series of challenges to test the numerical accuracy of different  BCS-BEC crossover theories.

\subsection{Hamiltonian and interpretation of the ground state wave function}

All discussions of detailed theory will be deferred to later sections of the Review, but for the purposes of an overview we next introduce
the underlying Hamiltonian.  As in all superconductors, it is assumed
that electrons are paired in the superconducting phase. This pairing
arises from an attractive interaction. In strict BCS theory, pairing
takes place only between electrons with opposite momenta
($\veck,- \veck$). More generally, in BCS-BEC crossover theory we
consider pairing between $\veck+\vecq/2$ and $-\veck + \vecq/2$, where the pair-momentum $\vecq$ can be arbitrary, but generally small (compared to $k_F$).
This pairing physics is described by the following Hamiltonian:
\begin{align}
\mathcal{H}  & = \sum_{ \veck \sigma} \epsilon_{ \veck } a^{\dag}_{ \veck \sigma} a^{\ }_{\veck \sigma}  \nonumber \\
& + \sum_{ \veck, \veck^\prime, \vecq } V_{\veck \veck^\prime}
a^{\dag}_{\veck+ \frac{ \vecq }{2}        \uparrow}
a^{\dag}_{ - \veck  +\frac{ \vecq }{2}    \downarrow}
a^{\ }_{ - \veck^\prime +\frac{\vecq}{2} \downarrow}
a^{\ }_{\veck^\prime+\frac{ \vecq}{2}     \uparrow}    \, ,
\label{BCS_H}
\end{align}
where $a^\dagger_{\veck \sigma}$ creates an electron in the momentum
state ${\veck }$ with spin $\sigma $, and $\epsilon_{\veck}$ is the
kinetic energy dispersion.
We assume a separable potential
$V_{\bf kk'} = U \varphi _{\veck}\varphi^{\ }_{\veck^\prime}$, where
$U=-|U|$ is the attractive coupling strength; the momentum-dependent
function $\varphi_{\veck}$ will determine the symmetry of the order
parameter. For a contact potential or on-site interactions, $\phik=1$,
whereas for $d$-wave cuprate superconductors,
$\phik=\cos k_x - \cos k_y$. To avoid this notational complexity here
we will drop $\phik$, and set the volume to unity in free
space. Similarly we choose the lattice constant to be $1$ for the lattice case.

In Eq.~\eqref{BCS_H} we have assumed spin-singlet pairing, which is
relevant for both simple $s$-wave and $d$-wave superconductors.  We do
not make any assumptions throughout this Review about the origin or the
detailed nature of the interaction, other than that it is attractive.
The energy dispersion $\epsilon_{\veck }$ can be associated either
with a lattice or a free gas.  We generally consider only a one-band
model (with the exception of Section~\ref{sec:quantgeometry} where
band topology plays a role), but this Hamiltonian can be extended to
include more bands and a finite range of interaction.  For the
$s$-wave case on a lattice, the interaction $V_{\veck \veck^\prime}$
in Eq.~\eqref{BCS_H}
corresponds to an attractive Hubbard model with on-site
interactions. We have found that the effect of a finite range is
generally not qualitatively important in the context of BCS-BEC
crossover.  In the $d$-wave case, $V_{\veck \veck^\prime}$ is in
general nonlocal in real space and should be regarded as an
approximation to the actual pairing interaction in real materials.

It is important to note that when we refer to finite-$\vecq$ pairing, this does \textit{not} refer to condensed Larkin-
Ovchinnikov~\cite{Larkin1965} or Fulde-Ferrell~\cite{Fulde1964} phases
but rather to non-condensed or thermally excited pair states. These
are to be distinguished from condensed pairs having zero center-of-mass momentum. We emphasize that BCS-BEC crossover
deals with superconductors that have strong pairing or strong ``glue".
This characterizes the interaction term
in the Hamiltonian, where it is assumed that the pairing strength $|U|$
is not small compared to the kinetic energy.  As a result of large
$|U|$, pairing and condensation will take place at different
temperatures.  In particular, at the superconducting transition
temperature $T_\text{c}$ there will be a finite number of non-condensed pairs present.

Note that $\mathcal{H}$ in Eq.~\eqref{BCS_H} is a many-body
Hamiltonian and there are many ways of solving it.  In this Review, and
as in the literature~\cite{Leggett1980}, we base our solution on a
variational ground state of the BCS form that was presented in
Eq.~(\ref{eq:1}).  By contrast with strict BCS theory we allow the
attractive interaction to be arbitrarily strong, assuming this does
not change the generic form of the variational wave function $\PsiBCS$.
We emphasize that $ \PsiBCS$ is not an exact solution of
Eq.~(\ref{BCS_H}), but rather an approximation that presumes that the
system does not make large excursions from BCS theory no matter how
strong the attraction is.  \textbf{Throughout this Review we adopt this
  particular version of BCS-BEC crossover theory and, unless indicated
  otherwise, all equations we present in this review are based on this
  particular ground state and its finite-temperature implications.}

We emphasize that the advantage of this approach to BCS-BEC crossover theory
is that we are dealing with a known ground state. This preserves the
fundamental way superconductivity has come to be understood.  Another
advantage of the BCS wave function is that these Cooper pairs form an
essentially ideal gas.  One can see this from the form of the BCS wave
function of Eq.~\eqref{eq:1}, which can be rewritten as
$\Psi^{\textrm{BCS}} \propto e^{b_0^\dagger } | 0 \rangle$ with the
composite bosonic operator
$b_0^{\dagger}=\sum_{\veck } (v_{\veck } / u_{\veck}) a_{\veck,
  \uparrow}^{\dagger} a_{-\veck, \downarrow}^{\dagger}$. Thus, this
condensate corresponds to a ground state containing bosons that
interact directly with the fermions and only indirectly with each
other ~\cite{Combescot2013,Combescot2017}.  This makes for a simpler
and more solvable many-body problem~\cite{Richardson1963}.

One could contemplate other ground states with a structure different
from the Gaussian-like $\PsiBCS$, in which one has a composite bosonic
operator in the exponent that involves four or more fermionic creation
operators~\cite{Tan2008}. Such approaches can be viewed as more
equivalent to a weakly interacting theory of bosons: Bogoliubov
theory.  But such a more complicated theory is not necessarily an
improvement as Bogoliubov theory for bosons is known to be
inappropriate at temperatures near $T_\text{c}$, or even well above $T=0$, as
it is strictly a low-temperature theory.

Nevertheless, the known weaknesses of the BCS-Leggett approach should
be clarified at this point. In particular, such an approach leads to
inaccuracies in numerical values of thermodynamic parameters
associated with the unitary gas.  One can in part attribute this to
the approximate treatment of the particle-hole channel for BCS-based
theories, which focus primarily on the particle-particle channel.
This is evident, for example, through the Bertsch parameter appearing
as the ground state fermionic chemical potential ratio, $ \mu/E_\text{F} $,
of the unitary Fermi gas.  This is found experimentally~\cite{Ku2012}
to be around $0.37$, whereas in the BCS ground state this parameter is
equal to $0.59$~\cite{Viverit2004}.

\subsection{Kadanoff and Martin interpretation: BCS theory as a Bose condensation of electron pairs}

Knowing the ground state still leaves the challenge of how to
introduce finite-temperature effects.  At this stage, to gain further
physical insight into BCS-BEC crossover theory, it is useful first to
revisit an approach due to Schafroth~\cite{Schafroth1955}.  Two years
before the BCS ground state of Eq.~\eqref{eq:1} was ever proposed,
Schafroth suggested a more expanded interpretation of
superconductivity. He argued that superconductivity could be thought
of as being associated with Bose condensation of an ideal charged Bose gas.
While most in the community view his scheme as appropriate to the
extreme BEC, often called the ``local pair limit", here we wish to
think about this approach to fermionic superconductivity more
generally, for all systems beyond the strict BCS limit.

Schafroth argued that condensation sets in at the transition
temperature $T_\text{c}$, where there are preformed electron pairs. The
expression for this temperature, following that of an ideal Bose gas,
is given by:
\begin{equation}
T_\text{c} = \left(\frac{2\pi}{\mathcal{C}}\right)
  \frac{n_\text{B}^{2/3}(T_\text{c})} { M_\text{B}(T_\text{c})} ,
\label{eq:2}
\end{equation}
(where $\mathcal{C} =\left[\zeta(3/2)\right]^{2/3}$ with the Riemann
zeta function $\zeta(3/2)\approx2.612$. Throughout this Review we set
$\hbar=k_\text{B}=1$, unless indicated otherwise.) The parameters
$n_\text{B}$ and $M_\text{B}$ represent the (3D) number density and
mass of the bosons.
We should view these as yet unspecified bosons as representing
fermion pair degrees of freedom so that
\begin{equation}
n_\text{B} \equiv n_\pair \quad \mathrm{and} \quad M_\text{B} \equiv M_\pair.
\label{eq:3}
\end{equation} 
Note that, at the time of the BCS discovery, there was some resistance
to Schafroth's notion that his approach had anything in common with
BCS theory. The key point that Schafroth emphasized is that there
must be a form of Bose condensation embedded in superconductivity
theory and this boson inevitably involves a pair of electrons.

Schafroth's work introduces an important question: what kind of
out-of-condensate boson or preformed pair is in fact compatible with
BCS theory?  The answer to this query would allow us to compute the
transition temperature, after establishing a precise meaning for
$n_\pair$ and $M_\pair$. Presumably because his work predated BCS
theory, Schafroth did not ascribe any complexity to these quantities,
which we now think must depend on both temperature and attractive
interaction strength. Importantly, because of the latter, we
inevitably have to deal with BCS-BEC crossover physics.

The challenge to quantitatively characterize these out-of-condensate
pairs at general temperatures $T$ was met in an important paper by
Kadanoff and Martin~\cite{Kadanoff1961}. Just as Eagles
~\cite{Eagles1969} and Leggett ~\cite{Leggett1980} recognized the
greater generality of the BCS ground-state wave function, Kadanoff and
Martin provided key insights into the finite-temperature physics of
BCS theory.  Their work was based on a systematic study of the coupled
equations of motion. This established how to characterize the
non-condensed pairs associated with BCS theory (through their
propagator or ``t-matrix").

Kadanoff and Martin made an important observation that related to the
Schafroth picture. They stated that ``\textit{Below [the transition]
  temperature... a nonperturbative, stable solution involving a Bose
  condensation of pairs can be derived within the pair correlation
  approximation..  which [approximation] is identical with the one
  proposed by BCS. .... that the superconducting transition is a Bose
  condensation phenomenon [was] originally proposed by Schafroth [and
  co-workers]."}

From their work, one infers that the BCS gap equation can be
reinterpreted as a BEC condition requiring that the non-condensed
pairs have zero chemical potential (that is, are gapless) at every
$T \leq T_\text{c}$. This Hugenholtz-Pines constraint~\cite{Hugenholtz1959}
is a generalization, as well, of the familiar Thouless
condition~\cite{Thouless1960}.  While in strict BCS theory, all
preformed pairs at the onset of the superconducting transition should
be viewed as virtual, it is reasonable to presume that once one enters
the BCS-BEC crossover regime, these non-condensed pairs are no longer virtual 
and their number and mass at general $T$ can be quantified
according to the prescription of Kadanoff and Martin.

The work we summarize here should be differentiated from other
approaches to BCS-BEC crossover, such as that of Nozi\'eres and
Schmitt-Rink and others
~\cite{Nozieres1985,SadeMelo1993,Pieri2004,Ohashi2002}. Their
finite-temperature analysis was presumably designed to accommodate
some of the physics of bosonic Bogoliubov theory for the fermion
pairs. In the NSR picture, which involves more strongly interacting
composite bosons than would be associated with a BCS-like ground
state, the bosonic degrees of freedom are
described~\cite{Nozieres1985} as: ``\textit{A bound pair [which] is a
  collective mode of the superfluid \dots $T_\text{c}$ thus results from
  thermal excitation of collective modes''}.  Their scenario can be
compared with other work~\cite{Tan2006,Pieri2005} that addresses the
extreme BEC regime and investigates the nature of that fermionic
ground-state wave function associated with a composite-boson
Bogoliubov picture (including Lee-Huang-Yang corrections).

\subsection{Mechanisms for driving BCS-BEC crossover}
\label{sec:IE}

An important aim of this Review is to communicate in physical terms
what BCS-BEC crossover is and what it is not.  More specifically we
ask: how do we know when a superconductor is promoted out of the BCS
regime and what are typical mechanisms for promoting it?

It is useful to establish the variables that quantify the size of the
deviation from BCS.  One of these, the ratio $T^*/T_\text{c}$, has already
emerged.  When this ratio exceeds unity the superconductor may no
longer be in the BCS regime.  Here, as defined previously, $T^*$
corresponds to that temperature at which a gap opens in the fermionic
excitation spectrum, while $T_\text{c}$ corresponds to the temperature for
fermion pair condensation.  Strong pairing is not uniquely implied by large
$T^*/T_\text{c}$, but the converse, however, is true. Notably
there can be other mechanisms for this spectral gap
opening.

By contrast the presence of a large ratio of the zero-temperature gap to $E_\text{F}$, $\Delta_0/E_\text{F}$, is more unambiguously suggestive of a system
that has been promoted out of the BCS regime.  Finally there is a
third, equally important parameter that quantifies the deviation from
BCS theory. This corresponds to the size of the Ginzburg-Landau (GL)
coherence length, which we define more precisely later in this
subsection.  When this is anomalously small, the system may be driven
away from the BCS regime.

What then are the mechanisms that are responsible for driving a
superconductor out of the BCS regime and into the BCS-BEC crossover regime?  We identify 3 main mechanisms:
low dimensionality, strong attraction, and low electronic energy
scales.

We begin with the issue of low dimensionality, which is known to
naturally introduce distinct energy scales $T^*$ and $T_{\BKT}$.
Notably, as stated by Kosterlitz~\cite{Kosterlitz2016} ``\textit{The
  onset of superconductivity in 2D \dots requires a pre-existing
  condensate or pairing of electrons.''} One can understand this by
noting that the underlying physical picture characterizing the onset
of two dimensional superconductivity (or the
Berezinskii-Kosterlitz-Thouless (BKT) superconducting
state~\cite{Kosterlitz1973,Berezinskii1972}) assumes the separation of
energy scales: phase coherence cannot occur until a pairing amplitude
is established.

An equally important aspect of superconductivity in 2D is that there
is a stronger tendency to pair.  In particular, in the low density limit
where there is a quadratic band dispersion near the conduction band
bottom, it follows that there is no critical value of the pairing
interaction that is required to form two-body bound states.  This is
in contrast to the situation in 3D. Hence the ``pairing glue'' in a 2D
superconductor need not be anomalously strong to promote the system
into the BCS-BEC crossover regime.  These observations may explain why
there are many 2D examples in the recent BCS-BEC crossover literature.

\begin{figure}
\centering
\includegraphics[width=3.2in]
{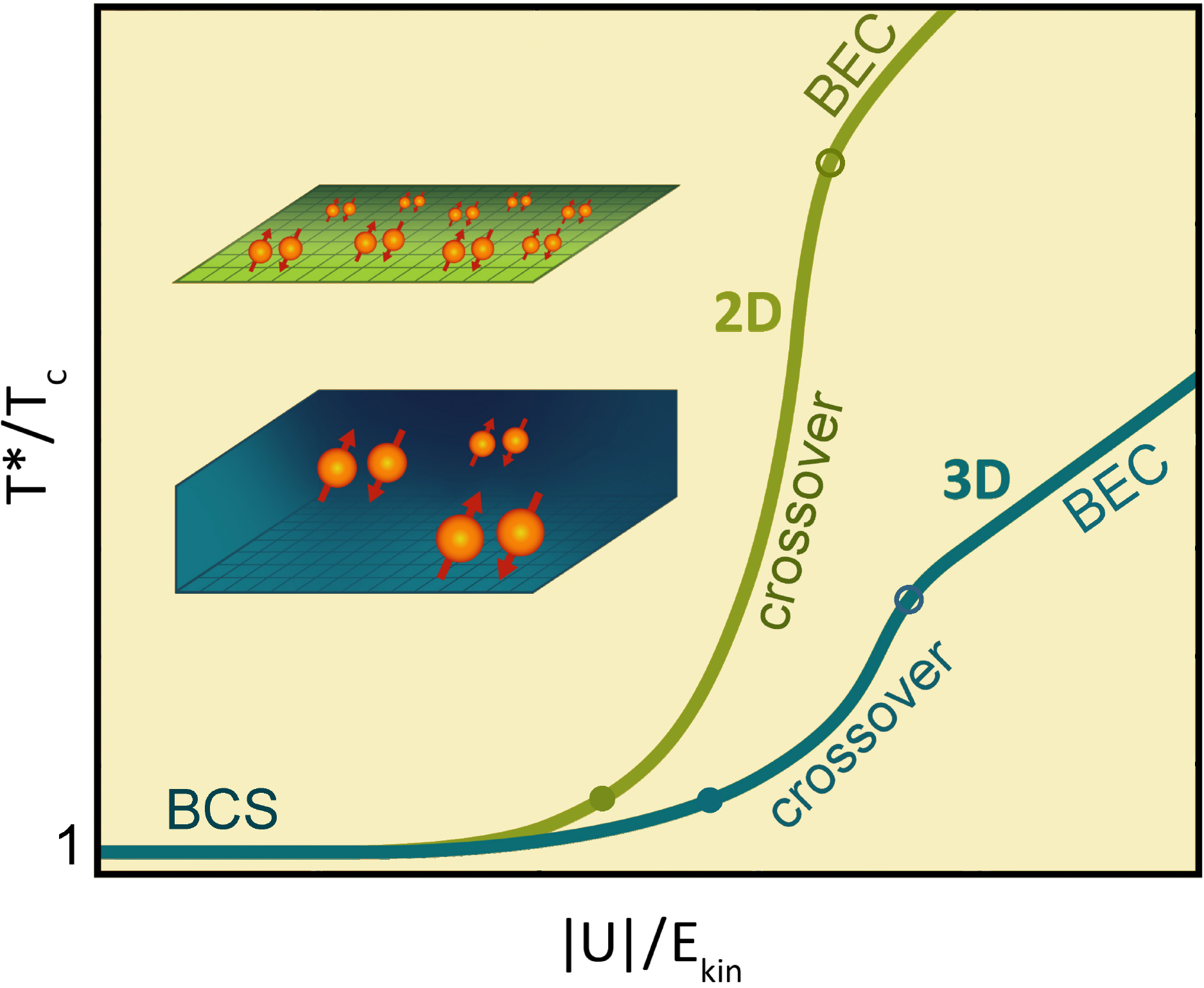}
\caption{Comparison of schematic phase diagrams in 2D and 3D for an
  attractive Hubbard model, based on plots of $T^*/T_\text{c}$ as a
  function of the dimensionless attractive interaction
  $|U| / E_\text{kin}$.  The onset of the departure from the BCS into
  the BCS-BEC crossover regime is determined from the point where the
  ratio $T^*/T_\text{c}$ slightly exceeds unity, as shown by the solid
  circles.  Thus, a relatively weaker attraction $|U| / E_\text{kin}$
  is sufficient to promote a 2D superconductor out of the strict BCS
  limit, as compared with 3D.  Reaching either of these onset values
  for $|U| / E_\text{kin}$ (\textit{i.e.,} the solid circles) can be
  achieved by increasing the attraction $|U|$, or decreasing the
  electronic energy scales $E_\text{kin}$.  The two insets represent
  schematically the number of pairs (or pair density) in the 2D sheets
  or 3D volumes at these onsets. Also shown is the transition to the
  BEC regime, indicated by open circles. For actual units on this
  figure see the inset in Fig.~\ref{fig:10}(a).  }
\label{fig:2}
\end{figure}

Figure~\ref{fig:2} provides a key summary of different mechanisms for
promoting a system out of the BCS regime. This figure quantifies the
values of the attractive interaction at which a given 2D or 3D superconductor
departs from BCS theory and enters into the BCS-BEC crossover regime as well as where it enters
into the BEC regime. Plotted on the vertical axis is $T^*/T_\text{c}$ (or
for the two-dimensional system $T^*/T_\BKT$).  The horizontal axis
indicates the strength of the dimensionless attractive interaction in
units of a characteristic electronic energy scale $E_\text{kin}$.

A key observation from this figure
is that a relatively weaker attraction $|U|/ E_\text{kin}$
is needed to
promote a 2D superconductor as compared with a 3D
superconductor out of the strict BCS limit. The point of departure from
the BCS regime is associated with the point where $T^*/T_\text{c}$
slightly exceeds unity, say, by 20\%. The figure is characteristic of the
intermediate- and low-carrier-density regimes.
The corresponding values of
$|U| / E_\text{kin}$
are indicated in the figure by the solid dots. 
This fraction can assume sufficiently large values as a consequence of a
very strong pairing ``glue'', i.e., associated with anomalously large
$|U|$.  We might speculate that this stronger pairing scenario applies, if
at all, to the cuprate superconductors. But the ratio can also be large
when the
characteristic electronic energy scales (called $E_\text{kin}$) become
anomalously small. This can occur through flat bands (because of small
hopping integral, called $t$,
or small bandwidth) or through low electronic densities (which reduce
$E_\text{F}$). We will see in this Review that both two dimensionality and/or
small electronic energy scales are likely responsible for the many
recent observations of BCS-BEC crossover superconductivity.

The fact that there is no critical value of the pairing required to
form bound states in a moderately low density 2D superconductor also
serves to interpret the illustrations to the left of the curves in
Fig.~\ref{fig:2}. These are schematic representations of the number of
pairs (or pair density, $n_\pair$) in the 2D sheets or 3D volumes at
the onset of the transition. For the same fixed attractive
interaction, these schematic figures emphasize that in 2D there is a
significantly higher density of pairs at $T_\BKT$ than for the
analogue in the 3D system.

We end this discussion by referring back to the GL coherence length
and showing that it provides a quantifiable measure of where a superconductor is within
the BCS-BEC crossover spectrum. This is based on a calculation of $T_{\BKT}$
rather similar to the Schafroth-like result in Eq.~(\ref{eq:2}) but
here for the 2D limit. This analysis is abbreviated here, by way of a
summary, and later discussed in more detail in Sec.~\ref{sec:twoD}.

We approach the BKT state from the high-temperature side and, thus,
will use the methodology advocated by the cold-atom
community~\cite{Jose2013,Hadzibabic2006,Prokofev2002}, where in atomic
Bose gases one finds some of the most convincing evidence for a
Kosterlitz-Thouless state.  Although originally much of this
literature was focused on BKT for bosonic superfluids, by extension to
fermionic superconductors and superfluids, one can deduce that this
transition temperature roughly scales as~\footnote{The proportionality
  constant between $T_\BKT$ and $n_\text{B}/M_\text{B}$ in
  Eq.~\eqref{eq:4} has an additional double-logarithmic
  dependence~\cite{Fisher1988} on $n_\text{B}$, which is very weak. }
\begin{equation}
T_\BKT \sim
  \frac{n_\text{B}(T_\BKT)} {M_\text{B}(T_\BKT)}\,,
\label{eq:4}
\end{equation}
where, again, these as yet unspecified bosons with (2D) number density
$n_\text{B}$ and mass $M_\text{B}$ represent pair degrees of freedom
as defined in Eq.~\eqref{eq:3}. It is important to note that a
fraction involving the same temperature-dependent terms $n_\pair(T)$
and $M_\pair(T)$ enters in both the 2D and 3D expressions for the
transition temperature. Here the omitted prefactor represents a
slightly more complicated term that will be discussed later in the
context of Eq.~(\ref{eq:21}) below.

These Schafroth-like expressions for the transition temperatures in 2D
and 3D (Eqs.~(\ref{eq:2}) and (\ref{eq:4})) then provide a simple
expression for the important superconducting GL coherence
length, $\xi_0^{\text{coh}}$; this is given by
~\cite{Boyack2019,Boyack2018}
$\hbar^2/[2M_\pair (\xi_0^{\text{coh}} )^2]= k_\text{B}T_\text{c}$, where we have
restored the Planck constant $\hbar$ and Boltzmann constant $k_\text{B}$. As
a result $\xi_0^{\text{coh}}$ depends only on the pair density
$n_\pair$ (presumed at the onset of the transition). Importantly this
coherence length reveals the location of a given system within the
BCS-BEC crossover:
\begin{equation}
k_\text{F} \xi_0^{\text{coh}} \propto (n/n_\text{pair})^{1/2}
\nonumber
\end{equation}
for the 2D case. Here $k_\text{F}$ reflects the total particle
density, $n$, and a similar expression (with the exponent of $1/3$)
can be obtained in the 3D case as well.  Since the number of pairs at
$T_\text{c}$ varies from essentially $ 0$ in the BCS limit to $n/2$ in the BEC
case, this provides a measure of where a given superconductor is
within the BCS-BEC crossover spectrum.  
Fortunately, this GL coherence length is rather widely discussed in
the experimental literature on superconductivity~\cite{Park2021,Suzuki1991,Nakagawa2021,Suzuki2022},
as it is accessible through the response to a magnetic field.
Because it has been measured in a large number of
systems which are viewed as candidates for BCS-BEC crossover, it will be addressed in some detail in this Review.

\section{Overview of BCS-BEC Crossover}

\subsection{Signatures of BCS-BEC crossover}
\label{sec:signatures}

Since the concept of BCS-BEC crossover is sometimes interpreted in
different ways in the literature it is important to emphasize what we
associate with the term ``crossover" in this Review. We consider here
solid-state superconductors (as distinct from atomic Fermi gases)
that are promoted out of the strict BCS regime through moderately
strong pairing interactions (or through a combination of the
mechanisms discussed in Section~\ref{sec:IE}). These interactions, in
turn, lead to emerging bosonic degrees of freedom which coexist with a
well-defined Fermi surface.  With ever increasing interaction strength,
the bosonic component will eventually become dominant leading to a
disappearance of the fermiology; here the system enters the BEC
regime. It is still an open question whether a BEC phase (with its
attendant very low transition temperatures) has ever been observed in
a solid-state system. While some researchers~\cite{Sous2023} have identified
crossover with the onset of the BEC regime, in this Review we adhere to
the conventional definition of ``BCS-BEC crossover'' emphasizing the
associated new and interesting properties, which are distinct from
those observed in either the BEC or BCS regime.

There are a number of signatures of BCS-BEC crossover, some of which
we discussed in the previous section and which we more precisely
quantify here.  Many of these features can have multiple
interpretations.  While the first three criteria in the list below are
necessary conditions, a conclusion in support of the appropriateness
of a BCS-BEC crossover for a particular superconductor often comes
from the preponderance of evidence, rather than from any ``smoking
gun'', single signature in this list.  One observes:
\begin{enumerate}
\item Large values of the normalized zero-temperature pairing gap
  $\Delta_0/E_\text{F}$, from $\approx 0.1- 1.0$.
\item The presence of a normal-state gap (or pseudogap) with onset at
  $T^*/ T_\text{c} \gtrsim 1.2$.
\item A moderately short coherence length that should be no longer
  than $k_\text{F} \xi_0^{\text{coh}} \sim 30$.
\item Enhanced superconducting fluctuation-like behavior, particularly
  in the response to a magnetic field (such as the Nernst effect and
  diamagnetic susceptibility), well above $T_\text{c}$.
\item A precursor downturn~\cite{Timusk1999,Boyack2021} in the
  temperature dependence of the resistivity around the gap onset
  temperature $T^*$.
\item The presence of bosonic (or pair) degrees of freedom above the
  transition.  The pairing gap and the bosonic degrees of freedom are
  indeed two sides of the same coin, although the latter aspect is more
  difficult to identify.
\item BCS mean-field-like relations that characterize the ratio of
  the ground-state excitation gap, $\Delta_0$, and the pairing onset
  temperature, $T^*$.
\item Two distinct energy gaps. In contrast to strict BCS theory, in
  the crossover regime, the gap associated with coherent
  superconducting phenomena which set in at $T_\text{c}$ is distinct from
  that associated with bosonic or pair excitations, which appear in the
  vicinity of $T^*$.
\item Normal-state experimental observations such as shot noise
  ~\cite{Zhou2019}, which are indications of $2e$ charge carriers.
\item The observation of BCS-like ``back bending''
~\cite{Kanigel2008}
 of the electronic band dispersion in the vicinity of but above $T_\text{c}$.
\end{enumerate}

\begin{figure*}
\centering
\includegraphics[width=5.in]{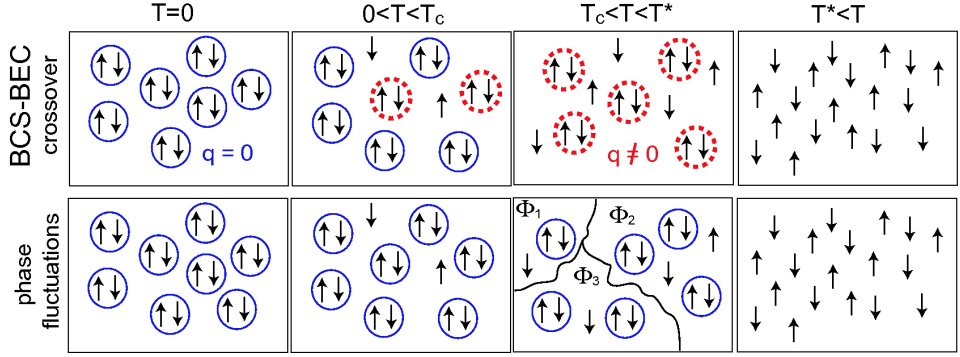}
\caption{Illustration comparing the 3D BCS-BEC crossover and
  phase-fluctuation scenarios. Throughout, blue closed circles, lone
  arrows, and dashed red circles represent condensed fermion pairs,
  unpaired fermions, and finite-momentum pairs, respectively. The
  crossover theory is distinguished by the presence of noncondensed
  pairs, whose center of mass momentum $\textbf{q}\neq 0$, for nonzero
  temperatures below $T^*$. The defining feature of the
  phase-fluctuation picture is the presence of different phase domains
  above $T_\text{c}$, indicated by the regions labeled with distinct phases
  $\Phi_{i}$.}
\label{fig:3}
\end{figure*}

\subsection{Analogies with an ideal Bose gas}

What is essential is that the treatment of BCS-BEC crossover, which
we present here, be compatible with generalized-BCS physics, both in
the ground state as well as at all temperatures $ T \leq T_\text{c}$.  Unlike
in strict BCS theory, in the crossover regime, bosonic degrees of
freedom or preformed pairs will be present already at the onset of condensation.
Their number progressively increases as the system evolves from BCS to
BEC. These normal-state pairs are associated with an excitation gap (or
``pseudogap") in the fermionic spectrum and in BCS-BEC crossover this
implies $\Delta(T_\text{c}) \neq 0$.
The gap size increases continuously starting at nearly $ 0$ in the BCS
regime.  The excited pair states involve a combination of two fermions
associated with momenta $\vect{k}+ \vect{q}/2$ and
$\vect{-k}+\vect{q}/2$ where, specifically, the pair momentum
$\vect{q}$ is non-zero.  Preformed pairs are necessarily distinct from
condensed pairs, for which $\vect{q} = 0$.

To understand these preformed pairs we present a simple figure based
on a rather close analogy to an ideal Bose gas.  The upper row of
Fig.~\ref{fig:3} is a schematic representation of the temperature
evolution of a BCS-BEC crossover superfluid. This shows that as
temperature decreases below an onset temperature $T^*$,
a new form of quasiparticle or excitation appears. These
non-condensed pairs are represented by dashed circles in red. At this
same temperature a pairing gap or pseudogap is present, which reflects
the fact that there must be an input of energy to create fermionic
excitations by breaking pairs. As temperature further decreases to
just above $T_\text{c}$, the number of these preformed pairs increases. Note
that, the figure shows that there are also a number of unpaired
fermions at the transition. The ratio of the boson to fermion number
continuously increases from BCS to BEC. In the BCS limit the number of
pairs at $T_\text{c}$ is essentially zero, while in the BEC limit this number
approaches $n/2$.

Below $T_\text{c}$, condensed pairs (solid circles in blue) appear. As
temperature is lowered further, non-condensed pairs gradually, (and at
$T=0$ completely), convert to the condensate. There are no
non-condensed pairs in the BCS-like ground state. Importantly, strict
BCS theory is the special case where $T^* = T_\text{c}$ and concomitantly
where the number of non-condensed bosons becomes arbitrarily small at
any temperature $T$. This signals that there is essentially no
pairing-related gap in the fermionic excitation spectrum at $T_\text{c}$.

\subsection{Contrasting the present pair-fluctuation and phase-fluctuation scenarios}
\label{sec:EmeryKivelson}

We emphasize that this pair-fluctuation picture of BCS-BEC crossover
is not the same as the phase-fluctuation scenario~\cite{Emery1995}.
There are similarities, but the contrast has been stressed
previously by Emery and Kivelson~\cite{Emery1995}, who describe the
phase-fluctuation scenario as follows: ``\textit{Our discussion attributes the properties of high-temperature superconductors to the 
low superfluid density \dots and not to a short in-plane coherence length and a crossover to real-space pairing}''.

The most significant differences would appear, then, to be attached to
the driving mechanisms (small superfluid density versus strong attraction)
behind the observed exotic normal states, as well as the pair ``size"
or in-plane coherence length. This can help experimentalists
distinguish between the so-called phase-fluctuation picture and
BCS-BEC crossover. A small coherence length or the observation of
concomitant, moderately large $\Delta_0/E_\text{F}$ similarly lends support
to the crossover scenario.

To compare these two scenarios we turn back to Fig.~\ref{fig:3}. In
this figure, the pair-fluctuation or BCS-BEC crossover picture in the
upper panel is to be associated with a new type of paired
quasi-particle (excited pair states) whereas the phase fluctuation
scenario in the lower panel relates to more collective behavior. In
this collective behavior, low carrier density is associated with poor
screening, which is then responsible for small phase stiffness. 
As a further point of contrast, it should be emphasized that all parameters
pertaining to the fermionic sector ($\Delta_0$, $T^*$, etc.) are
essentially absent in the phase-fluctuation scenario, as this theory
is an effective low-energy description of the bosonic degrees of
freedom once the fermions are integrated out.

At the same time, the deep BEC limit of the BCS-BEC crossover
scenario, where the fermions are essentially absent at $T_\text{c}$, will
have features in common with the phase-fluctuation scenario.
Similarly in 2D, where fluctuation effects become more pronounced, the
differences between the two approaches become more subtle, despite the
fact that this bosonic regime is driven by strong pairing ``glue"
rather than low carrier density. 

Finally, we emphasize that phase fluctuations themselves will be
present in the (usually narrow) critical region of temperatures near
$T_\text{c}$ in all superconductors, once one includes beyond-mean-field
effects, which are not addressed in this Review.

\subsection{Quantitative summary of the present theory}
\label{sec:Quant2}

It should not be surprising that accompanying the two forms of (red,
blue) quasi-particles in the upper panel of Fig.~\ref{fig:3} are two
different forms of fermionic excitation gaps: $\Delta_{\pg}$ and
$\Delta_{\sc}$. One can think of these as representing the
contributions from non-condensed and condensed pairs, respectively.
Indeed, their squares will turn out to be proportional to the number density of
these two types of pairs. 

A more detailed theory~\cite{Chen2005}, discussed in
Sec.~\ref{sec:details}, reveals that the gaps combine approximately in
quadrature in such a way as to yield the total, physically measurable
fermionic excitation gap called $\Delta(T)$. Thus
\begin{equation}
\Delta^2(T) = \Delta_{\sc}^2(T) + \Delta_{\pg}^2(T).
\label{eq:5}
\end{equation}
In this way, the total number density of pairs
which is proportional to $\Delta^2(T)$  will determine the energy that must be applied in
order to excite fermions.

A central consequence of this picture to be established below is that
\begin{equation}
\Delta^2(T) = \Delta_\text{BCS}^2(T)  \quad \textrm{for}  \quad  T \leq T_\text{c},
\label{eq:6}
\end{equation}
where $\Delta_\text{BCS}$ is the mean-field gap obtained in BCS
theory.  In this way, in the ordered phase, the total fermionic
excitation gap coincides with the results of strict mean-field BCS theory.

\begin{figure}
\centering
\includegraphics[width=2.8in,clip]{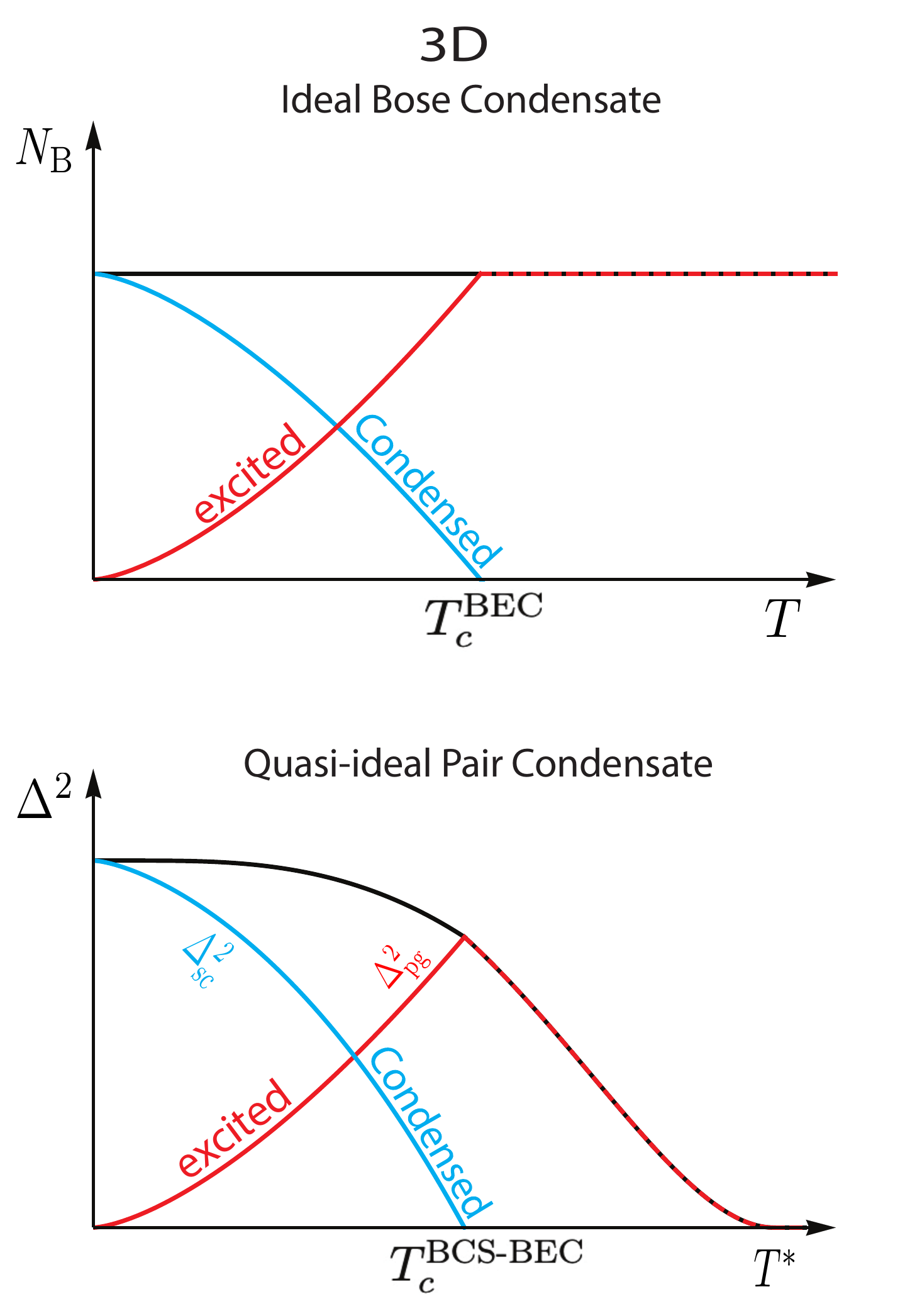}
\caption{Comparison of ideal-gas decomposition of the boson number,
  $N_\text{B}$, into condensed and excited contributions (upper panel)
  with the analogue decomposition for a fermionic superfluid (lower
  panel) which involves the square of the pairing gap $\Delta^2$ as a
  function of temperature $T$. $T_\text{c}^{\text{BEC}}$ and
  $T_\text{c}^{\text{BCS-BEC}}$ are the respective transition
  temperatures. This figure shows that the two gap contributions to
  $\Delta^2$, called $\Delta_{\pg}^2$ and $\Delta_{\sc}^2$, are
  closely analogous to their counterparts in the ideal Bose
  gas. Indicated schematically is how to arrive at the respective
  transition temperatures associated with the intersection of the
  ``excited" curve with either the total boson number curve (black
  line in the top panel) or the total $\Delta^2$ curve (black line in
  the bottom), which marks the onset of the condensate
  contribution.}
\label{fig:4}
\end{figure}

As shown in Fig.~\ref{fig:4}, the two contributions to $\Delta^2$,
called $\Delta_{\pg}^2$ and $\Delta_{\sc}^2$, play a similar role to
their respective counterparts in the ideal-Bose-gas scenario. This
latter theory considers a decomposition of the total number of bosonic
particles, $N_\text{B}$, in terms of those deriving from the excited
bosons $N^{\excited}$ and the condensed bosons $N^{\cond}$. As a
function of decreasing temperature, the former convert to the latter so
that there are no excitations in the ground state. The
temperature-dependent quantity $N^{\cond}$ is established by
evaluating the difference $N_\text{B} - N^{\excited}$.

In the crossover picture, as in an ideal Bose gas, the condensate contribution
$\Delta_{\sc}^2$ is obtained by subtracting the non-condensate piece
$\Delta_{\pg}^2$ from the total $\Delta^2$, approximated as
$\Delta_\text{BCS}^2(T)$ near but above $T_\text{c}$. This determines $T_\text{c}$
from the condition that the non-condensed contribution is no longer
sufficiently large to accommodate the full value of the mean-field gap
squared. Thus, there must be an additional contribution from the
condensate, $\Delta_{\sc}^2$. 

In this way, not only can one directly derive the
Schafroth expression~\cite{Schafroth1955} shown in Eq.~\eqref{eq:2},
but one can write this same equation in a more familiar way
from the perspective of BCS theory. In strict BCS theory, $T_\text{c}$ is
obtained from
\begin{equation} 
1 = \left. (- U) \sum_{\vect k} \frac{1 - 2f(|\xik|)} {2 |\xik |} \right|_{T=T_\text{c}},
\label{eq:7}
\end{equation}
where $U<0$ and $f(x)=1/(e^{x/T} +1)$ is the Fermi-Dirac distribution
function.  Here $\xi_{\vect k} = \epsilon_{\vect k} -\mu$ is the bare
fermion dispersion measured from the Fermi level.
It will be shown that, in the present BCS-BEC crossover theory, we
have a similar expression for the determination of $T_\text{c}$:
\begin{equation} 
1= (- U) \sum_{\vect k} \frac{1 - 2f(\widetilde{E}_{\vect k})}{2 \widetilde{E}_{\vect k}} \bigg\vert_{T=T_\text{c}},
\label{eq:8}
\end{equation}
where $\widetilde{E}_{\vect k} \equiv \sqrt{\xi_{\vect k}^2 + \Delta^2(T_\text{c})}$.

Thus, the central change from strict BCS theory (aside from a
self-consistent readjustment of the fermionic chemical
potential~\cite{Leggett1980}) is that $T_c$ \textbf{is determined in
  the presence of a finite excitation gap}, $\Delta(T_{c})$. Solving for
$T_\text{c}$ involves finding the point of separation between
$\Delta_{\pg}^2(T)$ and the mean-field gap $\Delta_\text{BCS}^2(T)$ as
a function of decreasing temperature, as shown in the bottom panel of
Fig.~\ref{fig:4}.

We now have two different equations, Eq.~\eqref{eq:8} and the
Schafroth expression in Eq.~\eqref{eq:2}, both of which determine the
transition temperature in the BCS-BEC crossover theory, and both are
intuitively quite reasonable. What is satisfying is to find that these
two equations are equivalent, provided one properly computes the
number of pairs and their mass. Thus, this meets the goal of
connecting a Schafroth-like approach to a more microscopic approach
based on BCS theory.  Schafroth's expression for $T_\text{c}$ in this
extended form is appropriate throughout the crossover, once the system
has emerged from the BCS limit so that $\Delta(T_\text{c})$ is no
longer strictly zero.

\subsection{Qualitative summary of BCS-BEC crossover}
\label{sec:Lexicon}

Before going into more technical details of the present BCS-BEC
crossover theory, as will be addressed in Section~\ref{sec:details},
we now consider some of the more obvious questions that can be raised
at this point.  One of the first issues that arises is to clarify
what is generic about BCS-BEC crossover theories.  We note that
BCS-BEC crossover theory belongs to the class of theories of strong-coupling superconductors. 
While there are a number of others in this class, what is essential is that this particular form of strong-coupling superconductivity is driven by charge $2e$ Cooper pairing. 
This differs from some of the alternative types of strongly
correlated superconductors: spinon-holon pairing~\cite{Lee2006},
kinetic energy driven superconductivity~\cite{Leggett1996},
superconductivity strongly coupled to antiferromagnetism
(``SO(5)")~\cite{Demler2004} and fractionalized electron
superconductivity~\cite{Senthil2000}. 

Moreover, within the BCS-BEC crossover class there are a number of
variants, some of which will be briefly reviewed in
Sec.~\ref{sec:alternative}.  Generically, a BCS-BEC crossover theory of
superconductivity represents an interpolation scheme between weak and
strong-coupling forms of $2e$-pairing-governed superconductivity.  In
the weak-coupling limit the fermions within a pair are very loosely
associated whereas in the strong-coupling limit they become tightly
bound. In between the two extremes, there is generally a smooth
crossover.  In all theories of the BEC regime in a lattice, the
fermionic chemical potential lies below the bottom of the
(non-interacting) conduction band.  These generic features are
illustrated in Fig.~\ref{fig:1}(a) which indicates how the transition
temperature and pairing onset temperatures smoothly vary between the
fermionic and bosonic regimes.

There are, however, a number of features that are \textit{not}
generic in the family of BCS-BEC crossover theories.  For example, not
all theories reproduce BCS theory in the weak-coupling limit. 
Indeed, even the ``BEC" limit has many different interpretations. 
Some would argue that the BEC limit should be that of a true weakly interacting
Bose system. Alternatively, in the present theory it is argued to be
distinctly different as this state is characterized through its
fermionic properties, even though a Fermi surface is no longer
present.  In such a BEC limit, for example, the fermionic pairing gap
parameter is large and temperature independent well above and below
$T_\text{c}$.  Among other features that are \textit{not} generic is the
presence in the intermediate-coupling regime of a pseudogap, which is
indicated in Fig.~\ref{fig:1}.  This pseudogap appears in some
crossover theories~\cite{Strinati2018}, but not in others
~\cite{Haussmann2009,Morawetz2011}.

More precisely, the pseudogap corresponds to a gap in the fermionic
excitation spectrum, which has a smooth onset at $T^*>T_\text{c}$.  The
pseudogap we consider here enters into the theoretical framework as a
distinct parameter $\Delta_{\pg}$ and is more
apparent~\cite{Levin2010}; in other approaches~\cite{Strinati2018} it
is only indirectly seen to be present through the behavior of the
fermionic spectral function. It reflects the fact that electrons are starting to pair up at $T^*$
and that breaking the pairs in order to create fermions will cost a
(gap) energy. There is no true ordering or broken symmetry that takes
place at $T^*$, only the onset of bosonic (pair) degrees of
freedom. Because of the pseudogap, superconductivity at $T_\text{c}$ will
occur in the presence of a finite fermionic excitation gap
$\Delta(T_\text{c})$.

Additionally, we argue that these pseudogap effects persist below
$T_\text{c}$ as they reflect the contribution of non-condensed pairs which
are continuously converting to the condensate as temperature is
lowered towards the ground state.  Below $T_\text{c}$ there is the additional
energy gap deriving from the order parameter, $\Delta_{\sc}$.  It is
often difficult to disentangle these two gap parameters, which reflect
the energies that must be input to break the non-condensed and
condensed pairs, and for many purposes they contribute additively in
quadrature.  Importantly, the pseudogap is not associated with
superconducting coherence and is not responsible for Meissner or
Josephson effects.

More concretely, this energy gap appears in both the charge and spin
channels and more generally in thermodynamics and transport in many
respects similar to the way the below-$T_\text{c}$ superconducting gap shows
up in BCS theory.  It enters, however, as a slightly rounded or
smeared gap structure in normal-state tunneling, and photoemission and
leads to a gentle onset of a decrease in entropy with decreasing
$T$. Importantly, it does not correspond to a true zero of the
fermionic spectral function but rather to a depression that appears
at energies around the chemical potential due to a finite lifetime of
the non-condensed pairs.

In the present approach, to a good approximation (see
Eqs.~(\ref{eq:14}) and (\ref{eq:18}) below) the electron spectral
function $A(\omega,\vect{k})$ depends on a self energy of the form
\cite{Maly1997,Chen2001}
\begin{equation}
\Sigma(\omega,\mathbf{k}) = \frac{\Delta_{\pg}^2}{\omega + \xi_{-\vect{k}} + i \gamma} + \frac{\Delta_{\sc}^2}{\omega + \xi_{-\vect{k}} }\,,
\label{eq:29b}
\end{equation}
which contains both gap parameters (here written for the $s$-wave case).
Note the presence of a phenomenological parameter $i\gamma$, which
reflects the fact that the non-condensed pairs have a finite lifetime
or are meta-stable.  Its magnitude is not particularly important.
Indeed, in the normal state this expression is associated with a
phenomenology widely used for the cuprates and introduced by
M.R. Norman and collaborators in their analysis of ARPES
data~\cite{Norman1998}.

Additionally, the pseudogap can be detected indirectly through bosonic
contributions that emerge as a result of the pairing of
fermions. These are generally associated with familiar fluctuation
transport signatures, as, for example, seen in a downturn in the DC
resistivity around $T^*$.

In this Review we aim to connect the BCS-BEC crossover scenario to
experiments.  There is a challenge here because the fundamental tuning
parameter $|U|$ of the BCS-BEC crossover is not accessible. This is in
contrast to the Fermi gases where the interaction strength can be
directly measured through a scattering length. What is most important
is that it can be reasonably straightforward to replace the attractive
interaction parameter, which always appears in traditional BCS-BEC crossover
calculations on a lattice, in favor of measurable variables. This
imposes a requirement on lattice crossover theories: a broad range of
phenomena must be able to be addressed, enabling connections to
multiple experiments. The phenomena of interest involve parameters
that scale directly or inversely with $|U|$. These are, for example,
$T^*/T_\text{c}$, $\Delta_0/E_\text{F}$ and $k_\text{F} \xi_0^{\coh}$.

How to interpret experimental observations is the final important
issue we consider in this qualitative summary section. In particular,
one needs to determine whether there are experimentally verifiable
or falsifiable conditions surrounding the applicability of BCS-BEC crossover.  We
identify qualitative trends that are seen through important
correlations.  These involve the fact that increases in $\Delta_0/E_\text{F}$
should be associated with increases in $T^*/T_\text{c}$, and that decreases
in the coherence length, through $k_\text{F} \xi_0^{\coh}$, should be
correlated with increases in $T^*/T_\text{c}$. In this Review these
correlations are represented in a more quantitative fashion by
detailed predictive curves. These are shown in a number of plots as in
Figs.~\ref{fig:8},~\ref{fig:10},~\ref{fig:LowDens1},~\ref{fig:11}, and
most importantly in Figs.~\ref{fig:30} and ~\ref{fig:34}, for example. 
Related issues have come up in experimental studies, as seen
for example in Fig.~\ref{fig:Mott1}. To address specific experiments,
these predicted associations, of course, have to be tested carefully
by changing an internal variable such as pressure or possibly doping
within the same superconducting family.

\subsection{Other theoretical approaches: addressing BCS-BEC crossover on lattices}
\label{sec:alternative}

As emphasized in the Introduction, this Review primarily focuses on
one particular theoretical approach to BCS-BEC crossover based on the
ground state of Eq.~(\ref{eq:1}). Nevertheless, for the sake of
completeness, it is useful to give an overview of some alternative
theoretical schemes in the literature that are particularly relevant
to solid-state systems.

We first note that there is significantly less literature on BCS-BEC
crossover theory in solid-state superconductors as compared to the
Fermi gases.  For these atomic systems this extensive effort has
been largely driven by experimental discoveries.  Review articles are
available which summarize different
variations~\cite{Levin2010,Chien2010} of a ``t-matrix approach" to
BCS-BEC crossover theory at finite temperature. Key aspects of these
comparisons will be briefly discussed in Sec.~\ref{sec:tmatrixlit},
albeit with an emphasis on applications to solid-state systems.  Among
the Fermi gas reviews are those from our own group~\cite{Chen2005}, from the
Camerino group~\cite{Strinati2018}, and the Munich
group~\cite{Zwerger2011}, as well as extensive overviews from Randeria
and Taylor ~\cite{Randeria2014} and Bloch and
co-workers~\cite{Bloch2008}.  What has not been as thoroughly reviewed
is the next generation research on crossover effects associated with
superconductors in the solid state.  Notable is a nice overview from
Loktev and co-workers~\cite{Loktev2001}, which covers early work
through 2001.

This section presents an overview of alternative theories 
of crossover in the solid state. A key point to note here is
that $T_\text{c}$ approaches zero in
the extreme BEC limit. This has to do with the fact that the hopping or 
kinetic degrees of freedom are associated with the
fermions. The ``composite bosons" do not directly hop on a lattice, even in the BEC
regime, as a consequence of the assumed form for the Hamiltonian
in Eq.~(\ref{BCS_H}).
This depression of $T_\text{c}$ in the BEC regime
coincides with the onset of negative $\mu$ 
or equivalently where $\mu$ falls below the band bottom. Indeed, the
transition to BEC can be seen in Fig.~\ref{fig:1}
to correspond to the onset of shoulders.

That $T_\text{c}$ in a superconductor
progressively decreases with stronger coupling
in the BEC regime was pointed out by
Nozi\'eres and Schmitt-Rink
and is reasonably straightforward to understand. The hopping of pairs
requires the individual hopping of fermions, and, when two fermions
are tightly glued together, this hopping is highly suppressed, leading
to the asymptotic behavior seen in Fig.~\ref{fig:1}(a).
More quantitatively, these authors showed
that this suppressed hopping of pairs varies as $t^2/|U|$, where $t$ is the
fermionic hopping matrix element and $|U|$ is the magnitude of the
attractive interaction.  

The contributions of
Nozi\'eres and Schmitt-Rink
~\cite{Nozieres1985} 
are considered ground breaking and it is fitting that we discuss
their work early in this section. Nevertheless,
they expressed some reservations which should be noted, as they
state that their particular ``\textit {continuum
  model \ldots provides an accurate description of the two [BCS-BEC]
  limits but [leads to] a failure for a lattice gas}''.  In hindsight,
this is probably an unduly negative assessment, but perhaps it bears
on the rather small body of literature applying NSR theory to solid-state superconductors.

\begin{figure*}
\centering
\includegraphics[width=6.0in]{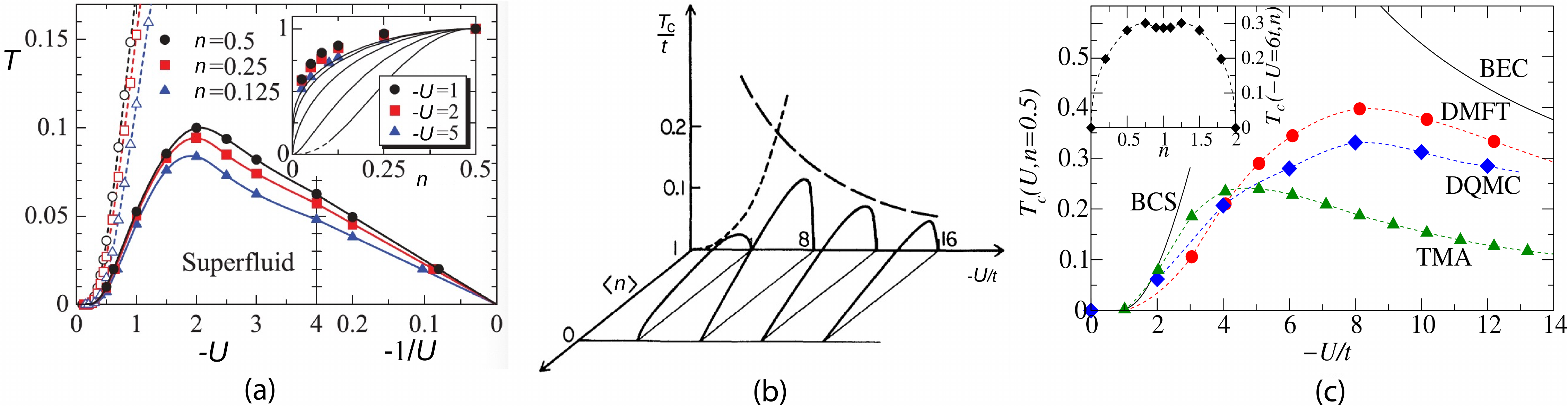}
\caption{Comparison of BCS-BEC crossover phase diagrams obtained from
  different theoretical approaches in the literature. All the diagrams
  are for a local attractive Hubbard model with the attraction
  strength $|U|$ on a lattice.  (a) Summary taken from
  \textcite{Koga2011} of dynamical mean field calculations.  Here the
  energy units are the half band-width associated with a Bethe
  lattice, having an infinite coordination number.  (b) Quantum Monte
  Carlo result for a 2D square lattice with a
  nearest neighbor hopping $t$, taken from~\textcite{Scalettar1989}.  (c) Comparison of $T_\text{c}$ calculated
  with different approaches in a 3D Hubbard model, taken from
  \textcite{Sewer2002}. }
\label{fig:1b}
\end{figure*}

Most of the canonical features in the lattice phase diagram, such as
those shown in Fig.~\ref{fig:1b} (panels (a) through (c)), including
this $t^2/|U|$ asymptote, can be obtained from different BCS-BEC
crossover theories.  These involve the t-matrix approximation (TMA)
based approaches (of which there are three main
categories~\cite{Levin2010,Chien2010} briefly discussed in
Sec.~\ref{sec:tmatrixlit}), dynamical mean field theory
(DMFT)~\cite{Georges1996,Koga2011,Peters2015,Lin2010,Kuchinskii2015,Kuchinskii2016,Bauer2009,Park2019,Sakai2015},
Quantum Monte Carlo simulations~\cite{Sewer2002}, functional renormalization
group~\cite{Strack2008}, as well as others.  Among these, the TMA
approach is principally analytical and, thus, provides more intuition
about the relevant physical processes behind the crossover, making it
the primary theoretical tool to be discussed in this Review.

We can understand why there is a relatively smaller body of analytical
literature on lattice BCS-BEC crossover theories as compared to the
Fermi gases. This is due in part to the fact that many of the sophisticated
and insightful field theory techniques, such as large-$N$ and
$\epsilon$-expansions~\cite{Veillette2007,Nikolic2007,Abuki2008,Nishida2006,Nishida2007a,Nishida2007b,Nishida2010,Nussinov2006},
are not directly adaptable to lattice systems. 
In the following we will summarize some of the DMFT and QMC
studies, highlighting a few prototypical phase diagrams shown in
Fig.~\ref{fig:1b}, which reflect a spectrum of different approaches in
the literature.  To begin, we note that Sewer, Zotos and
Beck~\cite{Sewer2002} have provided a very useful study of 3D
comparative crossover approaches that yield the phase diagrams shown
in Fig.~\ref{fig:1b}(c). These are in many ways similar to their 2D
analogues (see Fig.~\ref{fig:1b}(b) for Monte Carlo-based results).

DMFT studies of the attractive Hubbard model (addressing either the
ground state or the normal state) have been presented by \textcite{Keller2001}, \textcite{Garg2005}, \textcite{Capone2002}, and \textcite{Bauer2009}.
Some example phase diagrams~\cite{Koga2011} are presented in
Fig.~\ref{fig:1b}(a).  In DMFT, the attractive Hubbard model is mapped
to an impurity problem on a lattice, which typically has a dimension
that is effectively infinite.  In this infinite-dimension limit, the
fermionic self energy associated with pairing becomes a function only
of frequency.  As a result, computing the self energy can be reduced to
self-consistently solving a local impurity problem, for which one can
generally resort to various numerical methods.
The advantage of DMFT is that it may capture local dynamical quantum
fluctuations non-perturbatively, which can be important for a
quantitative accounting of the quasiparticle spectral function at
intermediate coupling ($|U|$ on the order of the bandwidth). 
On the other hand, DMFT is exact only in infinite dimensions because
it ignores both spatial fluctuations beyond mean-field level as well
as dimensional fluctuations.  Therefore, the DMFT results need to be
interpreted with care when making a quantitative comparison to other
approaches in three or two dimensions.

\textcite{Keller2001} have provided an interesting DMFT study
of the normal phase of the attractive Hubbard model showing that it is
a Fermi liquid at weak coupling but consists of bound pairs and
pseudogap physics at strong coupling.  Perhaps surprisingly, the
crossover between these two normal states may not be smooth at
temperatures lower than $T_\text{c}$, when the superconductivity is
suppressed. There are indications at these very low temperatures that
in this form of DMFT, a first order transition occurs in the
attractive Hubbard model between a thermally excited Fermi liquid
state and a thermally excited bound pair state as the attraction strength increases.

Figure \ref{fig:1b}(b) shows a Monte Carlo result for $T_\text{c}$ or
$T_\text{BKT}$ for an attractive Hubbard model on a 2D square lattice
with nearest neighbor hopping~\cite{Scalettar1989}.  At a generic
electron filling level, the overall shape of the $T_\text{BKT}$ vs $|U|/t$
curve looks rather similar to its 3D counterpart as shown in
Fig.~\ref{fig:1b}(c).  

It is notable that, in two dimensions, it is more straightforward to
arrive at a mean-field-level understanding of $T_\text{BKT}$ varying
from BCS to BEC (provided the lattice is away from half filling). An
illustrative example~\cite{Denteneer1993} is based on calculations of
the superfluid density or helicity modulus where one treats crossover
effects at the mean-field level. This can be done either within the
attractive Hubbard model or within its repulsive counterpart, obtained
by a particle-hole transformation on the bipartite lattice.  The
$T_\text{BKT}$ results calculated in this way are quite similar to
those shown in Fig.~\ref{fig:1b}(b).

For completeness, it is useful to highlight some additional literature
contributions that address the physics of BCS-BEC crossover for
fermions on a lattice.  Closely related to the NSR theory (which has
been mostly applied to the Fermi gas state) is work by Wallington et
al.~\cite{Wallington2000,Quintanilla2002}, who studied lattice crossover
theory using a functional-integral formalism, including Gaussian
fluctuations.  Their focus was on the effects of varying the symmetry
of the order parameter within the extended attractive Hubbard model.
Similarly Tamaki and co-workers~\cite{Tamaki2008} also addressed NSR
theory on a lattice providing an interesting comparison with other t-matrix theories.

It is useful also to summarize additional miscellaneous references
that may be of interest to the reader.  Zero-temperature approaches
mainly based on the BCS-like ground-state wave function in
Eq.~(\ref{eq:1}) are addressed by Pistolesi and
Nozieres~\cite{Pistolesi2002}, by  Herbut~\cite{Herbut2004}, by
\textcite{Andrenacci1999}, and by Volcko and
Quader~\cite{volcko2012}.
Similarly relevant to topics in the present Review are observations
about the contrast between $s$- and $d$-wave
superconductors~\cite{Loktev2001}, where it has been noted that in the
$d$-wave case moderate densities and large coupling suppress the BEC
region of the phase diagram, leading to a premature disappearance of
the superfluid phase deep inside the fermionic regime~\cite{Chen1999}.

Finally, by way of a digest of the more analytical theories of the
crossover (for the gas as well as lattice), we note that in describing
BCS-BEC crossover effects, it is tempting to introduce features of
Bose superfluidity. As in Bogoliubov theory this includes more direct
interaction effects between bosons or pairs of fermions.  In doing so,
one is saddled however with theoretical obstacles as finite-temperature effects are much more difficult to include properly in
Bose superfluids than in the BCS (fermionic) case. In strict BCS
theory the entire temperature range is accessible, whereas
in the Bose case one is restricted to the low-temperature regime. As a
consequence, in many BCS-BEC crossover approaches one can encounter
unphysical effects that are inherited from problems in theories of
Bose gases~\cite{Shi1998,Reatto1969}.  Among these are first order
jumps in thermodynamic properties
at $ T_\text{c} $ and violations~\cite{Haussmann2008} of the Hugenholtz-Pines constraint~\cite{Hugenholtz1959}.

\section{Detailed Microscopic Theory of 3D BCS-BEC Crossover Superconductivity at \texorpdfstring{$T \neq 0$}{}}
\label{sec:details}

Section \ref{sec:Quant2} provided a brief but reasonably complete
summary of results from the current formalism. In this section we
present additional details for the interested reader.

\subsection{Characterizing the bosons embedded in BCS theory}

Here we determine how to microscopically and quantitatively understand
the non-condensed bosons of the BCS approach using a slightly
different language~\cite{Chen2005} from that of Kadanoff and Martin.
We present the theory for the $s$-wave case, while the application to
$d$-wave superconductivity can be found elsewhere~\cite{Chen2000}. We
build on a centrally important observation: at any temperature in
which there is a condensate, the non-condensed bosons that are in
equilibrium with the condensate must have a vanishing chemical
potential:
\begin{equation}
\mu_{\pair} = 0 \quad \textrm{for}  \quad  T \leq T_\text{c}.
\label{eq:9}
\end{equation}
This statement is equivalent to the famous Hugenholtz-Pines
theorem~\cite{Hugenholtz1959}. How do we guarantee that the pair chemical
potential is zero? BCS provides us with an important
temperature-dependent self-consistency condition known as the gap
equation, valid for all $T \leq T_\text{c}$. 
This gap equation is given by
\begin{equation}
0=\frac{1}{U}+\sum_{\vect k} \frac{1 - 2f(E_{\vect k})}{2 E_{\vect k}},
\label{eq:10}
\end{equation}
where $E_{\vect k} = \sqrt{\xik^2 + \Delta^2}$ and $\Delta$ is the temperature-dependent pairing gap.

We argue that Eq.~\eqref{eq:10} should be incorporated in one way or
another to arrive at an understanding of pair excitations. This leads
us to constrain the form of the pair propagator $t(q)$ (or more
precisely the t-matrix) for the non-condensed pairs to satisfy
\begin{equation}
t^{-1}(q=0) \propto \mu_\pair = 0, \quad T \leq T_\text{c}. 
\label{eq:11}
\end{equation}
Indeed, Thouless has argued that a divergence of a sum of ``ladder''
diagrams (within a pair propagator) is to be associated with the
BCS transition temperature. Here we assert that this Thouless
condition can be extended to characterize the \textit{full}
temperature-dependent gap equation for all $T\leq T_\text{c}$, not just the
transition region. This constraint leads to a pair propagator of the
form~\footnote{A more systematic and first principles derivation of
  this t-matrix can be found using Eqs. (2.3-2.4), (2.7-2.8),
  (2.7$^\prime$-2.8$^\prime$) and (2.10) in \textcite{Kadanoff1961}.}
\begin{equation}
\label{eq:12}
t^{-1}(q)=\sum_{k}G(k)G_{0}(q-k)+U^{-1},
\end{equation}
whose diagrammatic representation is shown in Fig.~\ref{fig:tmatrix}.
In the above equation
$G_0(k)=\left(i\omega_n - \xi_{\vect k}\right)^{-1}$ and
$G(k) \equiv \left[ G_0^{-1}(k) - \Sigma(k) \right]^{-1}$,
corresponding to the bare and dressed fermion Green's functions,
respectively, with $\Sigma(k) =- \Delta^2 G_0(-k)$. We define
$k=(i\omega_n, \vect k)$ and $q=(i \Omega_l, \vect{q})$ as two
four-vectors with $\omega_n=(2n+1)\pi T$ and $\Omega_l=2 l \pi T$, and
$\sum_k $ is a short-hand notation for $ T\sum_n \sum_{\vect{k}}$,
with $\{n,l\}\in \mathbb{Z}$.

It is important in Eq.~\eqref{eq:12} to properly define the fermionic
chemical potential $\mu$.\footnote{To be consistent this requires
  setting $\textrm{Re}\Sigma (\vect{k}_\mu) = 0$, so that Hartree-like
  terms in the diagonal part of the self energy are absorbed into the
  chemical potential. Here $\vect{k}_\mu$ is the wavevector on the
  Fermi surface.} In this way one avoids unphysical effects that stem
from the asymmetric form of the t-matrix of BCS theory, involving
different spin states pertaining to dressed and bare Green's
functions. If care is not taken~\cite{Pini2019}, such calculations may
lead incorrectly to an artificial Fermi surface mismatch between the
two spin states and, thereby, regions of unstable
superconductivity in the phase diagram.

Importantly, Kadanoff and Martin~\cite{Kadanoff1961,Patton1971}
arrived at the same conclusion concerning the presence of both dressed
and bare Green's functions. As stated by Kadanoff and Martin:
``\textit{This asymmetry \dots has led several people to surmise that
the symmetrical equation \dots solved in the same approximation
would be more accurate. This surmise is not correct...}''.

\subsection{Determining the pair mass \texorpdfstring{$M_{\pair}$}{} and the non-condensed pair number density \texorpdfstring{$n_\pair$}{} for \texorpdfstring{$T \leq T_\text{c}$}{}}
\label{sec:MassNumber}

The fundamental quantities which determine the transition
temperature~\cite{Chen2005} in Eqs.~\eqref{eq:2} and \eqref{eq:4}
require that we determine $n_\pair$ and $M_\pair$. We argue that both of
these must depend on the BCS gap $\Delta$. In general t-matrix
theories the self energy is given by a convolution between a Green's
function and the t-matrix. Here this self energy due to non-condensed
pairs takes the form
\begin{equation}
\Sigma_{\pg}(k)=\sum_{q \ne 0}t(q)G_{0}(q-k). 
\label{eq:13}
\end{equation}
Note that the $q=0$ component of $t(q)$ (which corresponds to the
condensate) is necessarily excluded in the summation above. To proceed
further one adopts the so-called ``pseudogap (pg)
approximation''. This was motivated originally by detailed numerical
work~\cite{Maly1999,Maly1999a}. It should be emphasized that it is
appropriate at all $T$ below $T_\text{c}$. It also applies to a restricted
set of temperatures in the vicinity of but slightly above the
transition~\cite{Maly1999,Maly1999a} where $|\mu_{\pair}|$ is very
small. Since $|\mu_{\pair}| \approx 0$, $t(q)$ is strongly peaked
about $q=0$, so that the self energy can be approximated by
\begin{subequations} \label{eq:14}
\begin{align}
\Sigma_{\pg}(k)    & \approx - \Delta_{\pg}^2  G_0(-k), \label{eq:14a}\\
  \text{ with } \quad     \Delta_{\pg}^2
                   & = -\sum_{q \ne 0}t(q), \quad T \lesssim T_{c}.\label{eq:14b}
\end{align}
\end{subequations}
We emphasize that the above two equations constitute the
\textit{central} approximation made (for the sake of numerical
simplicity~\cite{Maly1999}) in the present theoretical framework. The
other crucial approximation is the adoption of Eq.~(\ref{eq:1}) as the
essential starting point.

We are now in a position to compute the pair mass and number
density. After analytical continuation,
$i\Omega_l \rightarrow \Omega + i 0^+$, we expand the (inverse)
t-matrix for small argument $q$ to find
\begin{equation}
t(\Omega, \vect{q}) = \frac {Z^{-1}}{\Omega - \Omega_{\vect{q}} +\mu_\text{\pair} + i \Gamma_{\Omega, \vect{q}}},
\label{eq:15}
\end{equation}
where $Z$ is a frequency- and momentum-independent proportionality
coefficient; the pair mass is contained in the pair dispersion\footnote{In quasi-2D, one may expand the pair dispersion as
  $\Omega_{\mathbf{q}} = \mathbf{q}_\parallel^2/ (2
  M_{\pair,\parallel}) + \mathbf{q}_\perp^2/ (2 M_{\pair,\perp})$,
  where the subscripts $\parallel$ and $\perp$ denote in-plane and
  out-of-plane components, respectively. Away from the long wavelength
  limit on a lattice, one can use a Bloch band dispersion instead of a
  simple parabola. An $\Omega^2$ term may be added to the $t^{-1}(q)$
  expansion for better quantitative accuracy.}
$\Omega_{\mathbf{q}} = \mathbf{q}^2/ (2 M_{\pair})$; the last term in
the denominator, $i \Gamma_{\Omega, \vect{q}}$, is frequency dependent
and describes the finite lifetime of the non-condensed pairs due to
decay into the two-fermion continuum. Defining the propagator for the
non-condensed pairs as $Z t(\Omega,\vect{q})$ and neglecting the
generally small dissipative term $i \Gamma_{\Omega, \vect{q}}$, one
can obtain the non-condensed pair density as
\begin{equation}
n_\pair =\sum_{\mathbf{q}} b(\Omega_{\mathbf{q}})
=Z\Delta_{\pg}^2,
\label{eq:16}
\end{equation}
which is naturally temperature dependent. Here, $b(x)=1/(e^{x/T}-1)$
is the Bose-Einstein distribution function.

We have asserted above that the total fermionic gap
$\Delta^2=\Delta_{\sc}^2 + \Delta_{\pg}^2$.  To complete the arguments
we now show that this derives from two self energy contributions ---
from the condensate (sc) and the non-condensate (pg):
\begin{equation}
\Sigma(k) = \sum_q  t(q) G_0 (-k  + q)
= \Sigma_{\sc}(k) + \Sigma_{\pg}(k).
\label{eq:17}
\end{equation}
Here, $\Sigma_{\sc}$ comes from the Dirac delta function piece of
$t(q)$ at $q=0$, i.e.,
$t_{\sc} \equiv t(q=0)= -(\Delta_{\sc}^2 /T) \delta(q)$. Using
Eq.~\eqref{eq:14a}, we then obtain
\begin{equation}
  \Sigma(k) \approx - (\Delta_{\sc}^2 + \Delta_{\pg}^2) G_0(-k) \equiv - \Delta^2 G_0(-k).
\label{eq:18}
\end{equation}
In this way, Eq.~\eqref{eq:5} results and we have
$\Delta^2=\Delta_{\sc}^2 + \Delta_{\pg}^2$.

\subsection{Establishing the form of \texorpdfstring{$T_\text{c}$}{}}

We approach $T_\text{c}$ from high temperatures, where
$\Delta_{\pg}^2=\Delta^2$ and $\mu_\pair < 0$.  As $T$ decreases,
$\mu_\pair$ increases, and Eq.~\eqref{eq:16} will be satisfied under
the condition $\Delta_\pg^2=\Delta^2$, at $T\ge T_\text{c}$.  The transition
temperature $T_\text{c}$ is reached when this is no longer possible; below this
temperature $\Delta_\pg^2$ can not accommodate the value of
$\Delta^2$, so that an additional contribution $\Delta_\sc^2$ is
needed. This occurs when $\mu_{\pair}$, as a function of decreasing
$T$, first reaches zero in Eq.~\eqref{eq:16}, from which one recovers
a Schafroth-like expression for $T_\text{c}$:
\begin{equation}
T_\text{c} = \bigg(\frac{2\pi}
{\mathcal{C}}\bigg)
  \frac{n_\pair^{2/3}(T_\text{c})} { M_\pair(T_\text{c})} , 
\label{eq:19}
\end{equation}
as was anticipated in Eq.~\eqref{eq:2}. While it was not recognized in
the original Schafroth calculations, on the right-hand side of
Eq.~\eqref{eq:19}, both $n_\pair$ and $M_\pair$ depend on $\Delta^2$,
and are therefore functions of $T$. Below $T_\text{c}$, Eq.~\eqref{eq:16} is
valid for non-condensed pairs with $\mu_\pair =0$ and $\Delta_\pg^2 < \Delta^2$.
Here the total pair density can be deduced to be $n_\pair^\text{total} = Z\Delta^2$.

\subsection{Alternative t-matrix approaches to BCS-BEC crossover}
\label{sec:tmatrixlit}

\begin{figure}
\centering
\includegraphics[width=0.9\linewidth]{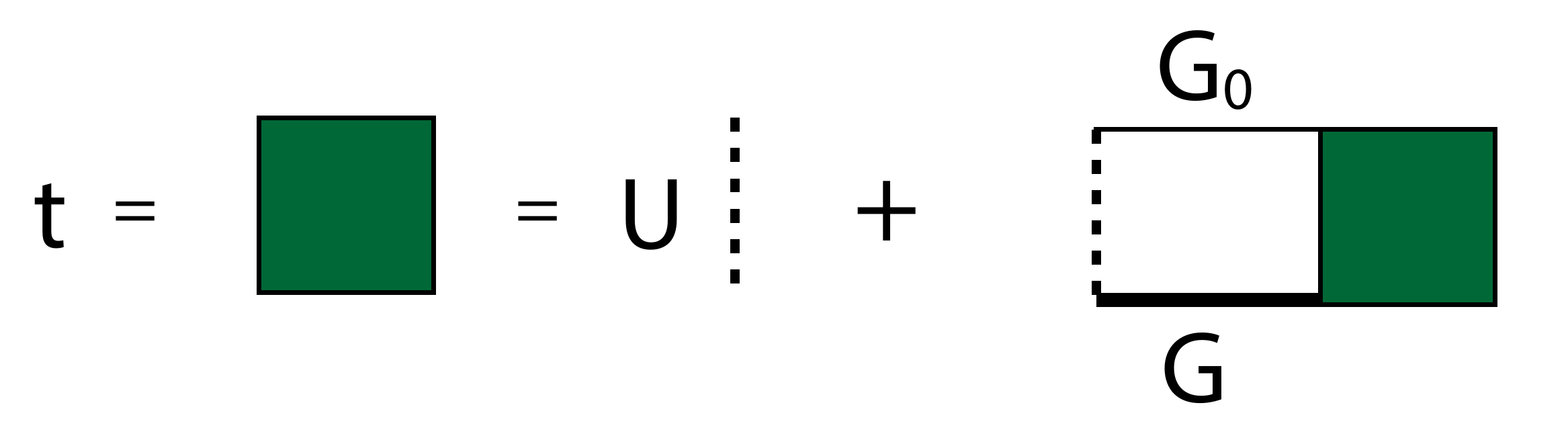}
\caption{The pair propagator of Kadanoff and Martin~\cite{Kadanoff1961}.  $U$ is the attractive interaction; $G_0$ and $G$ are the bare and dressed fermion Green's functions, respectively. }
\label{fig:tmatrix}
\end{figure}

From Fig.~\ref{fig:tmatrix} or equivalently Eq.~(\ref{eq:12}) one can
see that, within the BCS ground state based t-matrix approach to
BCS-BEC crossover, an asymmetric combination of dressed ($G$) and bare
($G_0$) Green's functions enters the definition of the t-matrix or
pair propagator.  As noted earlier, the connection between this particular combination
and BCS theory was first identified by Kadanoff and
Martin~\cite{Kadanoff1961}, using an equation of motion approach.
However, in general, one could contemplate other combinations of $G$
and $G_0$ in defining the t-matrix.
Except for the particular combination shown in the figure, the related
ground states are not as well understood~\cite{Diener2008}.  

The NSR
scheme is associated with two bare Green's functions.  The
self-consistent t-matrix approximation (SCTA), associated with two
dressed Green's functions, has been discussed by Haussmann and Zwerger
and their collaborators~\cite{Haussmann1993,Haussmann2007} in
applications to the Fermi gases and even earlier in the context of the
cuprates~\cite{Allen2001,Tchernyshyov1997,Micnas1995}.
It is also known as the Luttinger-Ward formalism~\cite{Haussmann2009}
or Galitskii-Feynman theory~\cite{Sopik2011}.  This $\Phi$-derivable
theory does possess an appealing simplicity as it readily satisfies
conservation laws, but this particular t-matrix theory will not satisfy
the equations of motion, e.g., those derived by Kadanoff and
Martin~\cite{Kadanoff1961}, as prescribed by the Hamiltonian.

Comparisons among these different t-matrix schemes have been
extensively discussed in the literature~\cite{Levin2010}.  Here, we
give a brief but critical summary, noting that it is useful to discuss
the comparisons first in the context of the Fermi gases and then turn
to the lattice case. While the differences among different schemes
might seem to be rather technical and therefore possibly minor, they
have led to significantly different qualitative physics.  
Among these
is the fact that the transition at $T_\text{c}$ is first
order~\cite{Pieri2004,Haussmann2007,Fukushima2007,Hu2007} in the
standard NSR based approaches as well as in the SCTA scheme. This
leads to unwanted features in the Fermi gas density
profiles~\cite{Perali2004} and temperature dependence of the
superfluid density~\cite{Fukushima2007}.  

The interested reader can
consult other references~\cite{Serene1989,Sofo1992} which address
other worries about the NSR approach.  Some additional concerns about
the SCTA scheme are the failure to satisfy the Hugenholtz-Pines
gapless condition~\cite{Haussmann2008}.  In this context it was also
noted by \textcite{Haussmann2009} that \textit{``a simple
  pseudogap ansatz for the spectral function~\cite{Norman1998} is not
  consistent with our results \ldots we do not observe a strong
  suppression of the spectral weight near the chemical potential."}
More generally, there is some controversy in the Fermi gas literature
~\cite{Zwerger2011,Morawetz2011,Tchernyshyov1997,Sopik2011} about the
presence or absence of a (pseudo)gap in this SCTA approach. 
Finally, we note that the principal weakness of the
BCS-Leggett approach is that it focuses on the pairing channel while
embedding all Hartree-like effects in an (effective) chemical
potential. This leads to numerical discrepancies of some significance,
particularly for the unitary Fermi gas.

In the lattice case an on-site attractive Hubbard Hamiltonian provides
a prototypical model for studying BCS-BEC crossover in the literature.
While in many ways t-matrix schemes involving all fully dressed
Green's functions ~\cite{Tamaki2008,Engelbrecht2002,Deisz1998} would
seem to be more complete, in this model, the nature of the (pseudo)gap
and whether it exists both above and below $T_\text{c}$ continues to be
debated in the lattice context as well
~\cite{Moukouri2000,Tchernyshyov1997,Micnas1995,tremblay2006,Allen2001}.
Indeed, a rather complete study of the associated excitation
spectrum~\cite{Micnas1995} for a conserving SCTA formalism shows
multiple, complex excitation branches.

A useful reference to consult~\cite{Tamaki2008} presents comparative
$T_\text{c}$ calculations for SCTA schemes along with the NSR approach
and with DMFT.  Here one sees that the transition temperatures in the
NSR scheme are significantly higher (particularly in the asymptotic
regime at large $|U|$) and this is attributed to the fact that this
approach may tend to underestimate the effects of an indirect
repulsion between fermion pairs.  All t-matrix approaches, in some
sense, ignore the
effects of direct inter-pair repulsion~\cite{Micnas1990}, but indirect effects
appear via the interactions with the fermions.  These observations may
bear on Haussmann's observation~\cite{haussmann1994} that the approach
to the BEC asymptote in the Fermi gas case should be from below and
not above, as found for example by NSR.

\section{Quantitative Implications for 3D Crossover Superconductors}
\label{sec:Quant}

\subsection{Two-gap physics present in BCS-BEC crossover}

It is important to understand the necessity of having two distinct
energy gaps in BCS-BEC crossover physics. These were illustrated in
Fig.~\ref{fig:4}. The recognition of these two distinct gaps is an
issue that bears on some of the interesting candidate materials that
are claimed to exhibit BCS-BEC crossover, as we discuss in this Review.

Indeed, one of the central ways in which these two gap contributions
are manifested has to do with the distinction between two classes of
experiments: these are associated with phenomena that reflect
superfluid coherence and those which reflect an excitation or pairing
gap. The superfluid density $n_s$~\cite{Chen1998,Chen2000} provides a
useful example, as it necessarily vanishes when coherence is destroyed
or equivalently when $\Delta_{\sc}=0$. But, notably, it also depends
on the total fermionic excitation gap $\Delta$ through the
quasiparticle energy $E_{\vect{k}}$:
\begin{equation}
\frac{n_s}{m}=\frac{2}{3}\sum_{\mathbf{k}}
\left(\frac{\partial\xik}{\partial  \mathbf{k} }\right)^2
\frac{\Delta^2_{\sc}}{\Ek^2}\left[\frac{1-2f(\Ek)}{2\Ek}+\frac{\partial f(\Ek)}{\partial \Ek}\right]
\label{eq:20}
\end{equation}
written here for an isotropic $s$-wave superconductor in 3D
with fermion mass $m$.

Similarly, it has been argued that Andreev scattering appears to
measure the gap associated with the order parameter as distinct from
conventional quasi-particle tunneling which measures the full pairing
gap, $\Delta$. This has been recognized for the
cuprates~\cite{Deutscher2005} as well as for twisted bi-layer
graphene~\cite{Oh2021}.

It is useful at this point to emphasize the fact that even though the
bosonic degrees of freedom may be viewed as ``quasi-ideal" within this
generalized BCS framework, in contrast to an ideal Bose gas this does
not compromise the existence of stable superfluidity.
Superconductivity is stable in this framework as it is to be
associated with the underlying fermionic degrees of freedom.

This analysis of the superfluid density provides a general template for other experiments
that reflect true long-range order in a superconductor. Its low-$T$ behavior has often been used to distinguish superconductors of different pairing symmetry, such as $s$- versus $d$-wave.
We end by noting that this intrinsic two-gap behavior appears to have
no natural counterpart in other preformed-pair scenarios (e.g., the
phase-fluctuation approach) for the pseudogap.

\subsection{Contrasting BCS-BEC crossover in \texorpdfstring{$s$}{}- and \texorpdfstring{$d$}{}-wave superconductors}
\label{sec:dwave}

\begin{figure}
\centering
\includegraphics[width=2.8in,clip]{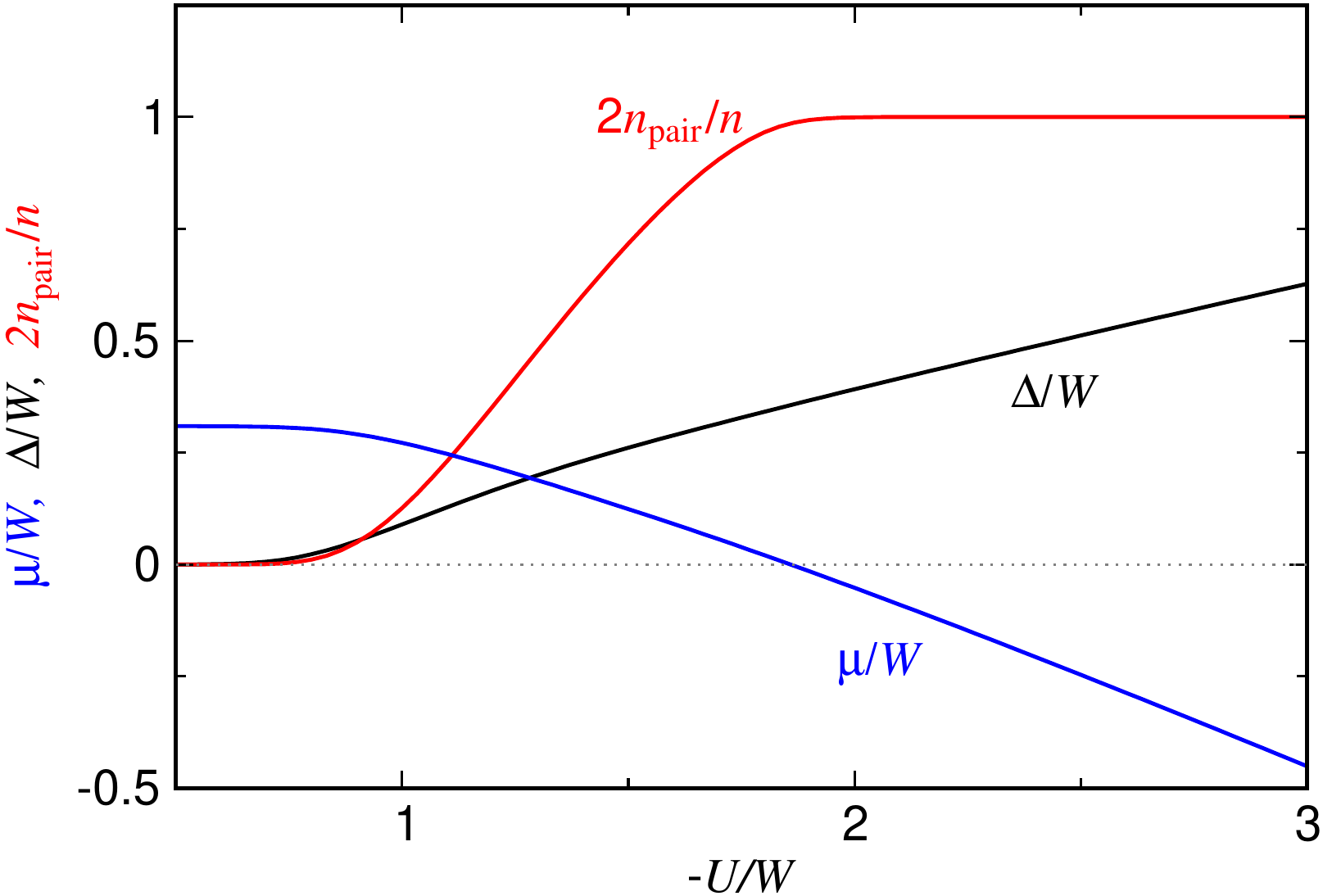}
\caption{Quantitative values of the parameters $\mu$, $\Delta,$ and
  the number of pairs $n_\pair$ at $T_\text{c}$ for the $s$-wave BCS-BEC
  crossover superconductor on a 3D cubic lattice in
  Fig.~\ref{fig:1}(a) as a function of the attractive interaction $U$
  (normalized by the half bandwidth $W=6 t$) with the electron density
  $n=0.1$ per unit cell. Here the normal-state electronic energy
  dispersion is
  $\epsilon_{\vect k}= 2 t (3-\cos k_x - \cos k_y -\cos k_z)$, where
  the lattice constant $a$ has been set to unity. }
\label{fig:5}
\end{figure}

A crucial feature of BCS-BEC crossover in superconductors (in either
2D or 3D) to be emphasized throughout this Review is that the
canonical plots of the phase diagram (based on the Fermi gases) do not
capture the physics of superconductivity in the solid state. For the
latter, as shown in Fig.~\ref{fig:1}(a), one finds $T_\text{c}$ follows a
superconducting dome as a function of variable interaction strength,
within the fermionic regime.  Thus, one should not infer, as is often
the case, that for solid-state superconductors in the BEC there is a
large and maximal transition temperature.

Figure~\ref{fig:5} provides more quantitative details on the key
energy scale parameters that enter BCS-BEC crossover for the $s$-wave
lattice case of Fig.~\ref{fig:1}(a).
The figure indicates the behavior of $\Delta$ and $\mu$ at $T_\text{c}$ in
units of a characteristic electronic scale (in this case corresponding
to the half bandwidth). These energies are plotted as a function of varying
attractive interaction strength, normalized to the half bandwidth
$ W = 6 t$, where $t$ is the hopping matrix element. Also plotted here
is the important parameter $n_\pair$ which corresponds to the number
density of pairs at the onset of the transition (normalized by $n/2$,
as determined from Eq.~\eqref{eq:16}).

In particular, one can glean from the plot of $n_\pair$ that the BEC
or $\mu=0$ transition is associated with the absence of fermions so
that only pairs are present ($n_\pair = n/2)$. More generally, one can
view the function $n_\pair$ as a kind of theoretical ``dial''
informing about where a given system is within the crossover. Tuning
the dial provides access to the counterpart values of $\mu$ and
$\Delta$ at $T_\text{c}$. When $n_\pair$ is essentially zero this corresponds
to the BCS case and when $n_{\pair}\approx n/2$ one enters the BEC
regime.

\begin{figure}
\centering
\includegraphics[width=3.0in]{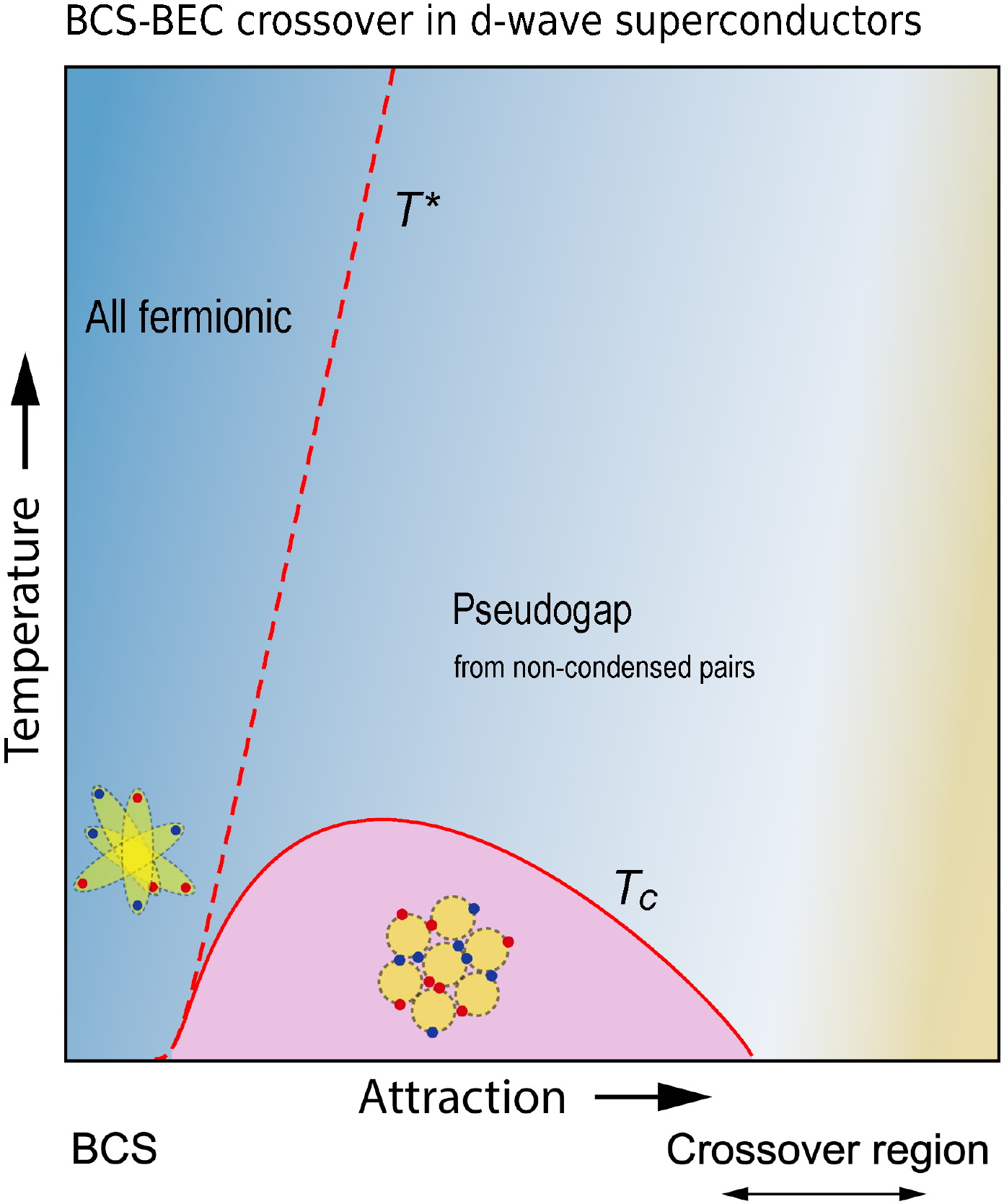}
\caption{BCS-BEC crossover phase diagram for a $d$-wave
  superconductor~\cite{Chen1999}.  This is for an attractive
  Hubbard-like interaction
  $V_{\bf k,k'} = U \varphi _{\bf k}\varphi^{\ }_{\bf k'}$, where the
  momentum dependent function $\varphi _{\mathbf{k}}$ possesses a 
  $d_{x^2 - y^2}$ symmetry.  This figure shows that this system (near
  half filling) has vanishing $T_\text{c}$ before the onset of the BEC
  regime. This behavior persists down to $n \simeq 0.1$. This figure
  is meant to be compared with the schematic $s$-wave case in
  Fig.~\ref{fig:1}(a). For $s$-wave symmetry the BEC regime is in
  principle accessible up to around a quarter filling. Actual units for the
  vertical and horizontal axes can be found in Fig.~\ref{fig:31} which
  corresponds to a very slightly modified band structure, specific to
  the cuprates.}
\label{fig:6}
\end{figure}

The crossover behavior for a $d$-wave superconductor is generally
different~\cite{Chen1999} and some aspects are additionally discussed
in Appendix~\ref{sec:AppB}. For definiteness, we consider here the
symmetry to be of the form $d_{x^2 - y^2}$, which is relevant to the
cuprate superconductors. The central contrasting feature is the
termination of $d$-wave superconductivity well before the BEC regime
is entered. This is found at all but very low electron densities and
derives principally from the fact that $d$-wave pairs have a more
extended size. As a result a pair-pair repulsive interaction which is
always present~\cite{Micnas1990} is sufficiently strong so that it
inhibits pair hopping and pairs become localized. And, importantly,
this happens in the fermionic regime, well away from where $\mu<0$.
Consequently, in the $d$-wave case, the BEC limit cannot
generally be accessed~\cite{Chen1999}. This important effect is
illustrated in the phase diagram shown in Fig.~\ref{fig:6}.

What this implies more concretely is that the $d$-wave system
undergoes a transition at moderately strong attraction, where
$T_\text{c} \rightarrow 0$. Here superconductivity continuously disappears,
albeit in the presence of a finite pairing gap $\Delta$ or finite
$T^*$. This has features that are suggestive of the widely discussed
``Cooper-pair insulator''~\cite{Hebard1990,Paalanen1992,Hollen2011}
or a pair density wave
alternative~\cite{Che2016,*Chen2019CPL,*Sun2022AdP,*Sun2022PRA}. But
the form of pair localization considered here pertains to a clean system and represents a
different mechanism, deriving from strong intra-pair attraction and
strong inter-pair repulsion, which inhibits pair hopping. Indeed, this same
localization has also been observed in cases where the band filling is
high in $s$-wave superconductors, as well as in 2D systems.  In these
instances it provides an interesting comparison, but is not to be
associated with strong disorder effects, which are known to drive a
superconductor-insulator transition in superconducting films
~\cite{Fisher1990,Hebard1990,Paalanen1992,Yazdani1995}.

\begin{figure}
\centering
\includegraphics[width=2.8in,clip]{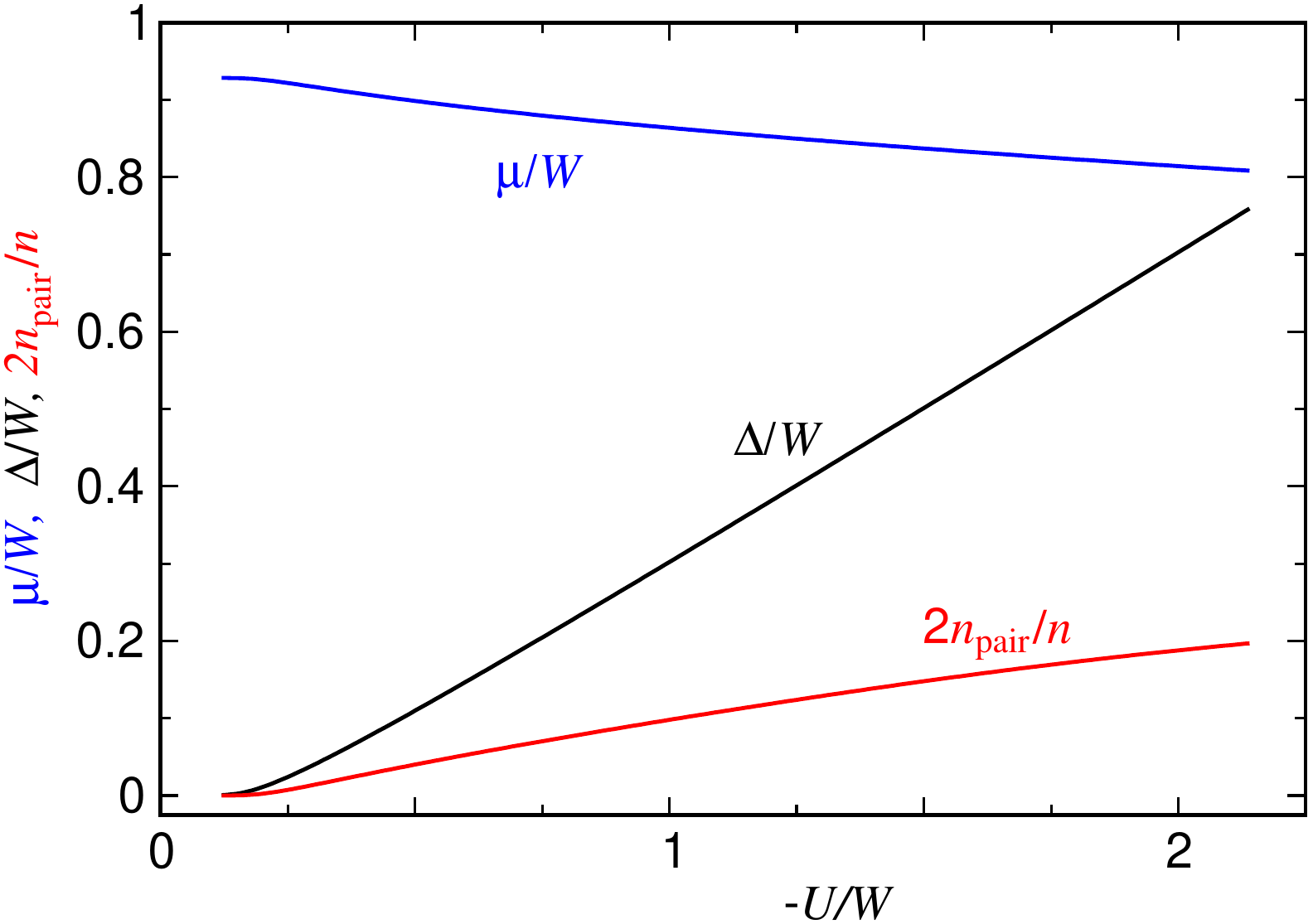}
\caption{Quantitative values of the parameters $\mu$, $\Delta$, and
  the number density of pairs $n_\pair$ at $T_\text{c}$ for the quasi-2D
  $d$-wave BCS-BEC crossover superconductor in Fig.~\ref{fig:6} as a
  function of the attractive interaction (normalized again by the half
  bandwidth $W$).  Here the normal-state kinetic energy dispersion is
  $\epsilon_{\vect k}= (4t+2 t_z) - 2 t (\cos k_x + \cos k_y) - 2
  t_z\cos k_z$ with $t_z/t=0.01$.  The electron density is $n=0.85$
  per unit cell. }
\label{fig:7}
\end{figure}

Figure~\ref{fig:7} provides more quantitative details on the
characteristic energy scale parameters that enter BCS-BEC crossover
for this $d$-wave lattice case ~\cite{Chen1999}. Plotted here is the
behavior of $\Delta$ and $\mu$ at $T_\text{c}$ as a function of varying
attractive interaction. Also indicated is the number of pairs (derived
from Eq.~\eqref{eq:16}), $n_\pair$, at the onset of the transition.

\subsection{The interplay of conventional fluctuations and BCS-BEC crossover physics: Normal-state transport}

The question of how conventional superconducting fluctuations relate
to BCS-BEC crossover physics continues to be raised in the
literature. In this regard it is interesting to note that the
treatment of preformed pairs presented here is closely related to
self-consistent theories of fluctuation superconductivity. In
particular, it represents a natural extension to arbitrarily strong
attraction of time-dependent Ginzburg-Landau-based transport
theory~\cite{Ullah1991} when the quartic terms in this free-energy
expansion are treated in a self-consistent Hartree-level
approximation~\cite{Ullah1991,Stajic2003,Tan2004,Patton1971}.  This
observation suggests that there is a continuous variation, associated
with an \textit{enhancement} of many transport fluctuation signatures,
as the coupling varies from weak to strong.

To address these issues more quantitatively, we note that dominating
transport in these more strongly correlated
superconductors~\cite{Boyack2019,Boyack2018,Boyack2021} is the fact
that there are now two distinct temperature scales which control
``fluctuation'' effects: $T_\text{c}$ and $T^*$. Transport is complicated
additionally by the fact that there are two types of quasiparticles:
fermions which experience the gap onset at $T^*$ where they thus 
generally become less conducting, and bosons whose presence is
expected to increase conductivity at temperatures somewhat below
$T^*$.  These two types of quasiparticles are represented
schematically in the upper row of Fig.~\ref{fig:3}.

The fermionic contribution has been
discussed~\cite{Wulin2012,Wulin2012a} in some detail both above and
below $T_\text{c}$.  The more familiar fluctuation contributions to bosonic
transport derive from the Aslamazov-Larkin~\cite{Aslamazov1968}
diagrams and are associated with a small pair chemical potential,
$\mu_\pair(T)$, which is found in the immediate vicinity of $T_\text{c}$. In
conventional superconductors, $\mu_\pair$ depends only on $T_\text{c}$, but
in the presence of more stable preformed pairs one expects that $T^*$
will play an important role. It is at this higher temperature that the
pair density vanishes; consequently, fluctuation effects are expected
to have some presence even at temperatures as high as $T^*$.

The above discussion leads one to conclude that, for more strongly
coupled superconductors, the nature of ``fluctuation'' effects
associated with $T^*$ in transport requires that one establish the
relative size of the contributions from the fermionic and bosonic
channels; as we have seen these generally introduce opposite
temperature dependencies in their conduction properties. Their
relative size depends on their relative scattering times.

Central to this comparison is the fact that the resistivity downturn,
a canonical signature of the pseudogap onset at $T^*$, is frequently
associated with the concomitant and rather ubiquitous large
normal-state resistivity. This ``bad-metal''
behavior~\cite{Gunnarsson2003,Boyack2021}
reflects a suppressed fermionic conduction channel. Importantly, bad
metallicity allows the bosonic conducting channel to become more
prominent and, for example, leads to a boson-related downturn near
$T^*$ in the resistivity which would otherwise be obscured by gap
effects in the fermionic spectrum.

We will see later in this Review examples of transport signatures
that are viewed as indicative of the presence of BCS-BEC crossover
physics. In addition to a resistivity downturn, these include enhanced
diamagnetism and Nernst signatures, albeit not all uniquely pointing
to a BCS-BEC crossover scenario.

\subsection{Relation between BCS-BEC crossover and the Uemura plots}

In an interesting series of papers,
Y. Uemura~\cite{Uemura1997,Uemura1989} has used muon spin resonance
($\mu$SR) experiments to establish a classification scheme for
superconducting materials. This classification, in effect,
distinguishes so-called ``exotic'' superconductors from conventional
superconductors. The $\mu$SR relaxation rates in these experiments
effectively measure the London penetration depth, which in turn
reflects the ratio of the number of superfluid electrons $n_s$ to
their effective mass $m$.  Notably, at sufficiently low temperatures,
these same two quantities help to determine an effective Fermi temperature.

Uemura used this analysis to suggest that ``unconventional''
superconductors are characterized by the proportionality
$T_\text{c} \propto T_\text{F}$, where $T_\text{F}=E_\text{F}/k_\text{B}$ is the Fermi temperature. This
observation, which follows from plots of the transition temperature
versus muon-spin relaxation rate, has led many to believe that a
dependence on a single parameter $T_\text{F}$ is suggestive of a
Bose-condensation description of exotic superconductors. Underlying
this inference is the behavior of the Fermi-gas phase diagram as
shown, for example, in Fig~\ref{fig:1}(b), where the asymptotic BEC
value of $T_\text{c}$ is given by $T_\text{c} \equiv T_\text{BEC} = 0.218 T_\text{F}$ in
3D.

\begin{figure}
\centering
\includegraphics[width=\linewidth,trim=0 0mm 0 0]
{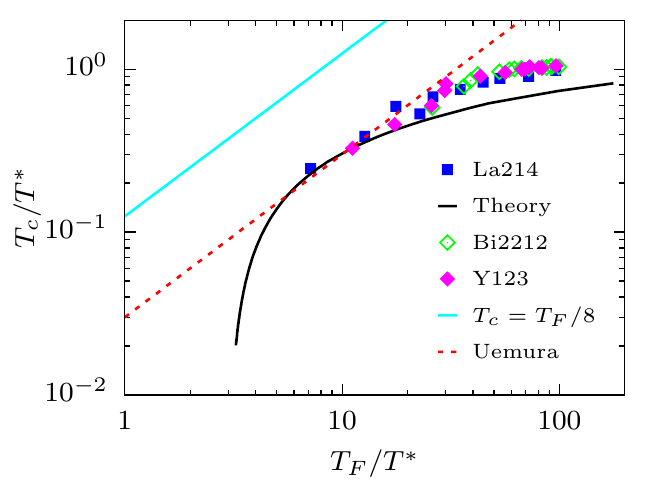}
\caption{A replot of results from Tallon and
  co-workers~\cite{Tallon2003}, which suggests a modification of the
  Uemura plot in which $T_\text{c}$ depends not only on $T_\text{F}$ but also on
  $T^*$. 
  This replotting yields a simple, complete scaling of cuprate
  transition temperatures for different hole concentrations. A BCS-BEC
  crossover theory curve (black solid line) for the quasi-2D $d$-wave
  case~\cite{Chen1999} is included here.  In the legend, La214: \LSCO;
  Bi2212: \BSCCO; Y123: \YBCO. The dashed line (labeled Uemura)
  corresponds to $T_\text{c}=0.03 T_\text{F}$.  }
\label{fig:8}
\end{figure}

In Uemura's analysis it would seem that there is a very large number
of superconductors belonging to the unconventional category, although
one should not presume that all of these are associated with Bose
condensation or BCS-BEC crossover. While focusing on a smaller subset
of just the high-temperature superconductors, Tallon and co-workers
~\cite{Tallon2003} have argued for an interesting and modified version
of the Uemura scheme which plots the ratio $T_\text{c}/\Delta_0$ versus
$T_\text{F}$, thereby introducing a second energy scale $\Delta_0$, which
reflects $T^*$. Figure~\ref{fig:8} shows the resulting rather universal scaling
of the cuprate data. The solid black line represents the $d$-wave
BCS-BEC crossover theory at moderate band filling which was discussed
above.

Such an analysis emphasizes that, for an arbitrary superconductor,
more relevant for establishing that a crossover picture is applicable
is showing the presence of distinct energy scales $T^*$ and
$T_\text{c}$. This is a necessary but not sufficient requirement. In the
crossover scenario a moderately
large value for $\Delta_0/E_\text{F}$ must simultaneously be present. 
In this way the Uemura plots have
elucidated a useful classification scheme, but we stress that one
should be cautious about inferring too strong a connection to BCS-BEC crossover.

It will be useful, thus, in this Review to show how to arrive at a
more discriminating procedure, inspired to some extent by
Fig.~\ref{fig:8}.  We will do so here, focusing on 2D superconductors
in the form of plots of $\Delta_0/E_\text{F}$ versus $T^*/T_\BKT$.  First,
however, one has to have a better understanding of 2D superconductivity.

\section{BCS-BEC Crossover Physics in the 2D Limit}
\label{sec:twoD}

\subsection{Overview of 2D theory}
\label{sec:twoD2}
In two dimensions there is no true condensation with off-diagonal long-range order. 
More quantitatively, in the language of a t-matrix
approach to BCS-BEC crossover, the chemical potential for pairs
$\mu_\text{pair}$ never reaches zero; this is effectively a
consequence of the Thouless criterion which provides a constraint on
the t-matrix.  A subtle issue that is pertinent here and in the
following discussion is that a fermionic system in either 2D or 3D
involves in some sense non-interacting bosons, but these
non-interacting pairs nevertheless support superconductivity only
because they interact indirectly through their underlying fermionic
nature.

In this Review we build on the cold-atom literature to address the BKT
phase transition~\cite{Kosterlitz1973,Berezinskii1972}. This focuses
on the approach from the high-temperature side and on bosonic degrees
of freedom or bosonic ``quasi-condensation'' (associated with
algebraic rather than long-range order). The onset of this transition
can be equivalently described as that of the onset of vortex-pair
binding and unbinding as in the original BKT papers; in this context
the role of superfluid phase stiffness is more apparent.

From the bosonic perspective, the BKT transition occurs when the de
Broglie wavelength is large and comparable to the inter-pair
separation, similar to a BEC transition in 3D. More precisely, this
transition arises when the temperature-dependent \textit{bosonic}
phase-space density reaches a critical value as was independently
established in famous papers by Hohenberg and Fisher~\cite{Fisher1988}
and also by Popov~\cite{Popov2001}. This leads to
\begin{equation}
T_\BKT = \bigg(\frac{2 \pi}
{\mathcal{D}_\pair^\text{crit}}\bigg)
  \frac{n_\pair(T_\BKT)} { M_\pair(T_\BKT)} , 
\label{eq:21}
\end{equation}
where $\mathcal{D}_\pair^\text{crit}$ is the critical phase space
density, which is essentially a constant and will be specified shortly.
Importantly, here we have replaced the number density and mass of true
bosons appearing in the \textbf{standard expression}
(Eq.~\eqref{eq:4}) by their counterpart values for a composite-boson
(or fermion-pair) system. In this way we see that the pair density and
pair mass play a similar role as in the 3D superfluid transition in
Eq.~\eqref{eq:19}.

Note that, since $n_\text{\pair}(T)$ is temperature dependent and
disappears at $T^*$, there is a significant difference between BKT
behavior in Bose and Fermi superfluids. That is, the latter will be
implicitly dependent on the two distinct temperature scales $T^*$ and
$T_{\BKT}$. Since $T_{\BKT} \le T^* $, the physical implications of
these two scales become apparent only when studying the BKT
transition, as we do here, by approaching the transition from the
normal state.

The most detailed numerical analysis of 2D atomic-gas condensates
focuses on the Bose gas in the weakly interacting limit and
provides~\cite{Prokofev2002} results for the critical value
$\mathcal{D}_\pair^\text{crit}$, which is given by
\begin{equation}
\mathcal{D}_\pair^\text{crit}= \ln (C/\tilde{g}),
\label{eq:22}
\end{equation}
where $\tilde{g}$ is a dimensionless coupling constant reflecting the
effective \textbf{repulsive} interaction between pairs.  Importantly,
the constant
\begin{equation}
C \approx 380
\label{eq:23}
\end{equation}
has been established~\cite{Prokofev2002} from Monte Carlo studies.  We
note that $\tilde{g}$ in Eq.~\eqref{eq:22} is, in principle,
dependent on the bosonic pair density, as shown in
\textcite{Fisher1988}.  However, this dependence is logarithmic, and
therefore weak, and can be neglected for most purposes because of the
large constant $C$.  This normal-state approach to the BKT transition,
using the phase space density, has been supported by numerous
experimental studies on atomic Bose gases~\cite{Jose2013, Tung2010,
  Clade2009}.

It is useful to compare this with the more familiar
expression~\cite{Nelson1977} for the same $T_\BKT$ in a
superconductor, when approached from the low-temperature superfluid
side. This provides a complementary interpretation.
\begin{equation}
T_\BKT =\frac{ \pi}{2} \rho_s(T_\BKT)
\equiv  \frac{\pi}{8}\left[\frac{n_s} {m} \right](T_\BKT),
\label{eq:24}
\end{equation}
where one introduces the temperature-dependent superfluid phase
stiffness $\rho_s(T)$, evaluated at $T_\BKT$, instead of the total
pair density as in the formula of Eq.~\eqref{eq:21}. In Eq.~\eqref{eq:24},
$n_s$ and $m$ are the superfluid density and effective mass of
fermions, respectively. To connect Eq.~\eqref{eq:21} to Eq.~\eqref{eq:24}, one
replaces $\mathcal{D}_\pair^\text{crit}$ with $4$ and converts from
pairs to fermions, following Halperin and Nelson~\cite{Halperin1979}.

It should be noted that there is a practical difficulty in using either of these formulations. We need phenomenological input to arrive at $\tilde{g}$ in Eq.~\eqref{eq:22}.
Whereas to apply Eq.~\eqref{eq:24}, one must approximate $\rho_s(T)$ by a suitably chosen (generally mean-field) expression~\footnote{This excludes using the present t-matrix theory, more precisely the 2D counterpart of Eq.~\eqref{eq:20}, where the superfluid density $n_s$ is necessarily zero
in 2D, reflecting the fact that simple bosonic condensation with long range order cannot occur.}.

\subsection{Procedure for determining \texorpdfstring{$T_\BKT$}{} in the Fermi gases}

The Heidelberg cold-atom group ~\cite{Murthy2015} has claimed that the
fits for their 2D Fermi gas data find a range of values for
$\mathcal{D}_\pair^\text{crit} = 4.9 - 6.45$~\cite{Ries2015,
  Murthy2015}. These values are close to but somewhat different from
values for atomic Bose gases, where the range is about $6-10$. In
general, $\mathcal{D}_\pair^\text{crit}$ depends on the non-universal
boson-boson interaction strength $\tilde{g}$, about which one has no
precise knowledge. However, a relatively small value of $\tilde{g}$ is
presumed in the theoretical
framework~\cite{Prokofev2002,Prokofev2001}, representing an
effectively weakly interacting gas. This would be expected in a BCS
ground state of composite bosons, as the bosonic degrees of freedom
enter this wave function in a quasi-ideal manner. For the analysis in
this Review, we adopt the value $\mathcal{D}_\pair^\text{crit} = 4.9$~\cite{Ries2015},
which turns out to best fit the data on Fermi
gases\footnote{It should be noted that this best fit case does
  presume a larger value of $\tilde{g}$ than would be expected for the
  weakly interacting case~\cite{Murthy2015}.}.

Therefore, based on experiments~\cite{Ries2015} in Fermi gases, the 2D
BKT superconducting transition is thus interpreted as a
``\textit{quasi-condensation}'' \textit{of preformed Cooper pairs}.
For application to 2D superconductors, more generally, the BKT
transition temperature is presumed to be:
\begin{equation}
\frac{n_{\pair}(T_{\BKT})} {M_{\pair}(T_{\BKT})} =    \left(\frac{4.9} {2\pi}\right)  T_{\BKT}     \quad \textrm{in 2D}. 
\label{eq:25}
\end{equation}

\begin{figure}
\centering
\includegraphics[width=3.3in,clip]{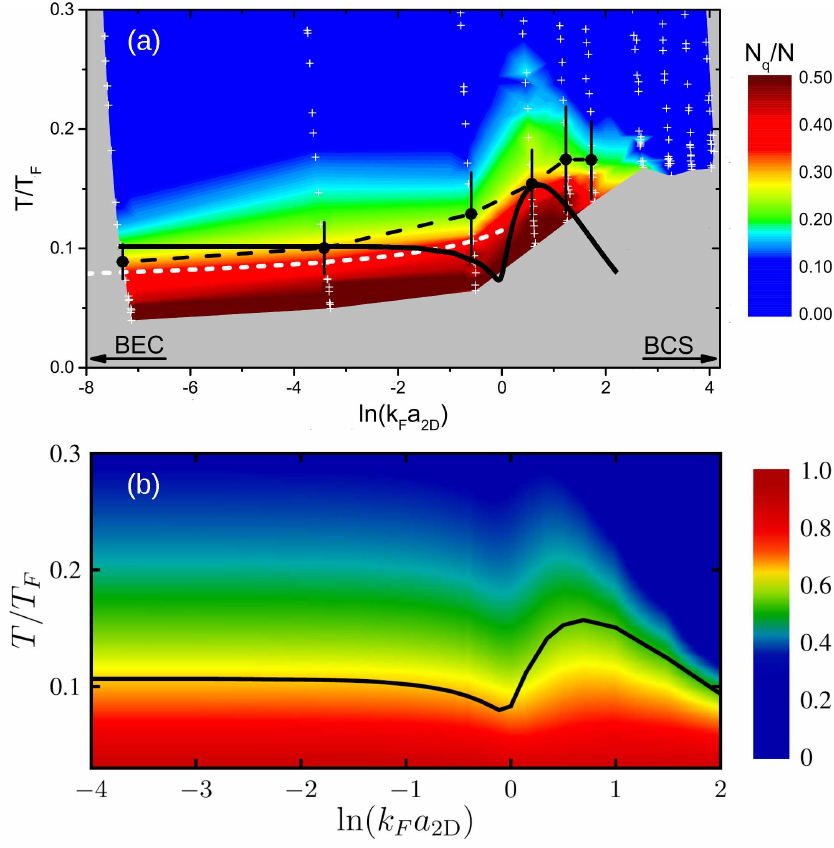}
\caption {(a) Comparison of theory \cite{Wu2015a} and experiment (adapted from \textcite{Ries2015}) for
  \textit{quasi-condensation} phase diagram of the strongly
  interacting 2D Fermi gas. The color variations reflect the
  normalized momentum distribution of pairs at low momentum $\vecq$,
  $N_\vecq/N$, which is used to quantify the quasi-condensate fraction.  (b) Theory results (with a trap
  included), taken from \textcite{Wu2015a}. Here the color variations similarly refer
  to the pair momentum distribution at low $\vecq$.  The estimated
  onset of the superfluid transition which derives from a rather
  abrupt change in $N_\vecq/N$ is indicated by the black solid line in
  both panels, dashed for experiment and solid for theory.  The white
  dashed line in panel (a) is a theoretical estimate for the BKT
  transition from \textcite{Petrov2003}. }
\label{fig:9}
\end{figure}

\begin{figure*}
\centering
\includegraphics[width=5.4in]
{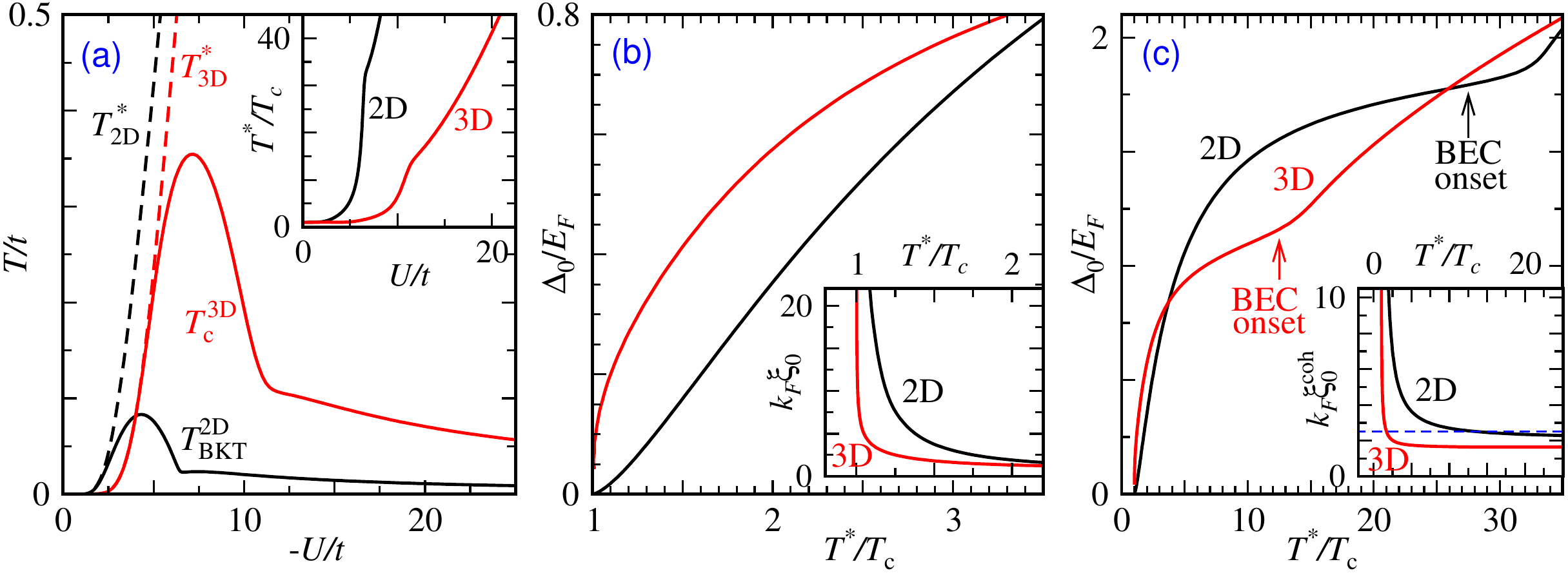}
\caption{Comparison of 2D and 3D transition temperatures as well as other properties in the BCS-BEC crossover scenario for a tight-binding $s$-wave superconductor at a low density $n = 0.1$. (a) Transition ($T_{c}$ or $T_\textrm{BKT}$) and pairing onset ($T^*$) temperatures, as a function of $-U/t$, the strength of the attractive interaction in units of the hopping matrix element $t$. The vertical axis in the inset quantifies the degree of departure from strict BCS (through the difference between $T^*/T_\text{c}$ and unity). (b)  Characteristic magnitude of $\Delta_0/E_\text{F}$ in 2D and 3D on a normalized scale, along with the pair size $\xi_0$ in the inset. (c) More extended view of the results in (b). Indicated here are the (rather high) critical values of $T^*/T_\text{c}$ at which the system crosses over to a BEC. The inset shows the behavior of the superconducting coherence length $\xi_0^{\text{coh}}$ which should be contrasted with the pair size $\xi_0$. The former reaches a finite saturation value in the BEC regime, while the latter continuously decreases towards zero.
}
\label{fig:10}
\end{figure*}

Experiments from the Heidelberg group~\cite{Ries2015,Murthy2015} on a
strongly interacting 2D Fermi gas use the momentum distribution to
establish the presence of a quasi condensate. This is based on
magnetic field sweeps which, through a Feshbach resonance, convert
pairs to deeply bound molecules. As shown in Fig.~\ref{fig:9}(a), in
this way one obtains a plot of the quasi-condensation transition
temperature as a function of scattering length or equivalently
variable interaction strength. Importantly, these measurements show
BKT signatures. An overlay of theory and experiment is shown in
Fig.~\ref{fig:9}(a), while Fig.~\ref{fig:9}(b) represents only the
theory~\cite{Wu2015a}.  It should be noted that there are
claims~\cite{Hazra2019} that the experimentally observed maximum,
which goes beyond $T_\text{F}/8$, could be an artifact of coupling to
a third dimension in the trap, although this issue, which pertains
exclusively to the 2D Fermi gas, has not been settled.

Subsequent experiments on the 2D gas~\cite{Hueck2018,Sobirey2021}
extended these measurements on trapped gases to accommodate a box
potential. Here an alternative methodology was used to obtain the
momentum distribution. These studies presented more direct
measurements of superfluidity, as distinct from 
quasi-condensation of pairs. Determination of one particular critical
temperature in the BEC regime yielded consistency with the experiments
of the Heidelberg group as a check.

\subsection{Quantitative description of BCS-BEC crossover in 2D and comparison with 3D} 
\label{sec:cohlength}

Equation~\eqref{eq:25} is adopted along with the results of
Sec.~\ref{sec:MassNumber} for $n_{\pair}$ and $M_\pair$ to
characterize $T_\BKT$ and other features of 2D
superconductors. Figure~\ref{fig:10} presents a comparison of
transition temperatures, pairing onset temperatures, pair size
~\cite{LeggettRuvalds}, gap size and coherence length in both two and
three dimensions for the $s$-wave case. In panel (a) one sees the
presence of a dome-like structure reflecting BCS-BEC crossover in the
solid state, which should be evident for $T_\text{c}$ or $T_{\textrm{BKT}}$.
This dome is well within the fermionic regime, where $\mu > 0$.  The
transition to the BEC regime with negative $\mu$ is also evident here as a
shoulder in each of the transition temperature curves.  There has been
some emphasis on bounds on the magnitude of the highest transition
temperature in these 2D systems~\cite{Hazra2019}, although one should
be cautioned that in a lattice system, these are less indicative of
the BEC limit, as the maximum is found in the fermionic regime.

The inset of Fig.~\ref{fig:10}(a) quantifies the important effect of
two dimensionality which was presented earlier in the schematic plot
shown in Fig.~\ref{fig:2}. This inset, representing moderately low
filling $n = 0.1$ per unit cell, shows that the deviation from BCS
behavior (associated with $T^*/T_\text{c}$ substantially above 1.0) occurs at
significantly smaller attraction for 2D as compared with 3D
superconductors.

We turn now to Figs.~\ref{fig:10}(b) and \ref{fig:10}(c) which are the
basis for more experimentally relevant studies. The main plots in
these two figures represent a natural extension of the Tallon-Uemura
scaling~\cite{Tallon2003}
in Fig.~\ref{fig:8}, but for the case of $s$-wave pairing in
both two and three dimensions. They show that the ratio of the two
distinct temperature scales $T^*/T_\text{c}$ or $T^*/T_\BKT$ (which are, in
principle, measurable), is correlated with the magnitude of the
$T\approx0$ value of $\Delta/E_\text{F}$ (which is also measurable).

The inset in Fig.~\ref{fig:10}(b) shows how the zero-temperature pair
size, $\xi_0$, varies as the system crosses out of the BCS
regime. Representing this crossover in the figure is $T^*/T_\text{c}$, chosen
as the horizontal axis. The pair size is a reasonably good indicator
of when the system is promoted out of the BCS regime. However, it can
be inferred from Fig.~\ref{fig:10}(c), (where the BEC onsets are
marked), that it does not display features at the onset of the BEC;
rather the pair size decreases continuously toward zero as this limit
is approached. Interestingly, in 2D the pair sizes for equivalent
$T^*/T_\text{c}$ are significantly larger than in the 3D case.

Finally, it should be emphasized that the pair size (which is
less accessible experimentally) and the coherence
length represent important but distinct length scales. The ``bare''
GL coherence length can be most readily obtained
experimentally from the measured slope of the upper critical field
$H_{c2}$ versus temperature $T$ plot
\begin{equation}
\frac{dH_{c2}}{dT} \bigg\vert_{T=T_\text{c}}= -\frac{\Phi_0}{2 \pi (\xi_0^{\text{coh}})^2 \, T_\text{c} }\quad
{ \rm{with} }
~~~\Phi_0 =\frac{hc}{|2e|},
\nonumber
\end{equation}
where $h=2\pi \hbar$. 
Here, the slope is evaluated at the zero field $T_\text{c}$, and
$\xi_0^{\text{coh}}$ is theoretically given by
~\cite{Boyack2019,Boyack2018}
\begin{equation}
\xi_0^\text{coh} =  \frac{\hbar}{\sqrt{2 M_\pair (k_\text{B} T_\text{c}) }}.
\label{eq:26}
\end{equation}
This quantity times the Fermi wave-vector is plotted in the inset in
Fig.~\ref{fig:10}(c).
From an experimental point of view there may be some advantage to
measuring and evaluating $\xi_0^{\text{coh}}$ in a somewhat different
way, just above $T_\text{c}$ in the normal state~\cite{Suzuki1991} as here
one avoids the rather challenging determination of $T_\text{c}(H)$, which
corresponds to a magnetic field broadened transition.

The coherence length has a distinct physical interpretation when we
make use of the expressions for the transition temperatures in
Eqs.~\eqref{eq:2} and \eqref{eq:21}.  First, define $k_\text{F}$ in terms of
the free and isotropic electron dispersion, so that
$k_\text{F} \equiv (3 \pi^2 n)^{1/3}$ or $(2 \pi n)^{1/2}$ in three and two
dimensions, respectively, where we use the same symbol $n$ to refer to
the appropriate fermion number density.  It follows, then, that
$k_\text{F} \xi_0^\text{coh}$ evaluated near the transition temperature
depends only on the normalized pair density, $n_\pair/n$. This leads
to
\begin{equation}
k_\text{F} \xi_0^{\text{coh}} = 1.6 (n/n_\pair)^{1/2}
\label{eq:new1}
\end{equation}
 and
\begin{equation}
k_\text{F} \xi_0^{\text{coh}} = 1.2 (n/n_\pair)^{1/3}
\label{eq:new2}
\end{equation}
for 2D and 3D respectively. 

We note that the above equations are relatively easy to understand
physically. The coherence length is a length scale representing the
effective separation between pairs. We find here, not surprisingly for
only weakly interacting pairs, that it relates to the density of
pairs. This is distinct from the pair size.  In BCS theory there are
almost no pairs present at $T_\text{c}$ and the length that represents their
average separation is necessarily very long. As pairing becomes
stronger more pairs form and their separation becomes shorter. On a
lattice, in the BEC regime their separation is bounded from below by
the characteristic lattice spacing and $\xi_0^{\coh}$ approaches an
asymptote set by the inter-particle distance as the system varies from
BCS to BEC.

Importantly, from plots of $n_\pair / n$ such as those in
Fig.~\ref{fig:5}, one sees that $k_\text{F} \xi_0^{\text{coh}}$ allows a very
useful and direct monitoring of the location of a system between the
BCS and BEC limits.  Notably, $k_\text{F} \xi_0^{\text{coh}}$ reaches a
finite lower bound at the onset of the BEC, given by
$k_\text{F} \xi_0^{\text{coh}} \approx 2.2$ for 2D and $1.5$ for 3D (for the
case of $s$-wave superconductors). That these saturation numbers
are of order unity is consistent with what has been anticipated by the
experimental community~\cite{Park2021}. 

We end this section with Fig.~\ref{fig:10}(c) which presents a
``zoomed out'' view of the main figure in Fig.~\ref{fig:10}(b). This
provides information about where one should expect the onset of the
BEC. Importantly, the BEC regime appears to be associated with very
large values of $T^*/T_\text{c}$. In this way, one might expect the BEC limit
to be rather inaccessible.

\subsection{Low carrier density in BCS-BEC crossover}

In this subsection we wish to clarify what one should expect when the
carrier density is dramatically reduced in a lattice superconductor.
For definiteness we will consider only two-dimensional systems here
and presume that ``low density" corresponds well below 1/4 filled
bands, say $n < 0.1$.

The notion that low carrier density promotes a system out of the BCS
regime dates back to Eagles~\cite{Eagles1969}.  Indeed, in the
literature it has been stressed~\cite{Kanigel2008} that when the band
is nearly empty it requires only a small change in the attractive
interaction to push the fermionic chemical potential below the
conduction band bottom; hence the BEC regime is more accessible at low $n$.

What is not so clear is whether low $n$ alone can increase the
magnitude of $T_\text{c}$ (or $T_\text{BKT}$) or not.  Also of
interest is determining whether or not at low densities the nature of
the underlying lattice dispersion becomes irrelevant. If so, this
would mean that the low-density system could be treated as a Fermi gas.

\begin{figure}
\centering
\includegraphics[width=3.0in]
{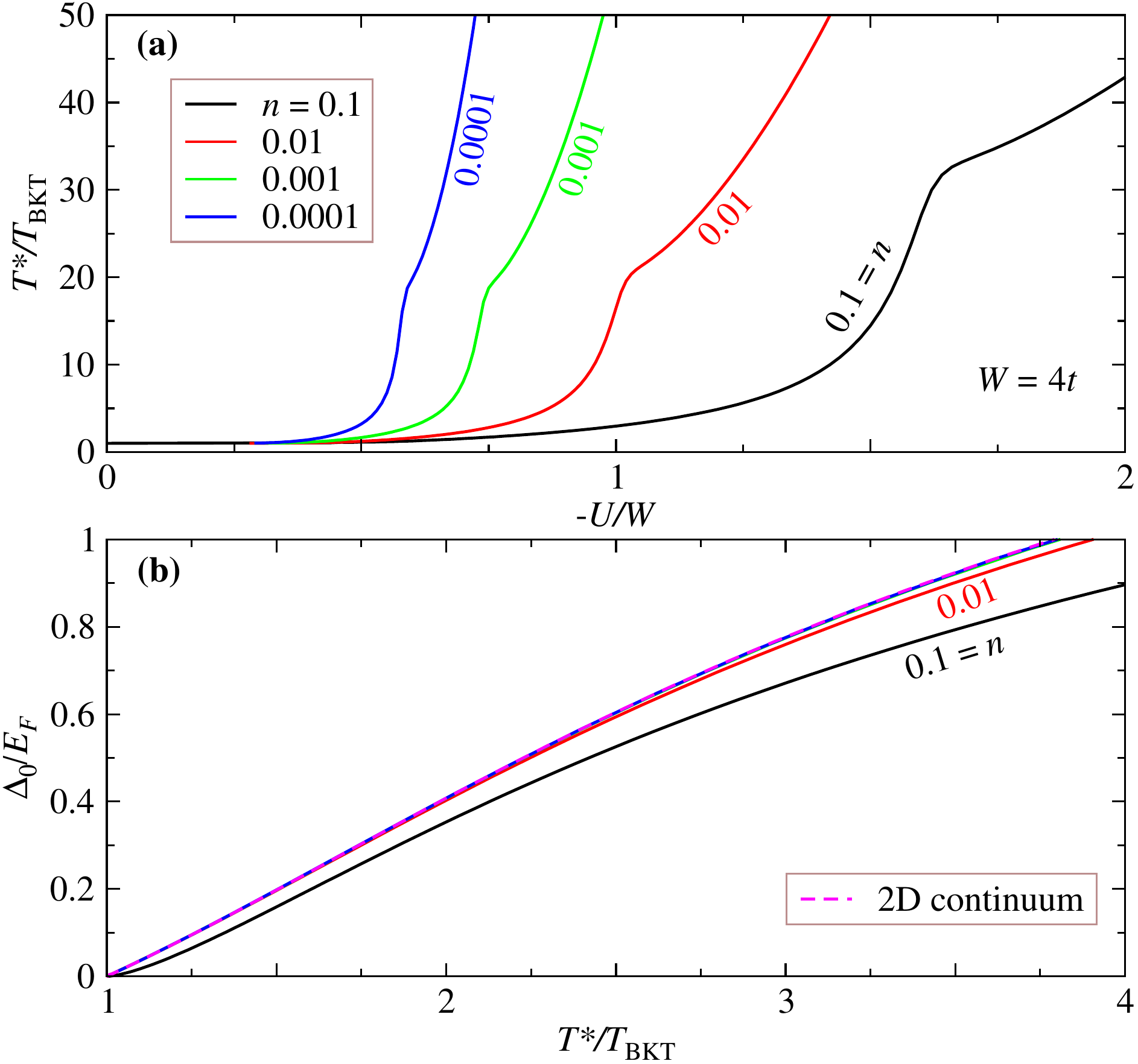}
\caption{(a) Plots of $T^*/T_\text{BKT}$ for an $s$-wave superconductor
  on a square lattice at different electron filling levels $n$, as labeled. The
  normal state band dispersion is
  $\epsilon_{\vect{k}}=4 t - 2 t ( \cos k_x + \cos k_y)$ with $t$ the
  hopping integral. $W=4 t$ is half of the band width. This panel
  shows that low density helps to more readily promote a given
  superconductor out of the strict BCS limit (where the ratio is
  unity). (b) The ratio of the zero-temperature gap to $E_\text{F}$,
  $\Delta_0/E_\text{F}$, versus $T^*/T_\text{BKT}$ for different $n$. This
  lower panel indicates that at extremely low densities and as long as
  $\mu/E_\text{F}$ is neither too small, nor negative, $\Delta_0/E_\text{F}$ plotted
  here is equivalent to the values obtained for a Fermi gas.  The
  sizable $\Delta_0/E_\text{F}$ is indicative of BCS-BEC crossover.  }
\label{fig:LowDens1}
\end{figure}

In the phase-fluctuation approach (of Sec.~\ref{sec:EmeryKivelson})
low density plays a rather dominant role~\cite{Emery1995}.  While this
scenario has been developed primarily for the cuprates, it can be
considered in a broader context, much as the BCS-BEC crossover
scenario is viewed as more generally applicable.  Indeed, one might
wonder if the two scenarios converge in the low carrier density
limit. We find that they do not.

In the phase fluctuation scenario it is emphasized that \textbf{low
 carrier density} is associated with both poor screening and small
phase stiffness or low superfluid density.  Small phase stiffness, in
turn, means that classical phase fluctuations of the superconducting
order parameter become more prominent. These fluctuations necessarily
lead to a more extensive (in temperature) ``critical regime".

\begin{figure}
\centering
\includegraphics[width=3.0in]
{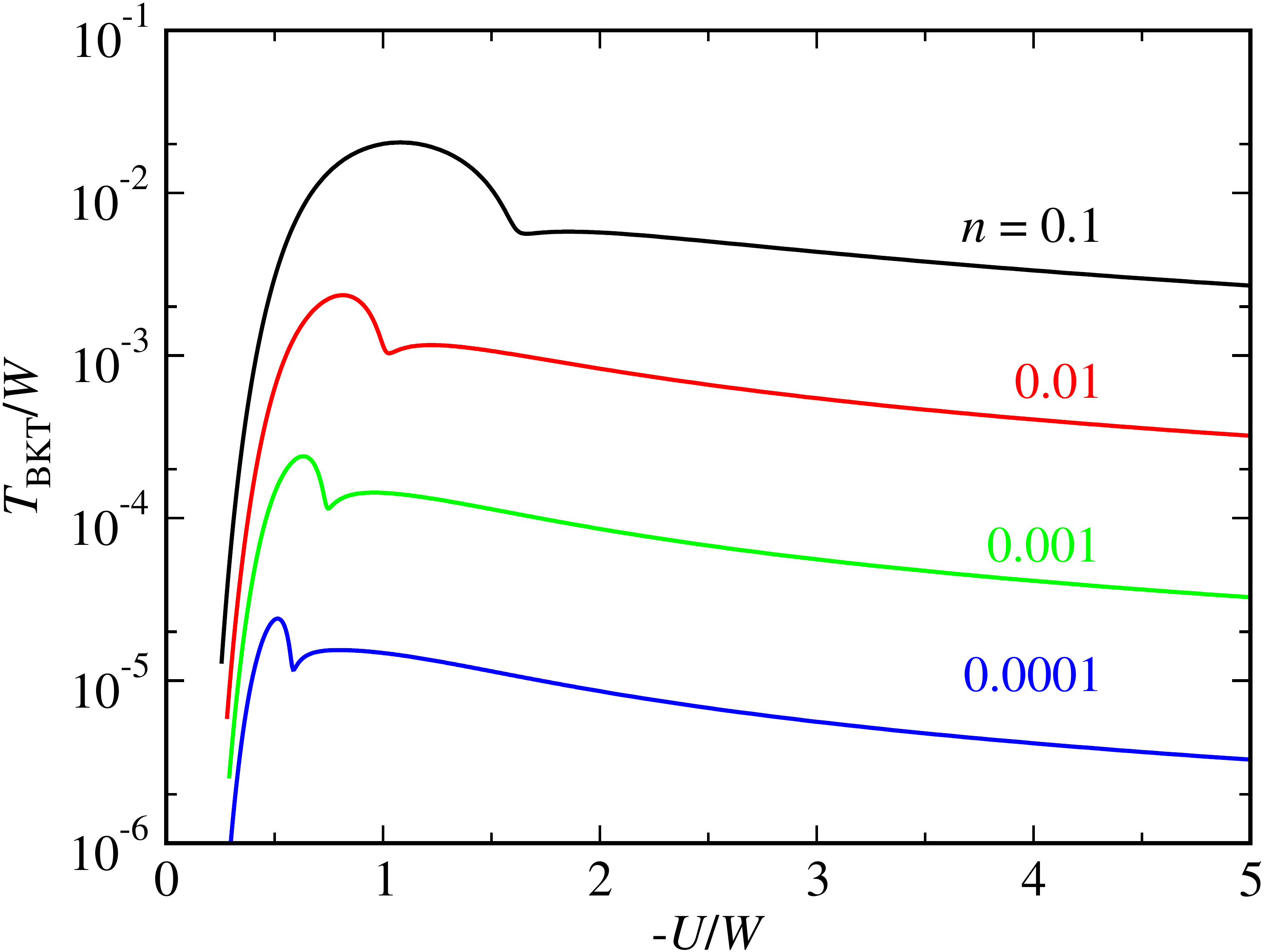}
\caption{$T_\text{BKT}$ as a function of $U$ on a semi-log scale for
  variable carrier density, showing a nearly universal shape but with
  a dramatically decreasing magnitude of the transition
  temperature. The model and dispersion are the same as in
  Fig.~\ref{fig:LowDens1}. The small dips in this figure are
  associated with the crossover to the BEC regime after which the
  canonical $t^2/|U|$ dependence is found for the transition
  temperature.  This dependence is a lattice effect, which persists
  even in the zero carrier density limit. Here $W$ is half the
  bandwidth.  }
\label{fig:LowDens2}
\end{figure}

To address to what extent this scenario is to be distinguished from
the low carrier density limit in BCS-BEC crossover it is useful to
determine what the implications for other properties are: namely, the
size of the transition temperature and of the coherence length along
with $\Delta_0/E_\text{F}$.  We refer to Figs.~\ref{fig:LowDens1} and
\ref{fig:LowDens2} to address these questions.

Figure \ref{fig:LowDens1}(a) presents a plot of $T^*/T_\text{BKT}$ as
a function of pairing interaction normalized to half of the
normal-state bandwidth for a range of different low densities. This
figure is in many ways similar to the inset of Fig.~\ref{fig:10}(a).
It shows that low carrier density does, indeed, promote the system out
of the BCS limit, where $T^* \equiv T_\text{BKT}$. One can determine
from the small kinks in the figure where the Bose condensation regime
sets in. It is evident that, as expected, low density makes this BEC
regime more accessible, as it occurs for smaller attraction strength.

An important message is contained in Fig.~\ref{fig:LowDens1}(b). This figure
shows that $\Delta_0/E_\text{F}$ remains comparably large at low and
relatively high densities for the same $T^*/T_\text{BKT}$. Thus
pairing remains strong and because of the large size of $\Delta_0/E_\text{F}$
(and small size of $k_\text{F} \xi_0^{\text{coh}}$, which is not shown), even
in the very low carrier density limit, it should be possible to
distinguish BCS-BEC crossover from a phase-fluctuation scenario.  It
should be emphasized, however, that the phase-fluctuation approach does
not address fermionic degrees of freedom, so that, strictly speaking,
the pairing gap is irrelevant.

Figure~\ref{fig:LowDens2} presents a plot of the normalized transition
temperature $T_\text{BKT}/W$ as a function of normalized interaction strength
for variable density.
One sees that all curves assume a fairly universal shape, but there is
a dramatic reduction in the size of the transition temperature
as the density is decreased.
One can glean from these observations a notable trend. In both
the case of changing dimensionality from three to two and changing carrier
density from moderate to very low, it follows that the superconductor
is more readily promoted out of the strict BCS regime. But at the same
time, the transition temperatures are significantly reduced.

Another important observation from Fig.~\ref{fig:LowDens2} is that the
effect of the underlying lattice structure is always present in the
BEC regime of the $T_\text{BKT}$ phase diagram~\cite{Chen2012}. In
particular, the $T_{\text{BKT}} \sim t^2/|U|$ asymptote at large $|U|$
persists all the way to the zero carrier density limit, so that a
Fermi-gas description of the phase diagram is not applicable.  At the
same time, interestingly, Fig.~\ref{fig:LowDens1}(b) indicates
$\Delta_0/E_\text{F}$ approaches its counterpart value for a Fermi gas.  This
occurs at extremely low densities but still in the BCS-BEC crossover
regime, where the strength of $|U|$ is such that the fermionic
chemical potential $\mu$ remains positive.

The small size of $T_\text{BKT}$ found here for BCS-BEC crossover at
low density, should not be surprising also from the perspective of the
phase-fluctuation scenario, as the transition temperature, even in 3D,
is governed by the small superfluid density.  But it is interesting to
note that there are instances in the literature when low carrier
density is found to be associated with an \textit{increase} in the
transition temperatures~\cite{Nakagawa2021}.  This would seem to
require that the pairing mechanism is assisted by low density. Although this is highly speculative, one might suspect that
when this occurs Coulomb interactions are driving the pairing and not undermining it.

\begin{figure*}
\centering
\includegraphics[width=6.9in, trim=0 5mm 0 0]{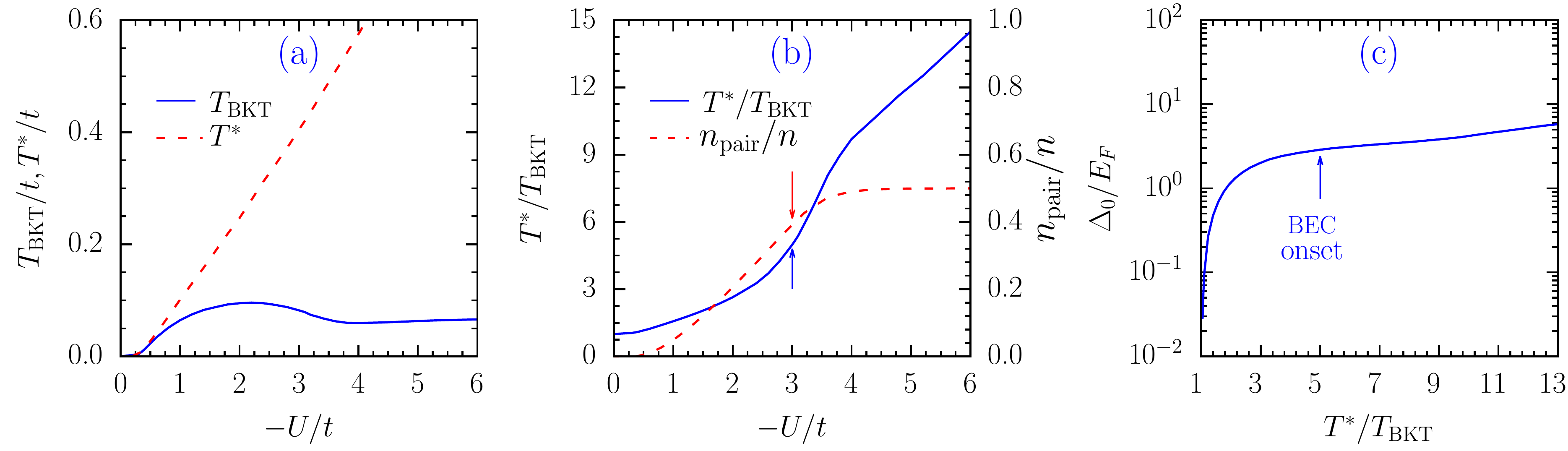}
\caption{Flat band and quantum geometric effects in the BCS-BEC
  crossover theory ~\cite{Wang2020} showing (a) $T_{\mathrm{BKT}}$ and
  $T^*$ for a 2D topological band structure.  (b) Plot of $T^*/T_\BKT$
  as well as the number of pairs as a function of attractive
  interaction strength. The BEC onset, determined from $\mu(T=0)=0$,
  is indicated by arrows. (c) Plots analogous to Fig.~\ref{fig:10}(c),
  but here the BEC appears with a similar $\Delta_0/E_\text{F}$ and
  considerably smaller $T^*/T_\BKT$. $T^*/T_\BKT$ is reduced by
  quantum geometric effects which substantially increase $T_\BKT$
  without affecting $T^*$.  This tight-binding band structure for a
  square lattice (with $t$ the nearest-neighbor hopping) leads to two
  energy bands whose conduction band width is approximately $ 0.2$ times the
  inter-band separation.  Here $n=0.3$ is the electron density per
  square lattice site. }
\label{fig:11}
\end{figure*}

\subsection{Topology and quantum geometry in BCS-BEC crossover}
\label{sec:quantgeometry}

In this Review, we will see that current experimental candidates for BCS-BEC crossover tend to have values of $T^*/T_\text{c}$  of the order of 2 or 3, and corresponding values of $\Delta_0/E_\text{F}$ on the order of $0.5$.  From Fig.~\ref{fig:10}(c), one might infer that these are not likely to be in the BEC regime. There is, however an exception having to do with flat-band, topological systems. These may be relevant to the recent discovery of 2D superconductivity in MATBG and MATTG where there are claims~\cite{Cao2018,Park2021,Kim2022} that these flat-band systems are somewhere between BCS and BEC (MATBG) or even beyond, that is,
within the BEC regime (MATTG). 
  
Experimentally, when twist angles in these graphene systems are associated with very flat bands, this seems  to correlate with the highest transition temperatures. There is, however, a subtle and important feature here. In flat-band superconductors, pair hopping, like single-particle hopping, is also suppressed~\cite{Torma2022,Peotta2015,Wang2020}. As a consequence, the pair mass $M_\pair$ becomes large and the superfluid stiffness is small. This would lead to a vanishing $T_\BKT$ in the extremely flat-band limit, were it not for multi-band/multi-orbital effects. Moreover, it has been emphasized~\cite{Peotta2015} that these latter interband contributions (which work to decrease the pair mass) can be amplified in the presence of nontrivial normal-state band topology. This occurs through so-called quantum geometric effects.

Such multiband effects have been incorporated into a 2D $s$-wave BCS-BEC crossover framework~\cite{Wang2020} where a phase diagram with the usual superconducting dome is found, as shown in Fig.~\ref{fig:11}(a). The model topological Hamiltonian yields two bands, whose conduction bandwidth is much smaller than the inter-band energy separation. The calculated phase diagram resembles that obtained from Monte Carlo results~\cite{Hofmann2019} using the same model Hamiltonian. 

Importantly, this phase diagram can be used to extract the ratio $T^*/T_\BKT$ along with the number of bosons $n_\pair/n$, as shown in Fig.~\ref{fig:11}(b); both these variables are plotted as a function of renormalized interaction strength. The quantity $n_\pair$ provides a ready indication of where the BEC sets in, as here $n_\pair$ first reaches $n/2$. 

At the transition point to the BEC regime (indicated by the arrows), the interaction strength $U$ is on the order of the entire  conduction band width. Correspondingly, $\Delta_0/E_\text{F} \sim 3$ as shown in Fig.~\ref{fig:11}(c), which is not so different from the single band result in Fig.~\ref{fig:10}(c). On the other hand, because of quantum geometry, $T_{\BKT}$ is substantially enhanced by inter-band effects while $T^*$ is almost unaffected, leading to a smaller and physically more accessible value of $T^*/T_\BKT \sim 5$.  This behavior is summarized in Fig.~\ref{fig:11}(c), where the BEC onset point is indicated by the arrow.  This provides a counterpart plot of Fig.~\ref{fig:10}(c) but here for a multi-band, topological case. We note that the value of $\Delta_0/E_\text{F}$ at the BEC onset is non-universal.  For a topological band structure with an extremely flat conduction band~\cite{Wang2020}, $\Delta_0/E_\text{F}$ can be as large as $30$. 

The above contrast leads us to the interesting conclusion that in the presence of flat bands and non-trivial band topology a BEC phase can potentially become more accessible, as it leads to a moderate size for $T^*/T_\BKT$.
We emphasize that these effects derive from the participation of more than one band in the superconductivity
and note, for completeness, that there are other, rather different approaches 
in the literature which also treat BCS-BEC crossover phenomena 
in multi-band systems both analytically~\cite{Chubukov2016,Tajima2021} and numerically~\cite{Loh2016}.

\section{Strongly Disordered Conventional Films: Two Energy Scales and a Pseudogap}
\label{Sec:Disorder}

We return to Fig.~\ref{fig:10}(b)  noting that this figure presents a unique signature of 2D pseudogap effects associated with a strong-pairing mechanism. It may seem surprising, but strong disorder can lead to similar pseudogap effects in 2D superconducting films ~\cite{Sacepe2010}. However, the parameters governing these dirty thin films are very different from those indicated in Fig.~\ref{fig:10}(b). In understanding the origin of this other pseudogap, it is important to recall that 2D superconductors have a propensity for manifesting a separation of the two energy scales $T^*$ and $T_\BKT$, which can be thought of as corresponding to the onset temperatures for amplitude and phase coherence, respectively.  As an important signature, those conventional superconducting films in which the two temperature scales are well separated due to disorder~\cite{Sacepe2010,Zhao2013} will have rather small values of $\Delta_0/E_\text{F} $. 

While the distinctions between the two scenarios for a pseudogap (strong pairing and strong disorder) should be obvious, a number of phenomenological similarities are rather striking. Most notable among these are the reported observations of charge $2e$ pairs~\cite{Bozovic2020,Bastiaans2021}, the contrasting behavior of Andreev and conventional tunneling~\cite{Oh2021,Dubouchet2019}, and the observations of boson or pair localization~\cite{Hollen2011,Chen1999}.

\begin{figure}
\centering\includegraphics{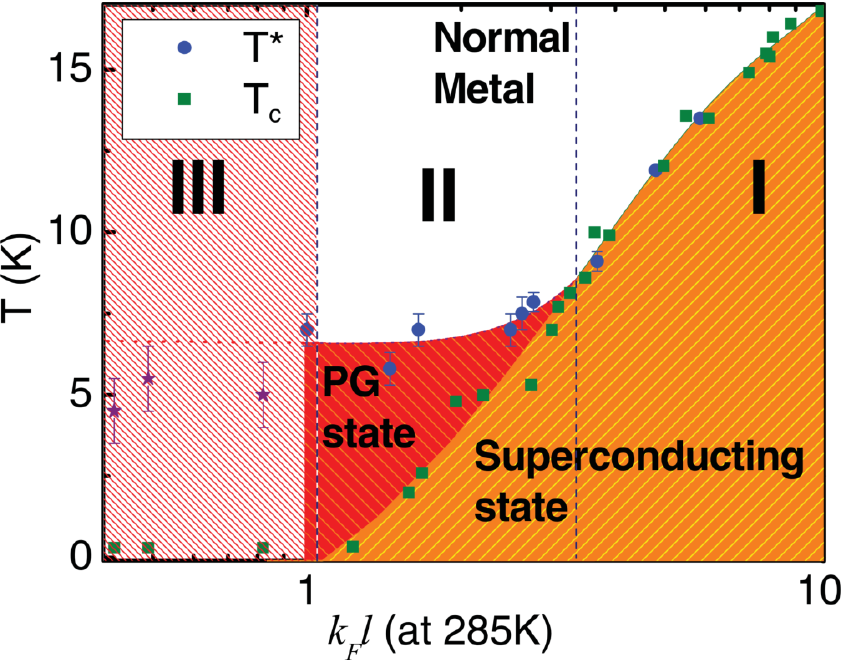}
\caption{Experimental temperature scales as a function of
mean free path $k_\text{F} \textit{l}$ in disordered NbN films. The value of $k_\text{F} \textit{l}$ is determined from resistance and Hall-coefficient measurements at $T=285$K.  With increasing disorder, or sufficiently small $k_\text{F} \textit{l}$, a pseudogap (PG) phase appears associated with $T^* \neq T_\text{c}$ in region II, while in region III, $T_\text{c}$ is zero although pairing likely persists in this insulating phase. From \textcite{Chand2012}.}
\label{fig:12}
\end{figure}

The behavior found rather generically for a highly disordered 2D superconductor is illustrated in Fig.~\ref{fig:12}, which represents an experimentally determined phase diagram~\cite{Chand2012} with temperature on the vertical axis and disorder measured through $k_\text{F} \textit{l}$ on the horizontal  axis. Here
$\textit{l}$ is the electron mean free path. In Fig.~\ref{fig:12}, the superconducting state is shown in orange, the pseudogap state in red, and the normal-state metal in white. Also indicated are the temperatures $T^*$ and $T_\text{c} = T_\BKT$.

There are three demarcated regions. At very small disorder (region I) a pseudogap is absent and $T^* \approx T_\text{c}$, while as disorder increases (region II), $T^*$ separates from $T_\text{c}$ and is relatively independent of the disorder strength, while the transition temperature (which is more sensitive to the undermining of coherence) rapidly decreases. Finally in region III, $T_\text{c}$ vanishes although there are indications that pairing persists. The two temperatures become distinct at a critical value of $k_\text{F} \textit{l}$.

These experiments on NbN are reasonably generic and similar observations have been made for TiN and InO$_x$ as well~\cite{Sacepe2010}, where the authors claim that a pseudogap appears to be present,  reflecting the existence of paired electrons above $T_\BKT$. Importantly, this pseudogap is found to be continuously and directly transformed into a superconducting gap below the transition.

\begin{figure}
\centering\includegraphics[width=2.5in,clip] 
{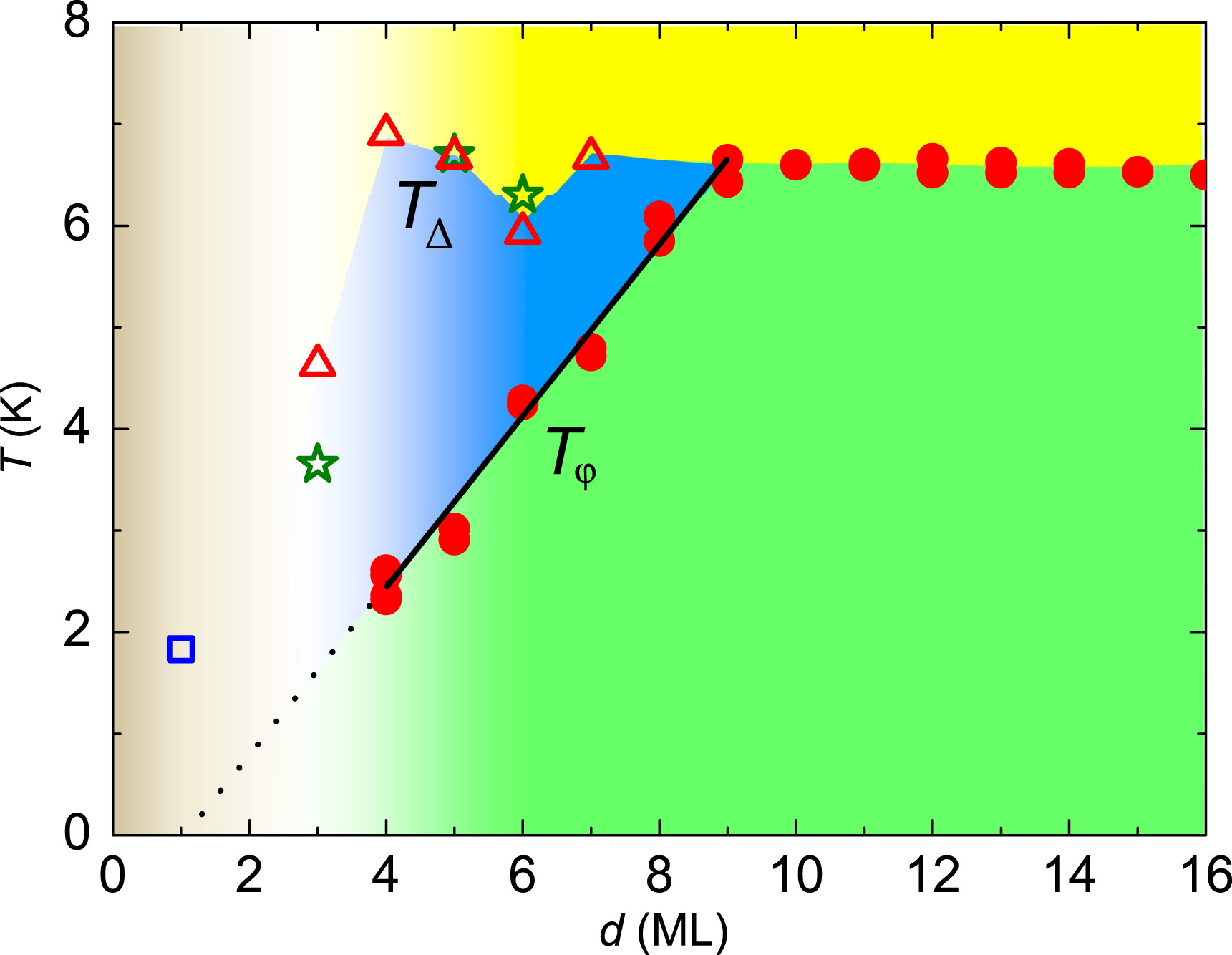}
\caption{Experimental behavior of characteristic temperatures $T^* \equiv T_{\Delta}$ and $T_{\varphi} \equiv T_\BKT$ as a function of the thickness $d$ of Pb films. 
``ML" stands for monolayer. A more extensive analysis of the resistivity (see text) suggests that 
the evident pseudogap effects, here and in the previous figure,
are likely associated with high disorder, rather than strong pairing correlations. From \textcite{Zhao2013}.}
\label{fig:13}
\end{figure}

An interesting set of parallel experiments~\cite{Zhao2013} shown in Fig.~\ref{fig:13} was performed on Pb films by a group at Tsinghua University, who determined the experimental phase diagram obtained by studying crystalline and atomically flat Pb films, now as a function of variable thickness. In Fig.~\ref{fig:13}, temperature appears on the vertical axis and increasing thickness on the horizontal axis. Here the superconducting state is shown in green, the ``fluctuating'' or pseudogap state in blue (where non-superconducting Cooper pairs are said to exist) and the normal-state metal in yellow. The solid circles represent superconducting or phase-coherent order, as determined by transport with an onset at $T_{\varphi} \equiv T_\BKT$; the open symbols represent the pairing transition ($T_{\Delta} \equiv T^*$), which is established by tunneling spectroscopy.

From Fig.~\ref{fig:13}, one can infer that the pairing temperature remains nearly constant with variable thickness, while the coherence temperature is strongly depressed. This appears to suggest that disorder may be playing a role~\footnote{Since $T^*$ essentially represents a mean-field transition temperature of an $s$-wave superconductor, this should satisfy Anderson's theorem~\cite{Anderson1959} of disordered superconductivity; $T^*$ is thus expected to remain relatively robust in the presence of weak disorder that does not break time-reversal symmetry, provided the effective pairing interaction is not strongly affected by localization effects.}, as supported by the sheet resistance data measured by the same group.

It is reasonably well established that, quite generally, $T_\BKT$ decreases with decreasing thickness in 2D films~\cite{Khestanova2018}, although there is no consensus on the extent to which disorder is the only relevant mechanism. The central point, then, is that pairs form at higher temperatures than those at which they exhibit superfluidity. Equivalently, at $T_{\varphi}$, while phase coherence is destroyed, the superconducting gap remains non zero. Note that for Pb, the two characteristic temperatures merge in the 3D regime, as is the hallmark of a ``conventional'' weak-coupling bulk superconductor.

\begin{figure}
\centering\includegraphics[width=2.5in]{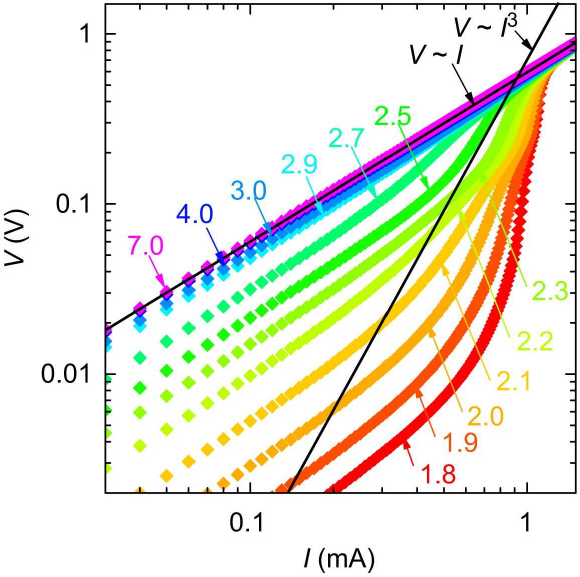}
\caption{$V$-$I$ isotherms on a log-log plot associated with the Pb films in the previous figure.  Each curve is labeled by its temperature (in units of Kelvin) and a straightforward analysis identifies $T_\BKT$ with the $V \propto I^3$ black line. One sees that the $V$-$I$ characteristics display a continuous evolution towards Ohmic behavior as the temperature is raised to the pairing onset temperature $T^*$, here identified to be $7\,$K (the $V\sim I$ black line) for a Pb film of a particular fixed thickness. From \textcite{Zhao2013}.}
\label{fig:14}
\end{figure}

A key finding of the Tsinghua group~\cite{Zhao2013} pertains to the voltage-current ($V$-$I$) characteristics, which provides an alternative method for deducing the pairing onset temperature $T^*$. We emphasize that this ``short-cut'' procedure should be applicable to all 2D superconductors. More precisely, the authors have shown that $V$-$I$ plots of this type can be used to \textit{simultaneously} measure the two important energy scales $T^*$ and $T_{\textrm{BKT}}$. This is illustrated in Fig.~\ref{fig:14} where voltage-current plots are presented for a range of different temperatures in one particular Pb thin film.

More specifically, it is well known~\cite{Halperin1979} that estimates based on $V$-$I$ curves allow one to determine the BKT transition, which occurs when the condition $V \propto I^{\alpha}$  is satisfied with a particular value of $\alpha = 3$. Importantly, the authors in the Pb experiments~\cite{Zhao2013} have pointed out that one can also obtain the pairing onset temperature, $T^*$, from $V$-$I$ plots. This is associated with the recovery of fully Ohmic behavior shown in Fig.~\ref{fig:14} by the $V\propto I$ black line.

While this observation could seem intuitively obvious, the authors have made the last point more convincing by accompanying their analysis with more direct measurements of the pairing gap through scanning tunneling microscopy (STM) experiments, which yield $\Delta(T)$ and hence $T^*$. We caution by noting that one should take care in establishing the ``Ohmic recovery'' temperature as it involves the behavior of the entire $V$-$I$ curve, for an extended range of $I$ above the critical current.

\section{Application of BCS-BEC Crossover in the Literature (Beyond  Fermi Gases)}
\label{sec:Application}

In this section we present summaries of experimental observations
concerning candidate systems for BCS-BEC crossover. We will show that
the majority of the candidates appear consistent with this scenario,
as they possess all or most of the (first three) discriminating
properties listed in Sec.~\ref{sec:signatures}.  These correspond to
(i) the observation of large $\Delta_0/E_\text{F}$, (ii) the presence of a
normal-state pseudogap so that $T^*/T_\text{c}$ is significantly above $1.0$,
and (iii) a moderately short coherence length,
$k_\text{F} \xi_0^{\text{coh}}$.  Also reported in a few cases is the
observation of enhanced superconducting fluctuation-like behavior in
the normal state, particularly in the response to a magnetic
field~\cite{Li2010,Proust2018}.

Notably, however, what is missing in a number of cases (particularly
for the organic superconductors~\cite{Suzuki2022} and the two twisted
magic angle graphene systems) is information about how the very
important temperature scale $T^*$ varies across their respective
$T_\text{c}$ domes. We note that in strictly 2D systems this appears
to be reasonably accessible should there be future measurements of the
$V\text{-}I$ characteristics. This capability was discussed in
Sec.~\ref{Sec:Disorder}, based on the Ohmic recovery temperature, which
effectively yields $T^*$.

Overall, what seems to be nearly universally observed in these
candidate BCS-BEC crossover superconductors is a large magnitude for
$\Delta_0/E_\text{F}$ and a relatively small size for the GL
coherence length $k_\text{F} \xi_0^{\text{coh}}$.  The focus on this last
quantity serves to emphasize the striking contrast with the Fermi
gases, where this coherence length is not as readily accessible.

Connections to more specific aspects of BCS-BEC crossover theory are
presented in Sec.~\ref{sec:Concl} via a summary figure
(Fig.~\ref{fig:30}) for all the candidate materials in 2D.  Unlike the
Fermi gases, where the magnitude of the attractive (Hubbard-like)
interaction can be quantified, here one has to circumvent this
parameter.  As will be shown, in the plot we address correlations of
$T^*/T_\text{BKT}$ and $\Delta_0/E_\text{F}$ instead.  Moreover, a related
plot that focuses on commonalities between the graphene and cuprate
families is Fig.~\ref{fig:34}.  While in Fig.~\ref{fig:30} a simple
tight-binding band structure is used for all candidate materials, we
argue in the spirit of this Review, the specific details of the
band structure are viewed as less important than distinguishing between
$s$- and $d$-wave pairing symmetries, or 2D and 3D systems or
addressing some of the more universal features of the crossover.

\subsection{BCS-BEC crossover in  2D organic conductors}

Over the years there have been observations that a class of quasi-2D
organic superconductors based on the BEDT-TTF molecule, of the type
$\kappa$-(BEDT-TTF)$_2$X, might have something in common with the high-temperature superconductors~\cite{Mckenzie1997}. 
Here, X is an inorganic anion and $\kappa$ denotes a particular packing arrangement
in the crystal. The basic unit here is a dimer consisting of two
BEDT-TTF molecules stacked on top of one another.  Upon binding with
the anion the dimer provides one electron to the anion leaving behind
a mobile hole.

The similarity with the cuprates has been based on the
observations~\cite{Imajo2021,Matsumura2022,Oike2017} of competing
metallic, insulating, superconducting and antiferromagnetic states in
the phase diagram, which is generally plotted as a function of
pressure.  As the pressure decreases (presumably in analogy to a
decrease in doping in the cuprates) the properties of the molecular
conductor (and its superconducting phase) deviate progressively from
those of a typical metal (and BCS superconductor).  Conversely with an
increase in pressure the behavior appears more conventional.

Of some interest is the case where X involves HgBr (more particularly
one studies $\kappa$-(BEDT-TTF)$_4$Hg$_{2.89}$Br$_8$) in the ``parent"
compound of these systems, which seems to exhibit features of a quantum
spin liquid~\cite{Suzuki2022,Powell2011}.  This quantum spin liquid is
associated with a frustrated spin configuration, often modeled
theoretically~\cite{Yokoyama2006} by a triangular Hubbard lattice.
Notably~\cite{Imajo2021,Oike2017,Matsumura2022} with varying pressure
this particular class of organic superconductors exhibits 
possible $d_{x^2-y^2}$ ordering, and transition temperatures as high
as $7\sim 10$ K, with suggestions of pseudogap behavior for $T > T_\text{c}$.  One
also sees an unexpectedly large slope for
$dH_{c2}/dT$ near $T_\text{c}$ in both fields in parallel with and
perpendicular to the two-dimensional conducting layers.  There is 
also a very wide region of fluctuating superconductivity above
$T_\text{c}$, along with a large superconducting energy gap.

\begin{figure}
\centering
\includegraphics[width=3.0in]{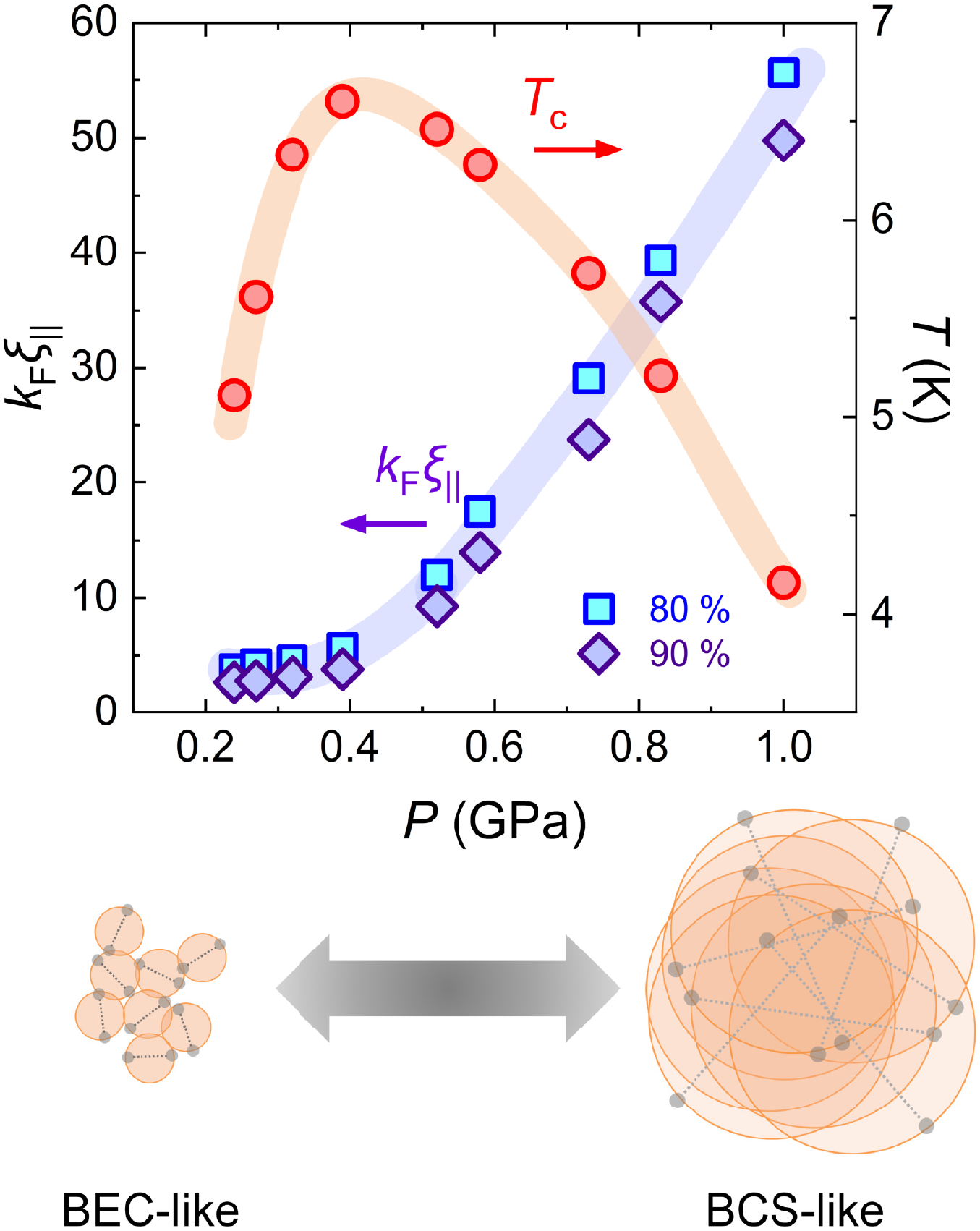}
\caption{Pressure dependence of the measured in-plane coherence length
  $k_\text{F} \xi_0^{\text{coh}}$ and superconducting transition
  temperatures in $\kappa$-(BEDT-TTF)$_4$Hg$_{2.89}$Br$_8$.  Here $k_\text{F}$ is determined by the Hall
  coefficient. If we assume that pressure scales
  inversely with the effective attractive interaction strength,
  the $T_\text{c}$ dome with overlain coherence length provides a
  rather ideal prototype for BCS-BEC crossover in the solid state. From
  \textcite{Suzuki2022}.}
\label{fig:Mott1}
\end{figure}

What is particularly relevant to the present Review is that recent
studies have more quantitatively addressed pressure variations in
$\kappa$-(BEDT-TTF)$_4$Hg$_{2.89}$Br$_8$ in the context of BEC-BCS
crossover.  It is presumed that pressure works to enhance the
itinerant nature of electrons through the increase of the transfer
integral $t$ between molecular orbitals, leading to a pressure
dependent band structure.  Thus, one might imagine in the context of
Fig.~\ref{fig:10} that variable pressure could cause a variation in
$T_\text{c}$ through the generic phase diagram parameter $|U|/t$; as
$t$ increases the dimensionless interaction strength decreases, thus
moving the system closer to the BCS regime.

Indeed, this is what is observed in Fig.~\ref{fig:Mott1}.  Of
considerable interest in this figure are the combined plots of the
in-plane coherence length $k_\text{F} \xi_0^{\text{coh}}$ and the transition
temperature.  Here, if we implicitly make the assumption that pressure
scales inversely with $|U|/t$, this provides a
pedagogical and rather powerful representation of BEC-BCS crossover.
This figure appears rather consistent with the various plots shown in
Fig.~\ref{fig:10} of the $T_\text{c}$ dome and the behavior of the
coherence length. Notably, for a $d$-wave gap symmetry, the smallest
value reached by $k_\text{F} \xi_0^{\text{coh}}$ will be significantly larger 
than for $s$-wave symmetry, since the BEC limit is
generally not reachable for these extended-size pairs. (See also
Fig.~\ref{fig:6}).

\begin{figure}
\centering
\includegraphics[width=3.0in]{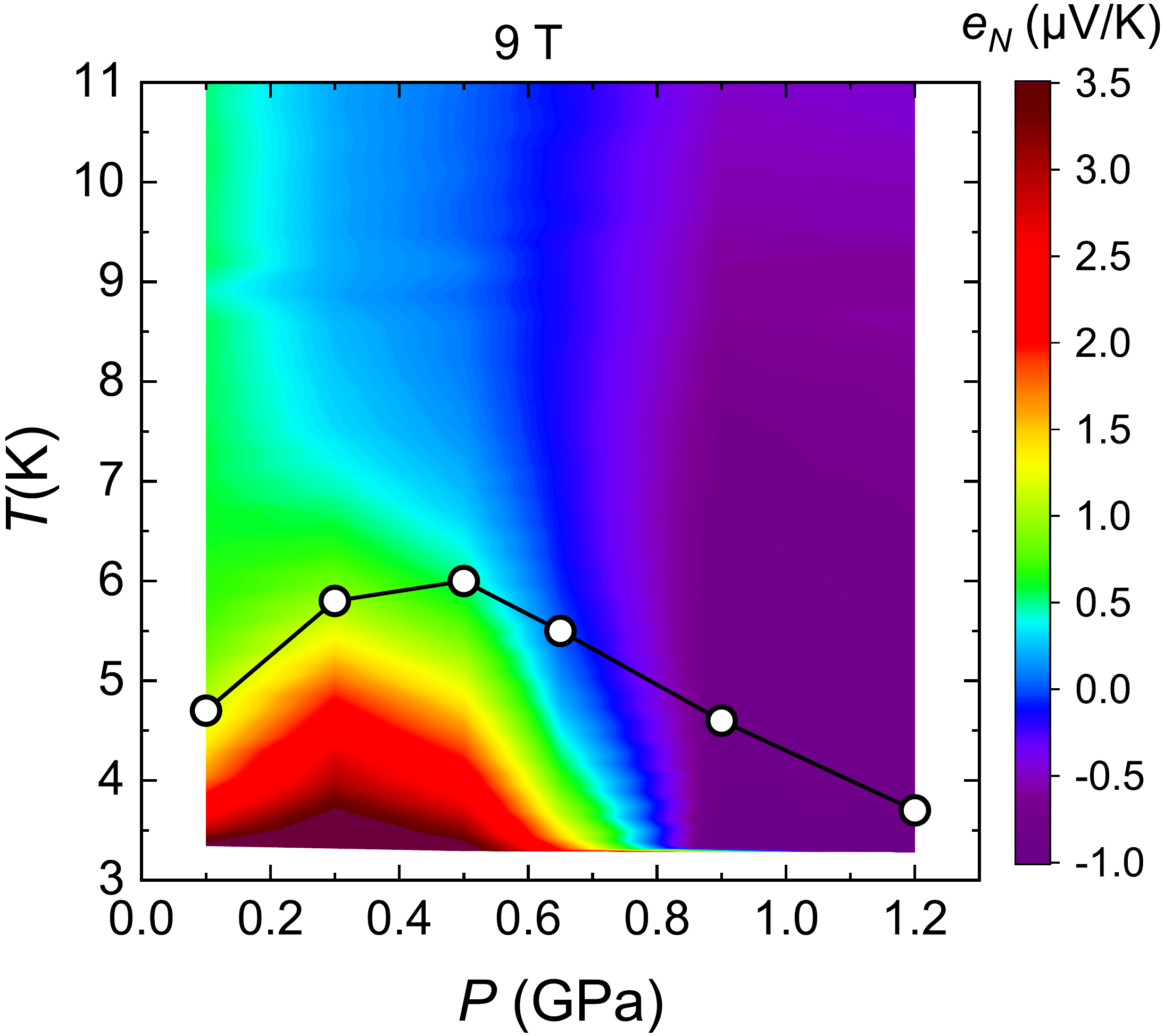}
\caption{Temperature-pressure plot of the Nernst signal
  $e_{\text{N}}$ for the organic superconductor at a
  magnetic field of 9~T. The white circles indicate the zero-field
  transition temperature $T_\text{c}$. As the system becomes more strongly
  paired with decreasing pressure a positive Nernst response is
  enhanced at temperatures far above $T_\text{c}$. From
  \textcite{Suzuki2022}.}
\label{fig:Mott2}
\end{figure}

Adding support to the picture of BCS-BEC crossover in this family of
organic metals are studies of nuclear magnetic resonance (NMR)~\cite{Mckenzie1997,Kanoda1996} and
the Nernst coefficient.  Interestingly in a closely related organic
superconductor~\cite{Mckenzie1998}, NMR experiments have provided
evidence for $d$-wave pairing as well as a pseudogap.

We turn next to the Nernst studies~\cite{Suzuki2022} in the HgBr
system, as shown in Fig.~\ref{fig:Mott2}.  In the strong-pairing regime, quite generally, as $T_\text{c}$ is approached from above, 
the Nernst coefficient acquires~\cite{Boyack2021} a large positive (magnetic-field dependent) value which peaks within the superconducting state
and subsequently falls below.  From Fig.~\ref{fig:Mott2} it can be
seen that the Nernst coefficient becomes anomalously large well above
the transition temperature, for low pressures where the
molecular superconductor is closest to the strong-coupling end of the spectrum.  This
enhancement of the standard Aslamazov-Larkin (AL) contribution is
expected~\cite{Boyack2021}. It reflects the fact that the
non-condensed pairs have a more extended temperature region where the
chemical potential of the pairs, $|\mu_\text{pair}|$ (which governs
the size of the AL contribution), becomes small. Such an enhancement
becomes more pronounced as the system deviates progressively from the
BCS regime.

In summary, these studies of $\kappa$-(BEDT-TTF)$_4$Hg$_{2.89}$Br$_8$
seem to suggest a welcome convergence between different schools of
thought for treating strongly correlated superconductors through the
concept of ``Mott-driven BCS-BEC crossover".  In the context of the
cuprates both the ``doped Mott insulator"~\cite{Lee2006} and the
``BCS-BEC" scenarios have been widely discussed. It would appear in
this organic superconductor system that both aspects are combined:
Mott physics may well provide the source of the pairing mechanism,
while BCS-BEC crossover appears relevant to the machinery.

\subsection{BCS-BEC crossover in the iron chalcogenides}

Considerable attention has been paid to superconducting properties of
the iron chalcogenides~\cite{Kasahara2016,Kasahara2014,Okazaki2014,Mizukami2021,Hanaguri2019,Shibauchi2020,Kang2020},
where there appears to be growing evidence that FeSe and isovalent
substituted FeSe$_{1-x}$S$_x$ and FeSe$_{1-x}$Te$_x$ may be in the
BCS-BEC crossover regime. These systems, in which the characteristic
electronic energy scales are anomalously low, appear to exhibit strong-pairing effects. 
This is not due to two dimensionality, nor is it because the
pairing ``glue'' itself is particularly large on an absolute
scale. Rather, the attractive interaction is large when compared to
the characteristic very low Fermi energies. Also present, and possibly
relevant are nematic effects~\cite{Shibauchi2020,Hashimoto2020}
associated with broken rotational symmetry (but preserved
translational symmetry). FeSe is a layered anisotropic material; it is
also a compensated semi-metal, with roughly equal densities of
electron and hole carriers. This leads to both electron and hole
pockets and a more complicated scenario for BCS-BEC crossover.

Adding to the support for a BCS-BEC crossover picture is the fact that
in the iron chalcogenides~\cite{Shibauchi2020} the characteristic
Fermi energies and zero-temperature gap magnitudes are comparable. STM and
STS experiments indicate gap sizes of the order of
$\Delta_1 \approx3.5$~meV and $\Delta_2 \approx2.5$~meV for the two
bands. From this it follows that the ratios of the pairing gaps to
transition temperatures ($T_\text{c} \approx 9$~K) in FeSe are large, of the
order of $2 \Delta_1/k_\text{B} T_\text{c} \approx 9 $ and
$2 \Delta_2/k_\text{B}T_\text{c} \approx 6.5$, well beyond the BCS value of
$3.5$. The Fermi energies associated with the two nearly cylindrical
Fermi surface sheets are anomalously small, of the order of
$E_\text{F}\approx 10 \sim 20$~meV for the hole-like Fermi
surface~\cite{Shibauchi2020}.  Importantly this leads to estimates of
$T_\text{c}/T_\text{F} \approx 0.04\sim 0.08$.
This analysis has led many to conclude that these superconductors are
well outside the strict BCS regime.

ARPES experiments~\cite{Hashimoto2020} on bulk FeSe show that rather
than the characteristic back-bending associated with conventional BCS
superconductors, there is instead a flat dispersion near
$\vect{k} = 0$, which appears to be more typical of the crossover
regime. This flat-band feature is even more enhanced with the addition of sulfur.

Of considerable importance is the characteristic correlation length
extracted from magnetic field data~\cite{Kasahara2014} which is argued
to be small, of the order of
$k_\text{F} \xi_0^{\text{coh}} \approx 1-4$. One can deduce from these
numbers that FeSe superconductors are most likely not in the BCS
regime. One should also compare with earlier theoretical estimates of
$k_\text{F} \xi_0^\text{coh}$, which found a BEC saturation value of
approximately 2 to 3 [Fig.~\ref{fig:10}(c)].  We caution, however,
that complementary diagnostic information comes from vortex imaging
using STM. This derives from the subgap fermionic states that are
inside the vortex core. The observation of Friedel-like oscillations
~\cite{Hanaguri2019,Chien2006b} suggests that fermionic degrees of
freedom are still present in bulk FeSe and thus these superconductors
are not yet in the BEC regime.

\begin{figure}
\centering\includegraphics[width=3.0in]
{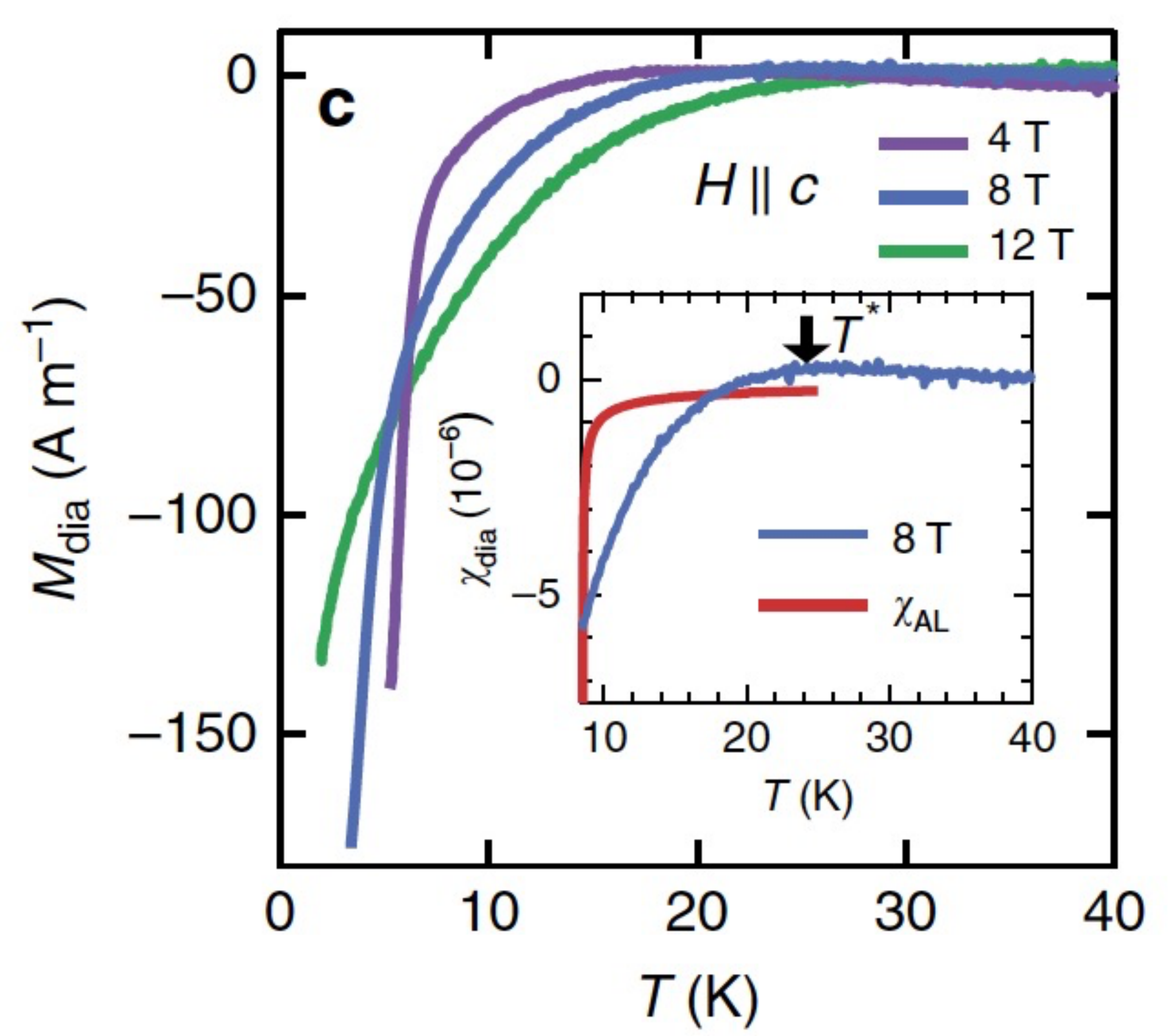}
\caption{Diamagnetic magnetization response in
  bulk FeSe as a function of temperature at different values of the
  applied magnetic field $H$. The inset presents a comparison of the
  diamagnetic susceptibility $\chi_{\text{diag}}$ with the predictions
  $\chi_{\text{AL}}$ of Aslamazov-Larkin (AL) theory~\cite{Larkin2009}, showing a very extended range of fluctuations. From \textcite{Kasahara2016}.}
\label{fig:15}
\end{figure}

Also notable is that there are enhanced superconducting fluctuation effects~\cite{Kasahara2016} in FeSe. This enables identification of a characteristic temperature $T^*$ where, in particular, diamagnetism sets in. Figure~\ref{fig:15} presents a plot of this ``unprecedented, giant'' diamagnetic response. The inset serves to emphasize the key point that the diamagnetic fluctuation regime in FeSe is considerably wider than predicted from the conventional fluctuation theory of Aslamazov and Larkin~\cite{Aslamazov1975, Larkin2009}. It is argued that this provides evidence for preformed pairs associated with BCS-BEC crossover, as fluctuation effects are expected to be amplified. Similarly, studies of the DC conductivity show that the expected 
downturn behavior is observed in the resistivity.  Additionally, NMR experiments~\cite{Shibauchi2020} show the expected suppression of $1/(T_1T)$ around $T^*$, although there seems to be~\cite{Hardy2019} none of these large fluctuation effects in the heat capacity.

\begin{figure*}
\centering\includegraphics[width=5.4in]
{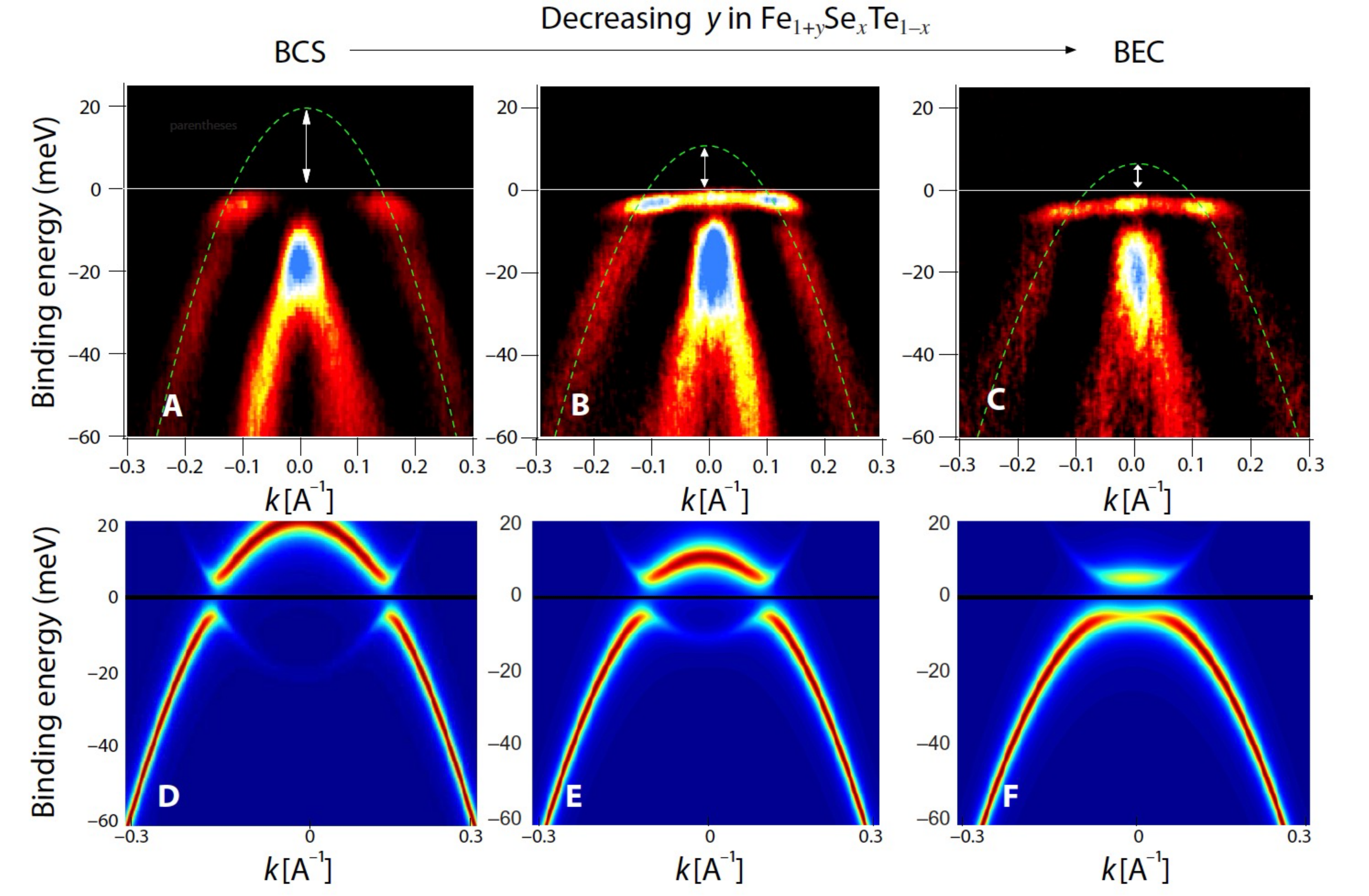}
\caption{ARPES signatures in Fe$_{1+y}$Se$_x$Te$_{1-x}$, where
  chemically doping the carrier concentration is through decreasing
  $y$. Shown in (A-C) are ARPES spectra for three
  samples in the order of decreasing $y$ from left to right. The green
  dashed line is the best fit to the data. Shown in (D-F) are theory
  plots using parabolic band dispersion and other model parameters. From \textcite{Rinott2017}.}
\label{fig:16}
\end{figure*}

There has also been a focus on crossover from BCS to BEC in a slightly
different iron chalcogenide~\cite{Rinott2017},
Fe$_{1+y}$Se$_x$Te$_{1-x}$, where chemically doping the carrier
concentration, through decreasing $y$, introduces an increased ratio
of $\Delta_0/E_\text{F}$, where $E_\text{F}$ is as small as a few
meV. Here, for example, there are claims\footnote{There are
  complications in this analysis due to the vicinity of a heavy
  $d_{xy}$ band, which may affect the interpretation.} based on figures
such as Fig.~\ref{fig:16} that as $\Delta_0/E_\text{F}$ increases, the
dispersion of the peak in ARPES evolves from the characteristic
back-bending behavior seen in the BCS regime to a BEC-like signature
with a gap minimum at $\vect{ k} =0$.

All of this would make a nice illustration of superconductivity in the
intermediate and even strong-coupling regime were it not for the fact
that STM/STS experiments do not support the existence of a
spectroscopic pseudogap~\cite{Shibauchi2020} in this class of
compounds. Understanding this behavior is still a work in progress; it
can be speculated that the multiband character of the iron
chalcogenides may be relevant here. Issues such as inter-band pairing
may also be playing an important role.

\subsection{BCS-BEC crossover in interfacial superconductivity}

A great deal of excitement has been generated recently in studies of
interfacial superconductivity~\cite{Richter2013, Zhang2019, Cheng2015,
  Kang2020, Caviglia2008, Reyren2007, Gariglio2015, Zhang2014a,
  Song2021, Han2021, Suyolcu2017, Gasparov2017, Wang2017, Rebec2017,
  Ge2015}, particularly involving the iron chalcogenide FeSe. Here one
sees an unexpected and dramatic enhancement of the pairing onset
temperature~\cite{Ge2015} in interfacial monolayer FeSe. While the
early literature~\cite{Rebec2017,Zhang2014a,Pedersen2020} did not
often distinguish this pairing gap onset from that of coherent
superconductivity, it is now becoming clear that this system is
associated with a large pseudogap, as well as a sizeable BKT
transition temperature.

Indeed, it was discovered in 2012~\cite{Wang2012} that
one-unit-cell-thick (1UC) FeSe grown on SrTiO$_\text{3}$ exhibits a
gap which survives up to $60\sim 70$~K. This remarkable gap onset
temperature is one order of magnitude higher than the $T_\text{c}$ of
bulk FeSe, and it has inspired an enormous effort to reveal the
mechanism driving the interfacial enhancement. Due to the extreme air
sensitivity, it has been challenging to perform traditional
resistivity measurements. FeTe-capping or \textit{in situ} transport
measurements have made it possible to characterize the $T_\text{c}$
from the resistivity transition. Among these measurements, except for
a singular study that reported a $T_\text{c}$ of 109 K, all other
transport studies reported a resistivity onset associated with
coherent superconductivity at $T\lesssim 45$~K.

\begin{figure*}
\includegraphics[width=5.4in,clip]
{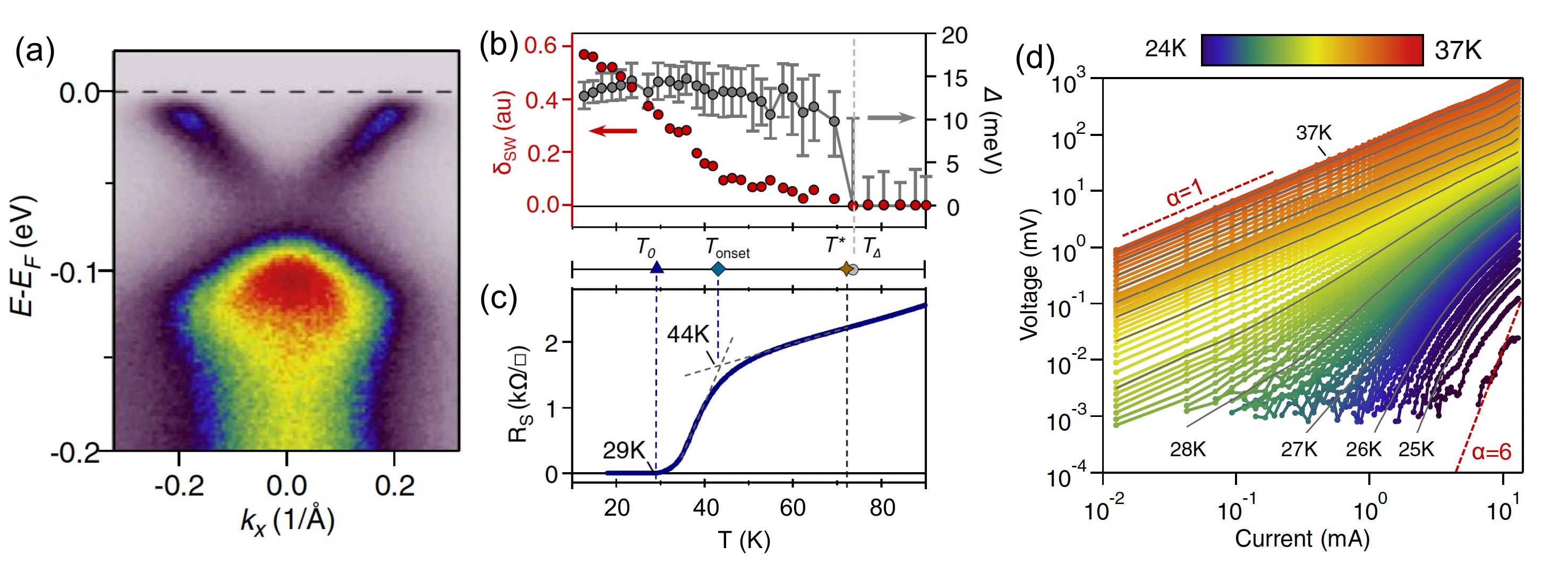}
\caption{Combined ARPES and transport studies on 1UC
  FeSe/SrTiO$_3$ showing (a) ARPES data near the M point of the
  Brillouin zone taken at 12K. (b) Extracted values of the gap
  $\Delta$ and spectral weights $\delta_{\text{SW}}$ at the Fermi
  level as a function of temperature. (c) Resistivity
  measurements. (d) Voltage-current relationship. Adapted from \textcite{Faeth2021}.}
\label{fig:17}
\end{figure*}

Recent work by one of the coauthors~\cite{Faeth2021} combined
\textit{in situ} ARPES and \textit{in situ} transport measurements to
simultaneously characterize the spectroscopic and resistive
transitions (Fig.~\ref{fig:17}). The former is sensitive to the
presence of a pseudogap that can be associated with pairing while the
latter probes superconductivity. The band structure of the 1UC FeSe is
simpler than in the bulk system. Only electron-like Fermi
surfaces are identified by ARPES near the Brillouin zone corners, with
a Fermi energy $E_\text{F}\approx 60$~meV~\cite{Liu2012}. An excitation gap
$\Delta \approx 15$~meV is observed at 12~K and persists up to 73~K.
This leads to a ratio of $\Delta/E_\text{F}$ of the order of 0.25. The
coherence length from vortex mapping is about $2$~nm~\cite{Chen2020},
which suggests $k_\text{F} \xi_0^{\text{coh}} \approx 4 $. This places 1UC
FeSe/SrTiO$_3$ firmly in the BCS-BEC crossover regime, but not yet in
the BEC.

\begin{figure}
\centering\includegraphics[width=3.0in]{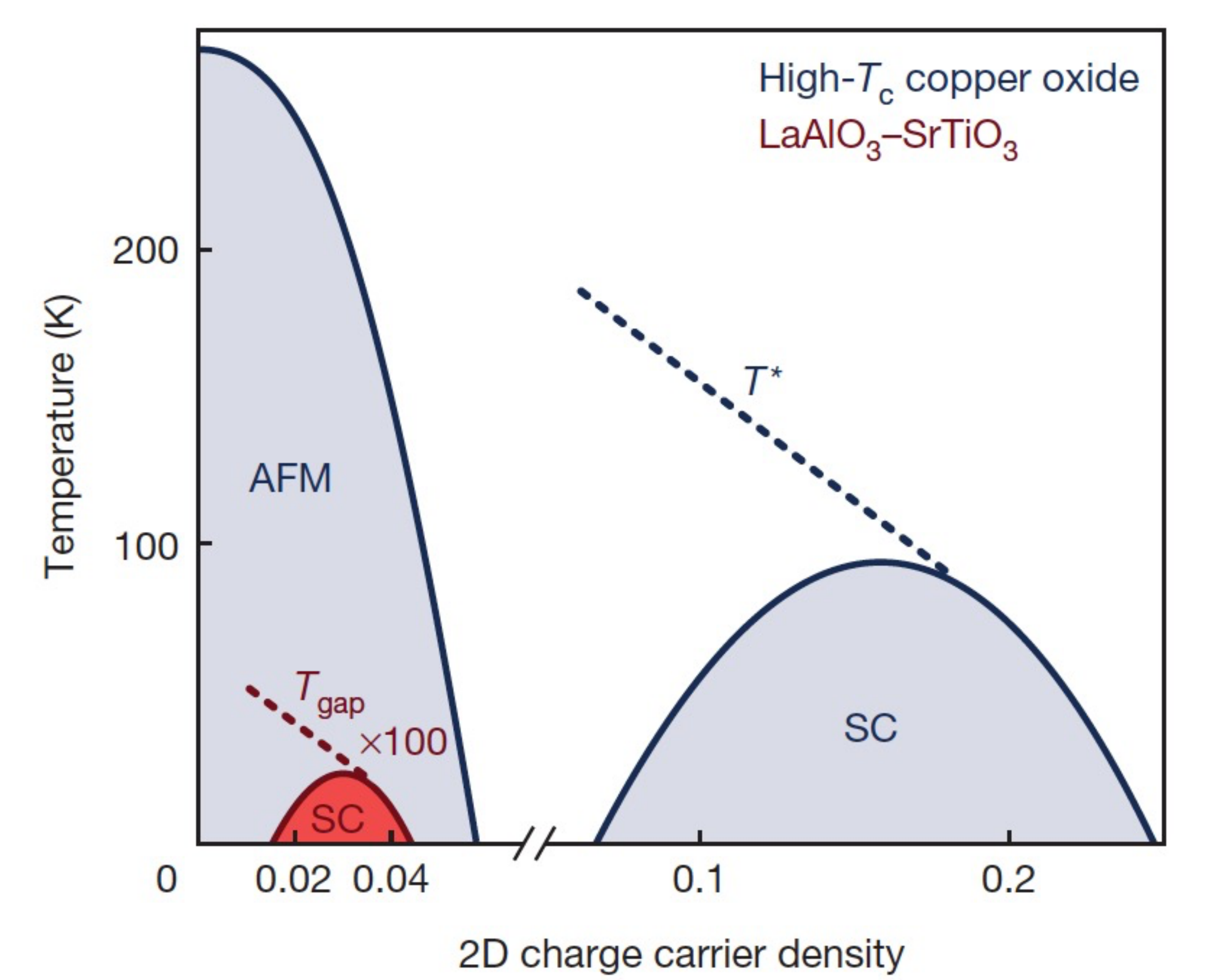}
\caption{Interface superconductivity in LaAlO$_3$-SrTiO$_3$ (shown in
  red) which is tuned with an electric gate field.  The figure represents a comparison between
  high-$T_\text{c}$ cuprate superconductors and the n-doped interface
  superconductors. In this figure the horizontal axis is the carrier
  density per unit cell.  SC: superconducting; AFM: antiferromagnetic.
  The endpoint of the LaAlO$_3$-SrTiO$_3$ SC dome on the underdoped
  side is a quantum critical point that separates the superconducting
  from an insulating phase~\cite{Caviglia2008}. From
  \textcite{Richter2013}.}
\label{fig:18}
\end{figure}

A second example of interfacial superconductors which has been interpreted in terms of a possible BCS-BEC crossover scenario~\cite{Bozovic2020} corresponds to a superconductor formed within the conducting 2D interface between two band insulators, LaAlO$_3$ and SrTiO$_3$. This belongs to the class of superconductors with anomalously low carrier density. Indeed, it is argued that this 2D superconductor is similar in many ways to the behavior in 3D doped \STO, and also has features of the high-$T_\text{c}$ copper oxides. The phase diagram ~\cite{Richter2013} shown in Fig. \ref{fig:18} is analogous to the cuprates in many ways; additionally there are claims of preformed pairs in both. In the two cases the gap onset temperature does not follow $T_\text{c}$ in the underdoped region but increases with charge carrier depletion. 

This heterostructural system is particularly useful as it can be tuned continuously through gating. There is a superconducting dome along with a pairing gap $\Delta$, which survives up to $T^* \approx 500$~mK~\cite{Richter2013} for the 2D carrier density $n\sim 0.02$ per unit cell.  At  $T=0$, $\Delta_0 \approx 65$~$\mu$eV. Moreover, with decreasing temperature, the pseudogap $\Delta_{\pg}$ evolves smoothly into the pairing gap within the superconducting phase. Also supporting the pairing-onset interpretation of $T^*$ is that the ratio of 
$\Delta_0$ to $T^{*}$ remains close to the BCS prediction; at more general temperatures the pairing gap follows the BCS-like mean-field temperature dependence.

Using an atomic force microscope (AFM) tip, the Levy group~\cite{Cheng2015} was able to draw single-electron transistors on the LaAlO$_3$/SrTiO$_3$ interface. Importantly, this enabled observation of preformed pairs which persist up to 900~mK, well above the transition temperature which ranges between $200\sim 300$~mK.

These temperature scales, however, pose some concerns about interpreting the nature of interfacial superconductivity in LaAlO$_3$/SrTiO$_3$. The Fermi energies of various $t_\textit{2g}$ bands have been characterized by soft X-ray ARPES~\cite{Cancellieri2014} and found to be around 50~meV for the $d_{xy}$ orbital band~\cite{Sulpizio2014,Pai2018}~\footnote{We note that in the literature it is still being debated whether the $d_{xy}$ orbital actively participates in the superconductivity or not (see, e.g., \textcite{Scheurer2015}). Using the $d_{xz}/d_{yz}$ orbital bands for $E_\text{F}$ would lead to a relatively larger $\Delta_0/E_\text{F} \sim 0.05$. Our choice of the $d_{xy}$ band for $E_\text{F}$ is based on the consistency between the estimated $\Delta_0/E_\text{F}$ and $k_\text{F} \xi_0^{\text{coh}}$.}, which leads to a rather small ratio of $\Delta_0/E_\text{F} \sim 10^{-3}$.

This observation, indicative of a more BCS-like system, appears
incompatible with a strong-pairing crossover scenario. Even more
persuasive of this incompatibility is the additional fact that the
measured coherence length is large, of the order of
$30\sim 70$~nm~\cite{FillisTsirakis2016}, leading to
$k_\text{F}\xi_0^{\text{coh}} \approx 30 \sim 70$. This is based on the previous
estimates in the literature for $k_\text{F} \approx 0.1$
\AA$^{-1}$~\cite{Pai2018}.

There is strong evidence that disorder effects~\cite{Chen2018} are
important in this interfacial superconductor.  In particular, it has
been shown in Ref.~\cite{Chen2018} that applying an electrostatic gate
voltage not only tunes the carrier density, but it can also
significantly modify the interfacial disorder via the mobility.
Nevertheless, it is somewhat difficult to associate a phase diagram
like that in Fig.~\ref{fig:18}, in which there is a $T_c$ dome while
$T^*$ is monotonic, with the effects of disorder.  This behavior of
$T^*$ can be contrasted with the disorder-induced pseudogap effects
discussed in Sec.~\ref{Sec:Disorder}.  While there is some
uncertainty, a reasonable conclusion is that disorder is relevant to
interfacial superconductivity in LaAlO$_3$/SrTiO$_3$, and a strong
pairing mechanism does not seem to be operative.  Possibly related to
these observations are theoretical calculations~\cite{Che2017}, albeit
for 3D $s$-wave systems, which reveal that disorder-induced
superconductor-insulator quantum phase transitions can occur in the
BCS regime; here the superconducting order is destroyed, leading to an
insulating phase that is caused by a residual pseudogap.

\begin{figure*}
\centering\includegraphics[width=6.in]{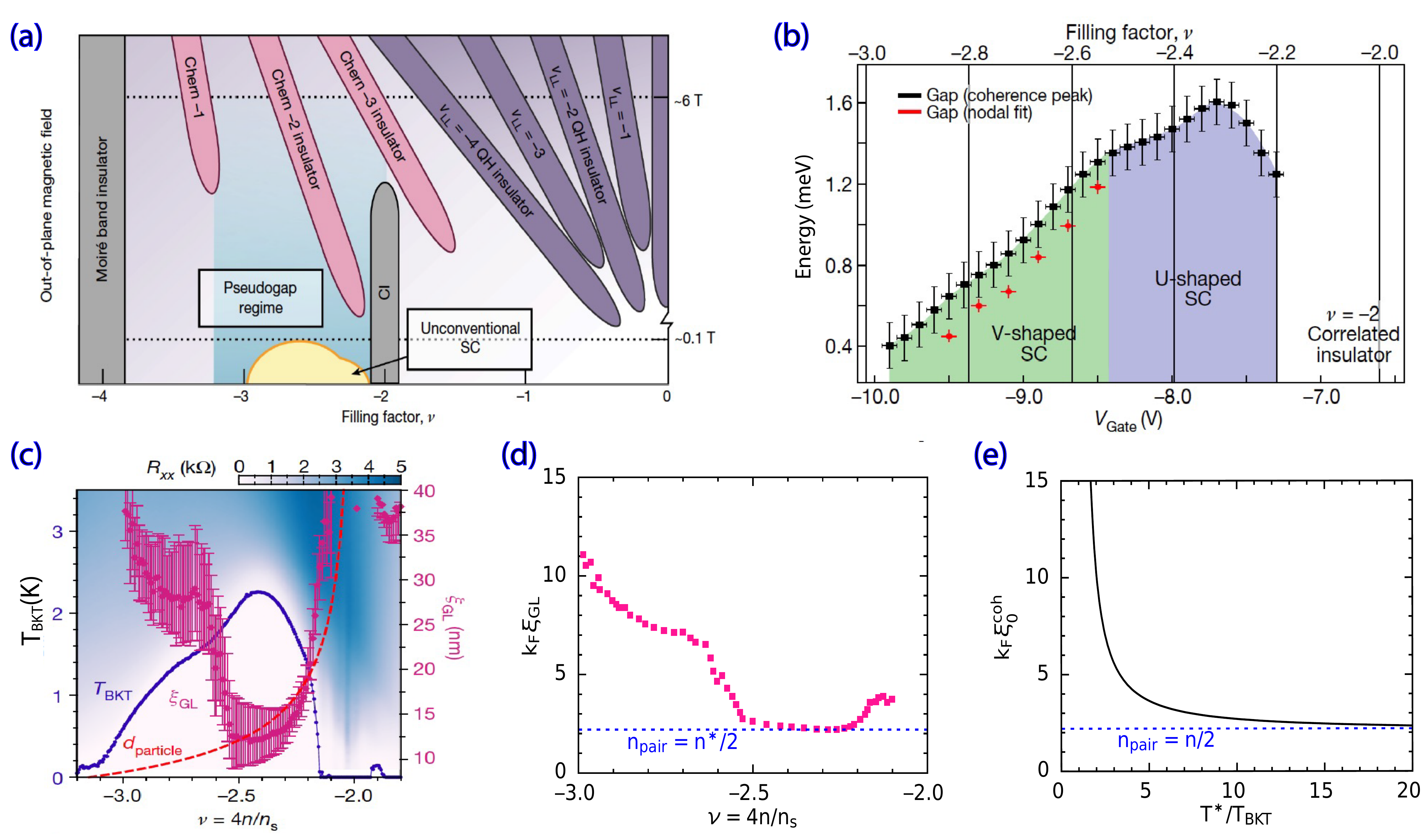}
\caption{Superconducting properties of MATBG and MATTG. (a) Phase diagram
of hole doped MATBG superconductors (SC), taken from \textcite{Oh2021}. The electron filling factor $\nu=4 n/n_s$, where $n$ is the carrier density defined by the applied gate voltage and $n_s$ is the corresponding $n$
when the lower four-fold degenerate Moire flat band is fully filled.  In this diagram, a very large pseudogap regime, indicated in light blue, is determined by combining conventional STM and point-contact Andreev tunneling spectroscopy. 
(b) Gap size $\Delta$ versus the gate voltage $V_\text{Gate}$ (and the filling factor $\nu$) for MATTG, taken from \textcite{Kim2022}. The $\Delta$ is measured from conventional STM tunneling at low temperatures. The data points are extracted from the separation between coherence peaks at the halfway point (black squares) and from a nodal gap fit (red dots). In the green and violet regions the $dI/dV$ curve exhibits a V shape and a U shape, respectively. (c) The $T-\nu$ phase diagram of MATTG at displacement field $D/\epsilon_0 = -0.5$ V nm$^{-1}$, along with the curves of the interparticle distance $d\equiv d_\text{particle}$ and the coherence length $\xi_\text{GL}$, taken from \textcite{Park2021}. Here $d_{\text{particle}} = 1/\sqrt{n^*}$, where $n^*$ is the effective carrier density that can be deduced from quantum oscillation and Hall density measurements. Note that $n^*$ is different from the density $n$. (d) Replotting of the $\xi_{\text{GL}}$ data from (c) in terms of the product $k_\text{F} \xi_\text{GL}$. To convert $n^*$ to $k_\text{F}$ we have used $k_\text{F} = ( 2 \pi n^*)^{1/2}$.
The blue dashed line shows the expected $k_\text{F}\xi_{\text{GL}}$ value when $n_{\text{pair}}$ saturates to $n^*/2$. 
(e) The product $k_\text{F} \xi_\text{GL}$ calculated theoretically as a function of $T^*/T_\text{BKT}$ for a 2D s-wave superconductor. 
In the theoretical calculation, $n^*$ is the same as $n$. 
}
\label{fig:12A}
\end{figure*}

\subsection{BCS-BEC crossover in magic-angle twisted bilayer and trilayer graphene}

There is growing support that MATBG~\cite{Cao2018} as well as MATTG~\cite{Park2021,Kim2022} superconductors exhibit BCS-BEC crossover features. Notably, these are very clean systems, associated with a BKT transition. One piece of cited evidence is based on the relatively large values of $T_\BKT/T_\text{F}$. These were reported in the initial groundbreaking paper~\cite{Cao2018} as well as in subsequent works ~\cite{Oh2021,Kim2022,Lu2019}. Such estimates are, in turn, based on $V$-$I$ plots which allow one to determine the BKT transition that occurs when $V = I^{\alpha}$ with a specific value of $\alpha = 3$. As a caution we note that the ratio $T_\BKT/T_\text{F}$ should not
be viewed as a proxy for the fraction of electrons involved in superconductivity;
in the BEC regime, this
parameter becomes very small.

More recent tunneling experiments~\cite{Oh2021} (which are summarized
in Fig.~\ref{fig:12A}(a)) on MATBG help to make the association with
BCS-BEC crossover stronger; they have presented clearer indications of
an extensive pseudogap regime in the phase diagram, as can be seen
from the figure. These STM experiments suggest~\cite{Oh2021} an
anomalously large value for the ratio
$2\Delta_0/(k_\text{B} T_\BKT) \approx 25$, which can be viewed as
representative of strong pseudogap effects, equivalently associated
with large $T^*/T_\BKT$. Adding support to a BCS-BEC crossover
scenario is the presence of another much smaller energy-gap scale
associated with point-contact Andreev tunneling, which is only present
in the ordered phase where there is phase coherence.

The results from this STM tunneling~\cite{Oh2021} provide a value for
$\Delta_0 \approx 1.4$~meV in MATBG. In an earlier section we pointed
out that $V$-$I$ measurements in 2D films can be used~\cite{Zhao2013}
for estimates of $T^*$.  One can infer from these data~\cite{Cao2018}
that $T^* = 3 \sim 5$~K, which is obtained from the Ohmic recovery
temperatures\footnote{Ideally one could arrive at more accurate
  numbers by making systematic $V$-$I$ plots over finely separated
  temperature intervals in order to more precisely establish the
  temperature for the Ohmic recovery, corresponding to $T^*$.}.  This
should be compared with the transition temperature $T_\BKT \approx 1$~K
and the Fermi energy of the bilayer system, which is estimated to be
$T_\text{F} \approx 20$~K~\cite{Cao2018}.  The resulting relatively large
ratios of $T^*/T_\BKT$ and $\Delta_0/E_\text{F}$ suggest that MATBG is a
superconductor in the intermediate BCS-BEC crossover regime.

Indeed, based on the claims~\cite{Oh2021} 
that MATBG has some similarities with high-$T_\text{c}$ superconductors, it
is striking to observe similar $T^*/T_\text{F}$ and $T^*/T_\text{c}$ values in
Fig.\ref{fig:34} (Appendix~\ref{sec:AppC}) for the underdoped cuprates
and (both) twisted graphene families of superconductors.
This figure addresses this similarity more quantitatively.

The situation for MATTG appears to be somewhat clearer and provides
more quantitative information. Some pertinent
results~\cite{Kim2022,Park2021} are summarized in Fig.~\ref{fig:12A},
where panels (c) and (d) address very useful coherence-length
experiments~\cite{Park2021} based on the magnetic-field dependence of
the superconducting transition temperature.  Fig.~\ref{fig:12A}(c)
shows this published data for $\xi_0^{\text{coh}}$ as well as the
inter-particle distance $d$ as a function of the band filling factor
$\nu$, along with
the transition temperature $T_\BKT$.  It should be noted that the
error bars are large here, indicative of the
experimental challenges encountered when deducing the
coherence length using resistivity measured at a finite magnetic field. Particularly in
2D and extreme type-II superconductors, this necessarily leads to very
broad transitions making it difficult to establish $T_\text{c}(H)$
without incorporating a fairly arbitrary standard for determining
where the transition is located.

The experimentally observed dimensionless product $k_\text{F}\xi_0^{\text{coh}}$
(Fig.~\ref{fig:12A}(d)) \footnote{ Note that the band degeneracy used
  in the conversion here is $2$, not the naive $4$. As supported by
  experiments, the spin-valley 4-fold degeneracy is broken to 2 at
  $-3 < \nu \lesssim -2$. } can be compared with the theory in
Fig.~\ref{fig:12A}(e), where $k_\text{F}\xi_0^{\text{coh}}$ is plotted as a
function of $T^*/T_\text{c}$. (This is similar to the inset in
Fig.~\ref{fig:10}(c)).  We note that the plot in Fig.~\ref{fig:12A}(d)
and the theory plot in Fig.~\ref{fig:12A}(e) are for different
horizontal axis variables; however, a direct association of the two
would allow one to relate the important ratio $T^*/T_\text{c}$ with the
filling factor $\nu$, hence completing the $T^*/T_\text{c}$ versus $\nu$
phase diagram.  From the data in Fig.~\ref{fig:12A}(d) it follows that for $\nu \gtrsim -2.5$ MATTG also
belongs in the intermediate BCS-BEC crossover regime.
 
Recent tunneling experiments~\cite{Kim2022} provide additional
important quantitative information about MATTG with a focus on the gap
energy scale as plotted in Fig.~\ref{fig:12A}(b) as a function of
$\nu$. These studies indicate $T^* = 7$~K at the $\nu$ value where the
gap is maximum. Additional parameters are:
$T_\BKT \approx 2.25$~K~\cite{Park2021} with the estimated Fermi
temperature given by $T_\text{F} \approx 30$~K.

Overall, there appears to be compatibility between the
$\xi_0^{\text{coh}}$ data from the MIT group and pairing-gap
experiments~\cite{Kim2022} shown in Fig.~\ref{fig:12A}(b). Making use
of the estimates of $E_\text{F}$ based on quantum-oscillation
experiments~\cite{Park2021} it follows that the ratio $\Delta_0/E_\text{F}$
exhibits a similar trend as $\xi_0^{\coh}$, changing from more
BCS-like behavior at $\nu \approx -3$ to characteristic crossover
behavior at $\nu \approx -2.2$.  We note that interpretations of these
tunneling experiments~\cite{Kim2022} have suggested that the BEC
regime is reached around the upper half of the $T_\BKT$ dome at
$\nu \gtrsim -2.5$, although it is not straightforward to reconcile a
BEC phase with the presence of coherence peaks in the tunneling data.

Finally, it should additionally be noted that the theory plot of the
coherence length in Fig.~\ref{fig:12A}(e) is for the $s$-wave case,
while the experimental data seem to suggest a nodal form of
superconductivity.  Some aspects of the crossover theory for an
anisotropic gap symmetry have been addressed in this Review (in
Sect.~\ref{sec:dwave})\footnote{In the single band $d$-wave case, the
  counterpart of the curve in Fig.~\ref{fig:12A}(e) looks
  qualitatively similar at very low density but will not reach BEC
  until a much larger $T^*/T_\BKT$. No BEC is found at high
  densities.}, but one might additionally expect that other
ingredients such as flat energy bands and quantum geometry (discussed
in Sect.~\ref{sec:quantgeometry}) may play an important role as well
in reaching an ultimate understanding of BCS-BEC crossover for MATBG
and MATTG.

\subsection{BCS-BEC crossover for 2D gated semiconductors}

\begin{figure*}
\centering\includegraphics[width=4.5in]
{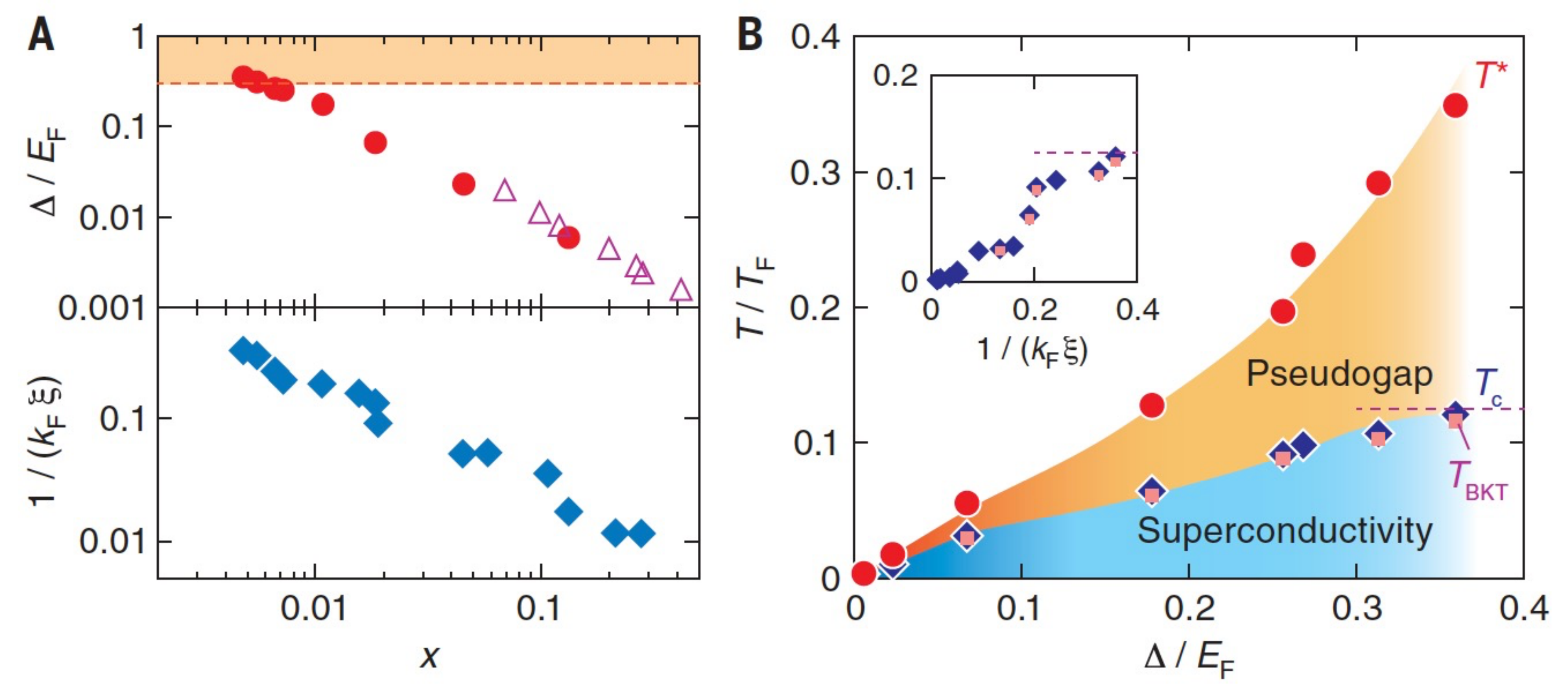}
\caption{Experimental data in electron-doped zirconium nitride chloride. The results shown here are  from tunneling spectroscopy and DC resistivity measurements. The transition temperature $T_\text{c}$ is defined as the midpoint in the resistivity curves, which is identified as $T_\textrm{BKT}$.  The (in-plane) coherence length $\xi=\xi_0^{\text{coh}}$ is determined from the temperature-dependent upper critical magnetic field measured near the zero-field $T_c$. From \textcite{Nakagawa2021}.}
\label{fig:20}
\end{figure*}

There has been recent interest~\cite{Nakagawa2021, Saito2016, Nakagawa2018} in a group of layered nitrides, Li$_x$ZrNCl, which are intrinsically semiconductors and exhibit superconductivity through Li-intercalated doping. These experiments impose control of the carrier density by use of ionic gating, which provides access to very low carrier density systems that are otherwise inaccessible. Concomitantly the varying carrier number enables a tuning of the weakly- to strongly-coupled superconducting regimes by controlling both the carrier density and simultaneously a dimensional crossover from anisotropic-3D to 2D. Both tunneling and resistivity measurements~\cite{Nakagawa2021} yield systematic information about the detailed phase diagram of this system.

The phase diagram~\cite{Nakagawa2021}, shown in Fig.~\ref{fig:20}, indicates a pronounced pseudogap regime established from $dI/dV$ measurements. This is particularly notable at low carrier densities, where the system is more two dimensional.  In particular, at extreme underdoping $T_{\BKT}$ shows a maximum of $19$~K. In the most underdoped sample probed, $\Delta_0/E_\text{F} \approx 0.3$, $T_\BKT/ T_\text{F} \approx 0.12$, and  $T^*$ is roughly $3 T_{\BKT}$ .

A summary~\cite{Nakagawa2021} of experimental observations is presented in Fig.~\ref{fig:20} as a plot in terms of $T/T_\text{F}$ vs $\Delta_0/E_\text{F}$ with data points indicating $T_\BKT$ and $T^*$. The pseudogap and associated $T^*$ were found to be largest
when the carrier number was lowest. Here, for these large gap systems (which are in the strong-coupling limit) one finds the smallest coherence length,
$k_\text{F} \xi_0^{\text{coh}} \approx 3$, as obtained from the upper critical fields. 
This suggests a system
that may be close to but not yet in the BEC regime.
In the opposite, highest electron doping regime one recovers more characteristic BCS behavior with $T_\textrm{BKT} \approx T^*$. We conclude that all of this adds up to a body of evidence that lends reasonably strong support to a BCS-BEC crossover description of these ionic gated superconductors.

\subsection{Magnetoexciton condensates with BCS-BEC crossover}

The concept of condensation based on particle-hole pairs~\cite{Comte1982,Combescot2017,Kohn1970} should be thought of as a very natural extension of particle-particle pairing in superconductors. Indeed one usually invokes the same ground-state wave function as in Eq.~\eqref{eq:1}, here modified by replacing one of the electron operators with a hole operator and presuming that the two are associated with different bands. This subject has generated considerable excitement as one could conceive of such condensation as taking place at very high temperatures. There are a number of subtle features, however, as the electrons and holes need to be sufficiently well separated so as to avoid recombination. Their number and effective masses also need to be equivalent, otherwise pairing can be impeded as this system behaves like a superfluid with population or mass imbalance.

An important configuration for arriving at exciton condensation involves quantum Hall fluids~\cite{Eisenstein2014,Eisenstein2019}, as was first implemented by Eisenstein \textit{et al.} in a GaAs/AlGaAs heterostructure. Here two thin GaAs layers are separated by the AlGaAs spacer layer, which serves to mitigate electron-hole recombination processes. Because  each layer forms a 2D electron gas, in the presence of a strong perpendicular magnetic field $B$, their energies are quantized into Landau levels (LL). These bilayer quantum Hall systems have the potential to realize novel quantum states that have no analog in a single layer. A relevant parameter for characterizing such states is $d/\ell_B$, where $d$ is the inter-layer spacing and  $\ell_B=\sqrt{\hbar/|eB|}$ is the magnetic length.

There has been a focus~\cite{Eisenstein2014} on the interlayer coherent state observed in the zero or small interlayer tunneling limit and at total electron filling fraction $\nu_{\mathrm{tot}}=\nu_{1}+ \nu_{2}=1/2+1/2=1$. Here, the electron filling fraction, $\nu_i = n_i (2\pi \ell_B^2)$, is defined for each individual layer with $n_i$ the electron density of the $i$-th layer. Important questions such as whether there is a quantum phase transition separating the large and small $d/\ell_B$ limits have been raised~\cite{Murphy1994, Bonesteel1996,Halperin1983,Moon1995}, although recently~\cite{Liu2022, Wagner2021, Sodemann2017} there has been the suggestion that the evolution of the state from the large to small $d/\ell_B$ might be understood as a crossover of BCS behavior to a BEC of magneto-excitons. 

This picture can be understood in terms of Jain's composite fermions (CF)~\cite{Jain2007}, where a CF can be roughly viewed as the original electron attached to two magnetic flux quanta ($2h/e$). In the extreme $d\rightarrow \infty$ limit, the two layers decouple and each of them has a LL filling fraction $\nu=1/2$, which can be described by a metallic state~\cite{Halperin1993} of either electron-like or, equivalently, hole-like CFs  with well defined Fermi surfaces.

At finite $d$ one can then consider electron- and hole-like CFs from the two different layers forming inter-layer Cooper pairs, i.e., magnetoexcitons. Importantly, it is reasonable to assume that their effective masses are equal near $\nu=1/2$, due to an approximate particle-hole symmetry. The pair formation is driven by an inter-layer attraction, $U$ which is derived from the original interlayer Coulomb interaction between electrons and holes, whose magnitude is $|U|\sim V_{\mathrm{inter}} \sim  e^2/ ( \epsilon d )$, where $\epsilon$ is the background dielectric constant\footnote{When $d\ll \ell_B$, the inter-layer interaction is actually governed by $e^2/(\epsilon \ell_B)$, not $e^2/ (\epsilon d)$. It should also be noted that the actual inter-layer interaction between CFs is not the same as $V_{\mathrm{inter}}$. Instead, it is mediated by an emergent Chern-Simons gauge field that makes the renormalized interaction highly frequency dependent~\cite{Halperin1993,
Bonesteel1996,Wang2014}. Here, we ignore these complications. }.  At the same time the parameter $E_{\mathrm{kin}}$, which represents the kinetic energy of a partially filled Landau state, is set by the intralayer Coulomb repulsion~\cite{Halperin1993}, $E_{\mathrm{kin}} \sim V_{\mathrm{intra}} \sim e^2/ (\epsilon \ell_B)$.

\begin{figure}
\centering
\includegraphics[width=2.5in]{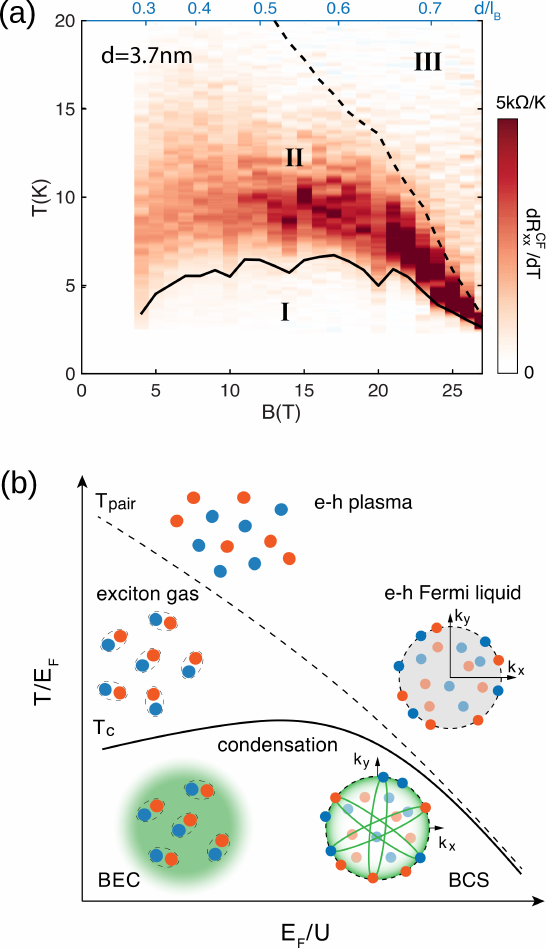}
\caption{BCS-BEC crossover for magnetoexcitons. The color coding in panel (a) is associated with the temperature derivatives of the measured longitudinal resistance in the counterflow configuration. Hall drag and counterflow resistances are used, respectively, to arrive at the pairing onset temperature $T^*$ (dashed line) and to infer $T_\text{c}$  (solid line), as a function of the ratio of effective attraction/kinetic energy (through the magnetic field $B$). (b) Schematic phase diagram expected for a magnetoexciton condensate. $T_{\mathrm{\pair}}$ is the same as $T^*$. From \textcite{Liu2022}.}
\label{fig:21}
\end{figure}

In this way the important ratio $|U|/ E_{\mathrm{kin}} \propto \ell_B/d$, which sets the scale of a BCS-BEC crossover can be tuned experimentally by varying either $d$ or $B$. Large $d$ or high magnetic fields corresponds to the BCS-like limit, while the more BEC regime is present at small $d$ or low magnetic fields (see Fig.~\ref{fig:21}). This BCS-BEC crossover picture is supported by recent measurements on graphene double-layer heterostructures~\cite{Liu2017,Liu2022}. Compared to the GaAs/GaAlAs double-layer experiments, this graphene bilayer system has an additional advantage as it allows the two graphene layers to be separated by a thin hexagonal boron nitride layer, which prohibits direct interlayer tunneling without introducing disorder.

Because the magnetoexcitons are neutral and cannot be probed in traditional electronic transport, two unconventional designs for resistance measurements have been employed to experimentally probe the magnetoexciton superfluidity via ``counterflow'' and ``drag'' experiments~\cite{Eisenstein2014}. Figure \ref{fig:21} presents a summary of the results from these measurements for the double-layer graphene system~\cite{Liu2022}. 

In the counterflow configuration electric currents in the two layers are of the same magnitude but flow in opposite directions. The absence of dissipation due to ``superfluidity'' is associated with a vanishing $R_{xx}^{\mathrm{counter}}$, which measures the longitudinal resistance. These experiments serve to determine the transition temperature $T_\text{c}$ (solid black line) in Fig.~\ref{fig:21}(a).

A striking signature of magneto-excitonic superfluidity is a quantized Hall drag resistance at low temperature in the ``drag'' configuration. Here the electric current is fed only to one layer, while the Hall voltage drops are measured in both layers, from which one can define the usual Hall resistance, $R_{xy}$, for the current-driving layer. One can also define a Hall drag resistance, $R_{xy}^{\mathrm{drag}}$, for the passive layer.

Both $R_{xy}$ and $R_{xy}^{\mathrm{drag}}$ are expected to be quantized to the same value $h/(e^2 \nu_{\mathrm{tot}})$ at low  $T$. As $T$ increases above $T_\text{c}$, $R_{xy}^{\mathrm{drag}}$ decreases monotonically. In \textcite{Liu2022}  the important temperature scale, $T^*$ is defined as the point where $R_{xy}^{\mathrm{drag}}$ drops to below $50\%$ of $h/(e^2 \nu_{\mathrm{tot}})$. This $T^*$ is plotted in Fig.~\ref{fig:21}(a) as the black dashed line. It is reasonable to associate the residual $R_{xy}^{\mathrm{drag}}$ at high temperatures with incoherent pair correlations between electron- and hole-like CFs. In this way one interprets $T^*$ as the onset of electron-hole CF pair formation. While there are some uncertainties in the definition of $T^*$, a clear separation of the two temperature scales, $T_\text{c}$ and $T^*$, is apparent from Fig.~\ref{fig:21}(a), which is to be compared to the schematic phase diagram sketched in Fig.~\ref{fig:21}(b). 

What is not as clear is whether at the lowest applied magnetic field $B\approx 5$~T the system has reached the BEC regime, as suggested by the schematic figure\footnote{Rescaling the measured $T_\text{c}$ of the top panel by $T_\text{F}$, which can be estimated as $e^2/(\epsilon \ell_B)$, and plotting the obtained $T_\text{c}/T_\text{F}$ as a function of $B$ shows that this ratio has not passed the point where it starts to decrease with decreasing $B$ even at $B\approx 5$T. It suggests that the system may still be in the crossover regime, not yet into the BEC, if we compare this trend of $T_\text{c}/T_\text{F}$ to that for a 2D electron gas in Fig.~\ref{fig:9}}. In comparing with a prototypical example of BCS-BEC crossover, as in the 2D electron gas, it is useful to establish the magnitude of the effective $\Delta_0/E_\text{F}$, which would be expected to become arbitrarily large in a more traditional BEC superconductor. On the other hand, exact diagonalization studies~\cite{Wagner2021} show that for the bilayer magnetoexciton system $\Delta_0 \lesssim E_\text{F}$. This contrast highlights some of the key differences between traditional superconductors and the magnetoexciton bilayer that one needs to bear in mind in the interpretation of the phenomenology. It is clear that quantification of the exact behavior of $T_\text{c}/T_{F}$, and other quantities characteristic of BCS-BEC crossover, for the entire range of $d/\ell_B$ from $\infty$ to $0$ requires further work, both theoretical and experimental.

One might speculate that, since one defining feature of the BEC regime is the disappearance of Fermi surfaces, a potentially useful future experiment is to directly probe the Fermi surface of CFs at $T_\text{c} < T < T^*$ for small $d/\ell_B$, using geometric resonance techniques as employed in the determination of the Fermi wave vector of CFs for the single layer $\nu=1/2$ state~\cite{Kamburov2014}.  Achieving a number of these goals seems promising given the high tunability of the bilayer graphene heterostructure, as demonstrated by the new generation of experiments~\cite{Liu2017,Liu2022}.

\section{Application to the Cuprates}

\subsection{Support for and counter-arguments against BCS-BEC crossover in the cuprates }
\label{sec:Counter}

The question of whether a BCS-BEC scenario is relevant to the cuprates is, like all aspects of the cuprate literature, a highly controversial one. Despite this controversy it is useful to let the reader independently judge; thus, here near the end of this Review article we discuss what the implications are of such a theory for the cuprates. We address aspects that are both consistent and inconsistent with the data.

There are claims in the literature that the cuprates are somewhere between BCS and BEC. We cite some of these here.
\begin{itemize}
\item From A. J. Leggett~\cite{Leggett2006}: ``\textit{The small size of the cuprate pairs puts us in the intermediate regime of the
so-called BCS-BEC crossover}.''
\item From G. Sawatzky and colleagues~\cite{Hufner2008}: ``\textit{High-$T_\text{c}$ superconductors cannot be considered as classical BCS
superconductors, but rather are smoothly evolving from BEC into the BCS regime}.''
\item From I. Bozovic and J. Levy~\cite{Bozovic2020}:  ``\textit{We show the likely existence of preformed pairs in the cuprates \dots The existence of preformed pairs is a necessary but not sufficient condition for BEC or for BCS-BEC crossover to occur.}
\textit{Indeed, since Fermi surfaces have been mapped out \dots this favors a picture in which pairing is relatively strong, pre-formed pairs first appear at $T>T_\text{c}$ \dots but copper oxides are still on the BCS side of the crossover.}''
\item From Y. Uemura~\cite{Uemura1997}: ``\textit{Combining universal correlations \dots and pseudogap behavior in the underdoped region, we obtain a picture to describe superconductivity in cuprate systems in evolution from Bose-Einstein to BCS condensation}.''
\end{itemize}

It should be noted that even if BCS-BEC crossover theory plays a role in the cuprate superconductors this will not address or elucidate a number of important issues which characterize their behavior and need to be understood in an ultimate theory. Among these is the pairing mechanism~\cite{Lee2006}, which remains unknown; also challenging is arriving at an understanding of the ``strange metal'' behavior including the linear temperature dependence of the resistivity, which is, indeed, very widespread among other strongly correlated superconductors~\cite{Varma2020}. Another puzzle is the distinct change observed in carrier concentration as a function of hole doping, which seems to correlate with the presence of a pseudogap~\cite{Proust2018}.
This appears consistent with recent ARPES claims~\cite{Chen2019} that the pseudogap suddenly collapses at a fixed hole concentration. 

We list next issues that have been raised to challenge the relevance of
BCS-BEC crossover theory for the cuprates.  Examples are the
following:

\begin{enumerate}
\item  Current cuprate experiments
show no signs of a chemical potential $\mu$ which is near or below the band bottom, as might be expected in the BEC regime. This would show up in ARPES experiments.
\item $T_\text{c}$ and $T^*$ are observed to vary inversely in the underdoped regime.
Some have argued that if $T^*$ were related to preformed pairs, then
as pairing becomes stronger both $T_\text{c}$ and $T^*$
would tend to increase together.
\item One finds~\cite{Vishik2018} that a number of (but not all) superconducting fluctuation phenomena appear only in the immediate vicinity of $T_\text{c}$, well below the pseudogap onset temperature $T^*$.
\item There are multiple signatures of ``a nodal-antinodal dichotomy"~\cite{Hashimoto2014},
corresponding to a different behavior of the $d$-wave
energy gap along the nodal
and anti-nodal directions. This is widely interpreted to
mean that rather than preformed pairs, another (unspecified) ordering must be
responsible for the pseudogap, which is mostly confined to the anti-nodes.
\item There are ARPES experiments~\cite{Hashimoto2010} which indicate that at higher temperatures in the normal state, but well below $T^*$, the fermionic dispersion shows disagreement with the characteristic energy dispersion associated with BCS-like quasi-particles.
\item There are other indications~\cite{Ghiringhelli2012} of additional ordering associated with the pseudogap phase, quite possibly with an onset associated with its boundary~\cite{Xia2008, Zhao2017}.
\item There are claims~\cite{Tallon1999} suggesting that quantum critical behavior is present so that $T^*$ actually vanishes beneath the superconducting dome; this is inconsistent with a BCS-BEC crossover picture, in which $T^*$ is necessarily larger than $T_\text{c}$.
\end{enumerate}

Of this list of 7, the last two seem to be most challenging for the
BCS-BEC crossover scenario, while the first 5 are not necessarily so, as will be
discussed in this section and the Appendices.  Attributing the cuprate
pseudogap to preformed pairs as distinguished from a competing order
parameter is admittedly highly controversial and this should not be
viewed as a central component of this Review, which is focused
principally on non-cuprate superconductors. Nevertheless, for
completeness, it is useful to present the predictions concerning the
cuprates which derive from one particular pre-formed-pair scenario ---
a BCS-BEC crossover perspective.
The discussion presented here and in
Appendices~\ref{sec:AppB} and \ref{sec:AppC} should be viewed as a
catalogue summary of some relevant theory literature.  The interested
reader can consult the cited papers to obtain more details.

\subsection{Experimental evidence that BCS-BEC crossover may be relevant to the cuprates} 

All indications are that, if this scenario is relevant to the cuprates, these superconductors are on the BCS side and well away from BEC. This is 
consistent with the claims in a recent paper~\cite{Sous2023}, although these authors adopted
a different definition of ``crossover" associating it with proximity to a BEC. Indeed, there are several experiments that stand out as providing among the strongest support for a BCS-BEC-crossover-like description of the copper oxides.

\begin{figure}
\centering\includegraphics[width=2.5in,clip]{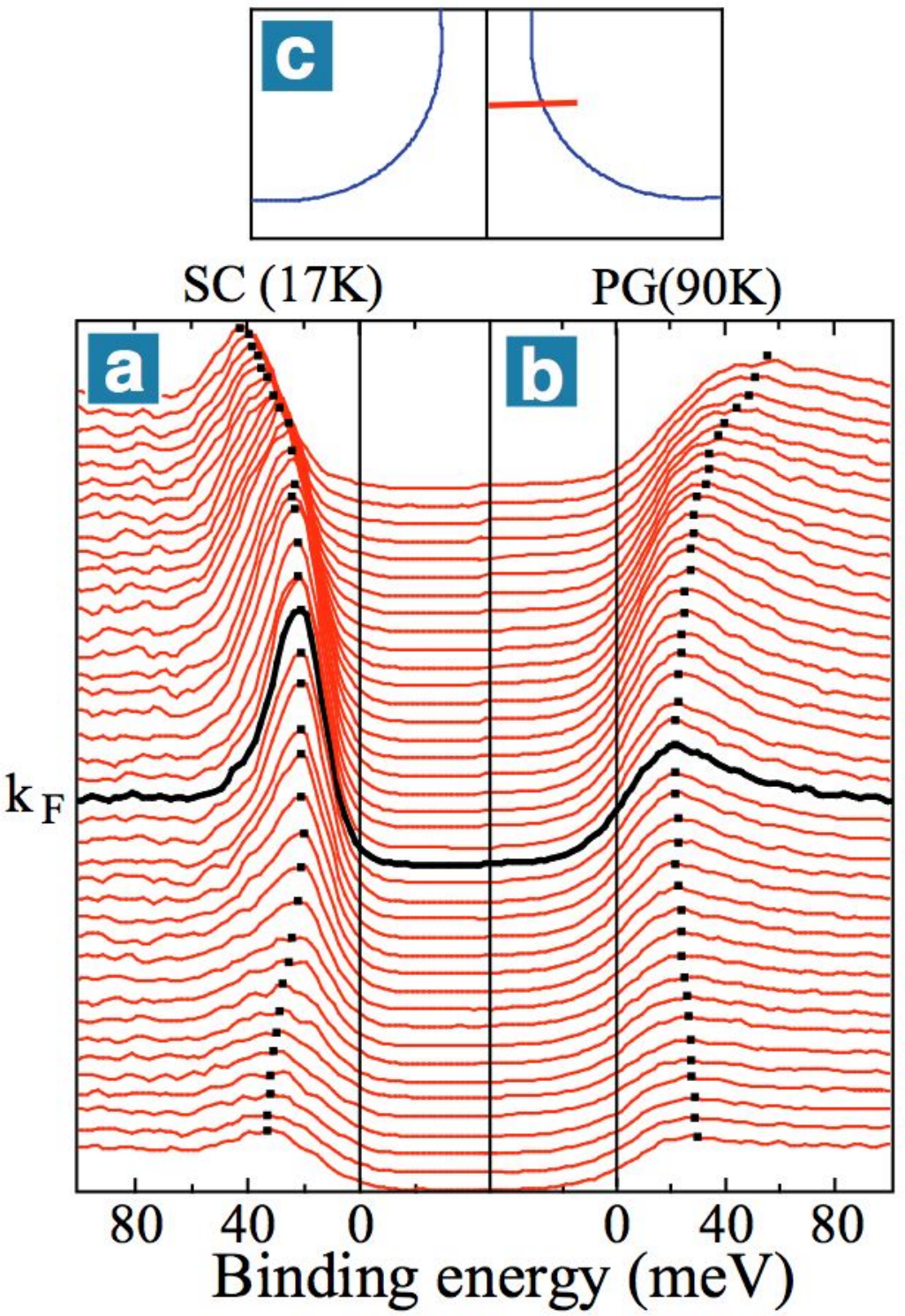}
\caption{Experimental pseudogap ARPES data showing backbending of the dispersion in the normal state (b), suggestively similar to that in the superconducting state (a). From \textcite{Kanigel2008}.}
\label{fig:22}
\end{figure}

ARPES measurements~\cite{Kanigel2008} reveal a Bogoliubov-like dispersion in part of the Brillouin zone that is away from the nodal Fermi-arc region. 
Importantly this is observed slightly above $T_\text{c}$, as shown in Fig.~\ref{fig:22}. It is highly unlikely, and indeed inconsistent with the theory we are discussing
(see Eq.~(\ref{eq:14}) which bears on point 5 in Sec.~\ref{sec:Counter}), that this Bogoliubov dispersion continues up to much higher temperatures, near the onset of the pseudogap. Indeed, there are studies~\cite{Hashimoto2010} that suggest this characteristic back-bending dispersion is absent well below $T^*$. But in the normal state, not too far from $T_\text{c}$, these experiments~\cite{Kanigel2008} provide indications that the presence of a pseudogap is associated with the same fermionic quasi-particles as are found in the ordered phase.

In a similar vein a smooth evolution of the measured ARPES
excitation gap around the antinodes as the temperature is varied
from above to below $T_\text{c}$ lends some support to
the crossover picture. 

An additional, conceptually simple experiment involves STM studies that compare the ratio of the zero-temperature pairing gap to $T^*$. 
This ratio appears to be very close to the expected mean-field result~\cite{Kugler2001,Oda1997}. This associates the ratio of 
$\Delta_0$ and $T^*$ in an analogous fashion as for the BCS prediction of $\Delta_0$ and $T_\text{c}$, and for a $d$-wave case.

There are additional classes of experiments that constitute less direct support, but are worthy of note and will be discussed in this section. These involve 

(i) recent shot-noise measurements~\cite{Zhou2019}, which provide a more direct and quantitative signature of pairing above $T_\text{c}$. Through pair contributions to tunneling these shot-noise experiments~\cite{Zhou2019} indicate that ``\textit{pairs of charge $2e$ are present in large portions of the parameter space dominated by the pseudogap.}'' We caution here, however, that evidence~\cite{Bastiaans2021} of 2$e$ pairing may be found in the pseudogap phase of highly disordered, presumably weakly coupled 2D superconductors. In this way, 2$e$ pairing is a necessary but not sufficient effect to establish BCS-BEC crossover.

(ii) Also relevant is the two-gap
dichotomy~\cite{Hashimoto2014,Hufner2008} in which there are
distinctive temperature dependencies of the ARPES- or STM-associated
gaps in the nodal and anti-nodal regions. In the BCS-BEC crossover
scenario this two-gap behavior derives from the simultaneous presence
of condensed and non-condensed pairs.

(iii) Additionally, an observed downturn~\cite{Timusk1999} in the DC
resistivity near or below $T^*$ seems most naturally to be associated
with the contribution from bosonic transport or from preformed
pairs. Indeed this small downturn feature is often used as the
canonical signature of $T^*$.

(iv) Lending some support to the crossover picture is the behavior of
the GL coherence length in the cuprates, which is still
not firmly established, as it turns out to be quite difficult to
measure due to vortex liquid effects.  Some indications of 
behavior that is rather similar to that found in the organic 2D superconductor~\cite{Suzuki2022} 
can be seen from
Figure 14(a) in \textcite{Suzuki1991}. This is measured above $T_\text{c}$ in the normal state.

(v) Finally, there is a notable similarity between many properties of
a single layer cuprate material and that found for its counterpart in
bulk systems~\cite{Yu2019b}; this would seem to be compatible with the
similarity contained in Eqs.~\eqref{eq:2} and \eqref{eq:4}.

We will discuss some of these experiments in the following subsections.

\begin{figure}
\centering
\includegraphics[width=2.5in,clip]{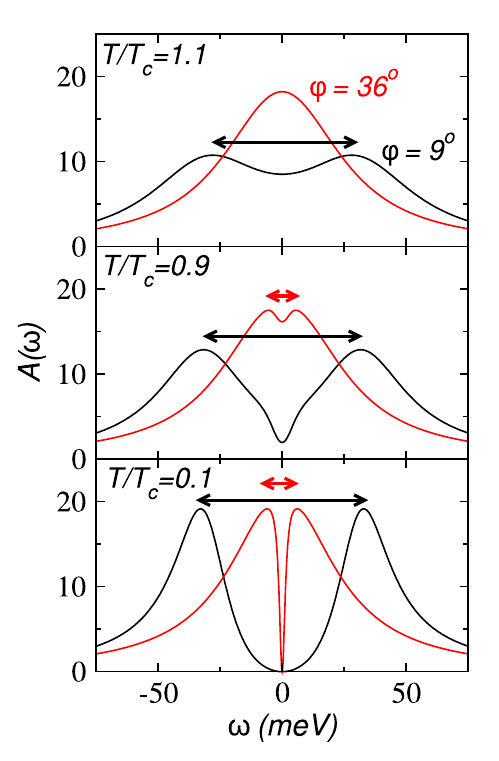}
\caption{Calculated spectral function
  $A(\omega,\varphi)$ at $T/T_\text{c}=1.1, 0.9, 0.1$ (from top to bottom)
  for $\varphi=9^\circ$ (black) and $\varphi=36^\circ$ (red), taken from \textcite{Chien2009}. Black
  and red arrows indicate the size of the spectral gap, which is measured
  in ARPES. $\varphi$ is defined in Fig.~\ref{fig:25}. }
\label{fig:23}
\end{figure}

\begin{figure*}
\centering
\includegraphics[width=5.5in,clip] {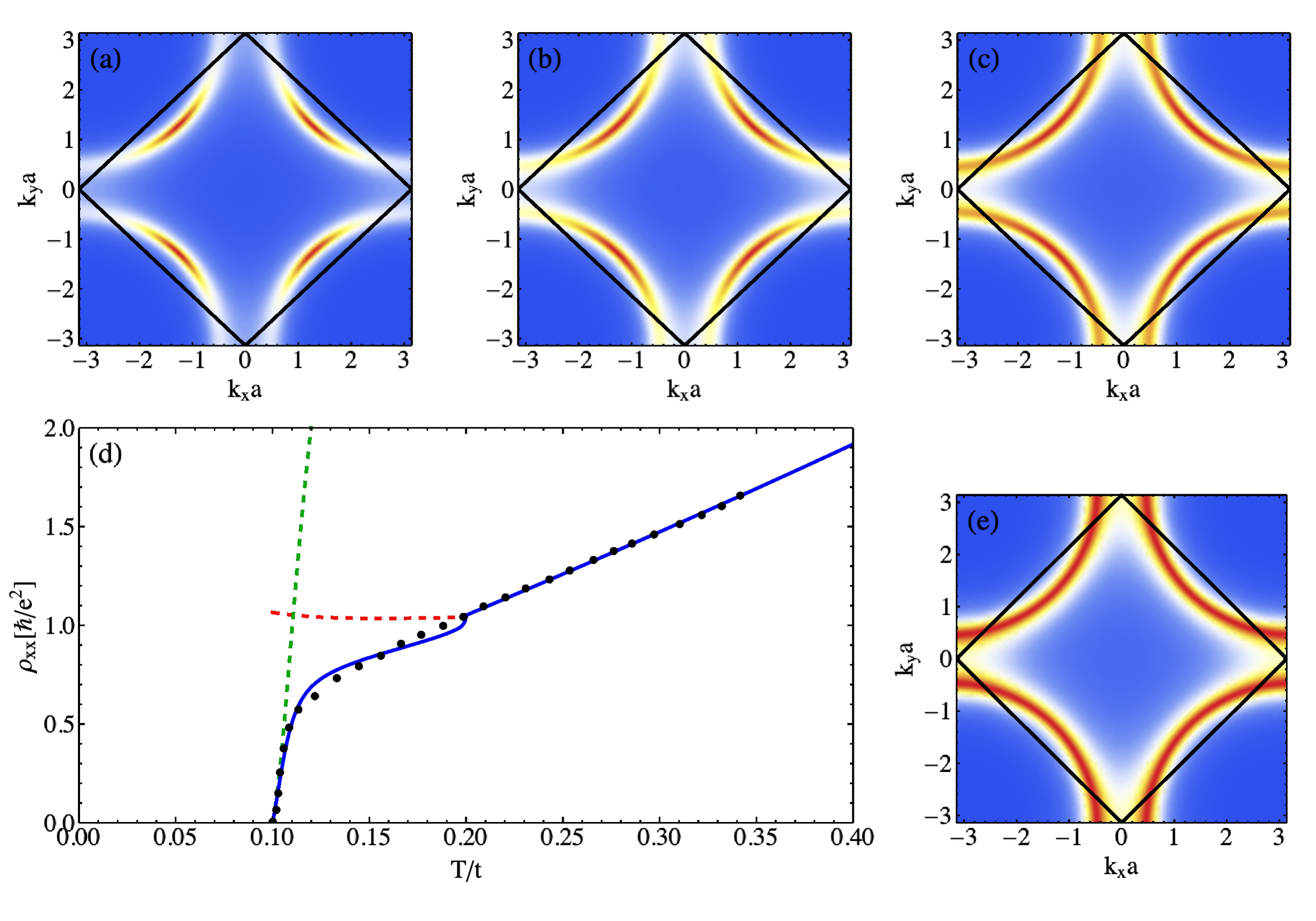}
\caption{Calculated behavior of the cuprate resistivity and
  temperature evolution of the Fermi
  arcs. ``Bad-metal'' behavior is important here as
  the small conductivity in the fermionic channel enables the bosonic
  downturn in the resistivity to be more evident. Panels (a)--(c) and
  (e): representative spectral function $A(\omega=0,\vect{k})$ for
  temperatures (a) $T/t = 0.11$, (b) 0.15, (c) 0.18, and (e)
  0.23. Here, $T_\text{c}/t = 0.1$ and $T^{*}/t = 0.2$, where $t$ is the
  nearest neighbor hopping integral. In panel (d) black dots are
  experimental data for an underdoped \BSCCO~\cite{Watanabe1997}. The
  solid and dashed lines are theoretical fits. Blue solid line:
  calculated total $\rho_{xx}$. Red dashed (dark-green dashed) line:
  fermionic (bosonic) contribution to $\rho_{xx}$. From \textcite{Boyack2021}.}
\label{fig:24}
\end{figure*}

\subsection{The spectral function: distinguishing condensed and non-condensed pairs}

We first address the so-called ``two-gap
dichotomy''~\cite{Hashimoto2014, Hufner2008}, which pertains to the
behavior of the spectral function where it should be clear that
$d$-wave pairing plays an important role. In the BCS-BEC crossover
scenario~\cite{Chen2005} the fermionic self energy, which is measured
in the spectral function, has two contributions from non-condensed (pg)
and condensed (sc) pairs:
\begin{equation}
\Sigma(\omega,\mathbf{k}) = \frac{\Delta_{\pg,\mathbf{k}}^2}{\omega + \xi_{-\vect{k}} + i \gamma} + \frac{\Delta_{\sc,\mathbf{k}}^2}{\omega + \xi_{-\vect{k}} } +i\Gamma_0.
\label{eq:29}
\end{equation}
This same spectral function appeared earlier as Eq.~(\ref{eq:29b}),
but here we emphasize the momentum dependence associated with
non-$s$-wave pairing, and, as customary, we add an additional
phenomenological lifetime $\Gamma_0$ arising from incoherent,
single-particle scattering processes.
It might be noted that because of these two components, this BCS-BEC
crossover scheme has Green's functions that are similar to those in a
highly regarded cuprate theory often called the ``YRZ''
theory after the authors Yang, Rice, and Zhang~\cite{Rice2011}. In the
BCS-BEC crossover scenario one finds Fermi arcs whereas YRZ
incorporates Fermi pockets~\cite{Scherpelz2014}.

In the normal state, a form~\cite{Chen2008} similar to
Eq.~(\ref{eq:29}) was shown~\cite{Norman1998} to provide a reasonably
good fit to ARPES data and insights into the Fermi
arcs~\cite{Kondo2015}.
How do the Fermi arcs originate? One should note
that the non-condensed pairs have a finite lifetime, in contrast to the
condensate. This is particularly important for the case of $d$-wave
pairing. If we consider cooling from above to below $T_\text{c}$, we see that
the onset of the condensate gap $\Delta_{\sc}$ in the fermionic
spectral function is more dramatic in the nodal region where there is
no normal state background gap already present. By contrast, in the
antinodal region the onset of $\Delta_{\sc}$ on top of a large
$\Delta_{\pg}$ has very little impact. Thus, as illustrated below, it
is the temperature dependence of the nodal gap that reflects the
onset of the ordered state.

\begin{figure*}
\centering
\includegraphics[width=6.0in,clip] {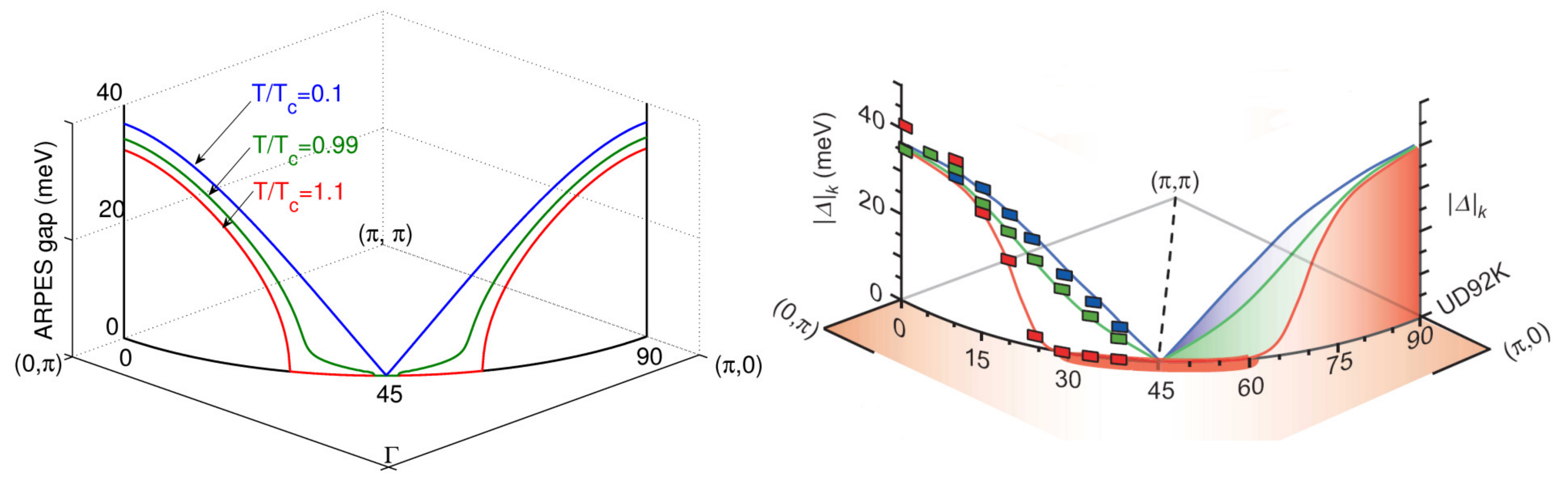}
\caption{Inferred ARPES gaps as a function of $\veck$ in one quadrant
  of the Brillouin zone.  Fermi arcs (associated with $d$-wave
  pairing) appear on the Fermi surface near the nodal direction around
  $\varphi=45^\circ$. Comparison of theory  from \textcite{Chien2009} on the
  left with experiment from \textcite{Lee2007} on the right.}
\label{fig:25}
\end{figure*}

More quantitatively~\cite{Chen2008,Chien2009}, one defines the spectral (or ARPES) gap as one half of the peak-to-peak separation in the spectral function. Figure~\ref{fig:23} illustrates the temperature evolution of the spectral function for $\varphi=9^\circ$ (close to the antinodes in Fig.~\ref{fig:25}) and $\varphi=36^\circ$ (close to the nodes) at varying $T/T_\text{c}$ from top to bottom.  Above $T_\text{c}$ (top panel) the well understood behavior~\cite{Chubukov2007,Kanigel2008,Norman2007} sets the stage for the normal phase which underlies the superconducting state in the next two panels. At this temperature, $T/T_\text{c}=1.1$, one sees Fermi arcs in the Brillouin zone. Here the spectral function is gapless on the Fermi surface near the nodal direction while it is gapped in the vicinity of the anti-nodal direction. The Fermi arcs derive from the presence of a temperature independent broadening term $\gamma$ in $\Sigma_{\pg}$. When $T$ is slightly below $T_\text{c}$ (middle panel), a dip in the spectral function at $\varphi = 36^\circ$ suddenly appears at $\omega=0$. At this $\varphi$ the underlying normal state is gapless so that the onset of the additional component of the self energy via $\Sigma_{\sc}$ with long-lived pairs leads to the opening of a spectral gap. 

By contrast, the presence of this order parameter is not responsible for the gap near the anti-nodes ($\varphi = 9^\circ$), which, instead, primarily derives from $\Delta_{\pg}$.  Here the positions of the two maxima are relatively unchanged from their counterparts in the normal phase.  Nevertheless, $\Delta_{\sc}$ does introduce a sharpening of the spectral function, associated with the deepening of the dip at $\omega=0$. When $T\ll T_\text{c}$ (lower panel), pairing fluctuations are small so that $\Delta(T) \approx \Delta_{\sc}(T)$ and one returns to a conventional BCS-like spectral function with well established gaps at all angles except at the precise nodes.

\subsection{Transport in the cuprates}

That the cuprates are highly resistive or bad metals~\cite{Gunnarsson2003}  is important for understanding their transport properties. This is what allows the boson-related downturn for transport at $T^*$ in the resistivity, a canonical signature of the pseudogap onset~\cite{Timusk1999}, to become evident (see Fig.~\ref{fig:24}).  This would otherwise be obscured by gap effects in the fermionic spectrum. The fits to the longitudinal DC resistivity shown in Fig.~\ref{fig:24}  are based on a phenomenological model~\cite{Boyack2021} for the pair chemical potential ($\mu_{\pair}$) which incorporates the standard fluctuation 
behavior for $T\gtrsim T_c$, 
given by $\mu_\text{\pair} \approx (8/\pi) (T_\text{c}- T)$.
Here, however, one includes
$T^*$ and higher temperature effects through
a natural interpolation by associating $T^*$ with the temperature where the number of pairs vanishes. This leads to a consolidated form:
\begin{equation}
\mu_{\pair}   =  \frac{8}{\pi} (T^*-T_\text{c}) \ln\frac{T^* -T}{T^*-T_\text{c}}.
\label{eq:28}
\end{equation}

This form for $\mu_{\pair}$ leads to fits to the resistivity, $\rho(T)$, and its downturn in Fig.~\ref{fig:24} which are not unreasonable; also emphasized here is the presence of ``Fermi arcs'', which additionally help to reveal bosonic transport by suppressing the gap in the fermionic spectrum. With the same parameters one can arrive at some understanding of the Nernst effect~\cite{Boyack2021}. However  there are problematic issues concerning the Hall coefficient~\cite{Boyack2021, Geshkenbein1997} and the thermopower, which affect essentially all theoretical attempts to understand these cuprate data and make a direct comparison difficult between theory and experiment.

Indeed, there is a sizeable literature dealing with the Hall coefficient in the underdoped regime~\cite{Rice1991,Hwang1994,Lang1994, Samoilov1994,Jin1998,Konstantinovic2000, Matthey2001,Ando2002,Segawa2004}. Among the most serious problems is that the measured $\sigma_{xy}$ appears to be not as singular near $T_\text{c}$ as is predicted by Gaussian fluctuation theories, where the expected singularity is stronger than in $\sigma_{xx}$. This is presumably associated with the experimental observation that $R_{\text{H}}\propto \rho_{yx}$ starts to drop with decreasing $T$
slightly above $T_\text{c}$~\cite{Lang1994,Jin1998}  and can even change its sign as $T$ decreases towards $T_\text{c}$.

Similarly, the normal-state thermopower in underdoped cuprates~\cite{Munakata1992, Huang1992,Fujii2002, Badoux2016, CyrChoiniere2017} (at $T\simeq T^*$) is positive in the experiments for the samples with the largest pseudogap. This is opposite to the usual band-structure predictions, and also opposite to the sign of the Hall coefficient. Given these problems for the thermopower and Hall coefficients, comparisons between experiments are best addressed in the case of the Nernst coefficient.

\subsection{Quantifying the Fermi arcs}

Understanding and quantifying the Fermi arcs has become an important
issue in the cuprates.  In addition to ARPES experiments the existence
of Fermi arcs appears to have been independently established in STM
data as well \cite{Pushp2009,Lee2009}.  The right panel of
Fig.~\ref{fig:25} presents gaps extracted from ARPES
data~\cite{Lee2007} for a moderately underdoped sample. The three
different curves correspond to three different temperatures with the
same legend as that in the left panel (representing the results of
theory). Importantly one sees a pronounced temperature dependence in
the behavior of the ARPES spectral gap for the nodal region (near
$45^{\circ}$) as compared with the antinodal region (near 0 and
$90^{\circ}$), where there is virtually no $T$ dependence. The left
panel presents the corresponding theoretically predicted behavior, which
exhibits some similarities.

\begin{figure}
\centering
\includegraphics[width=3.0in]{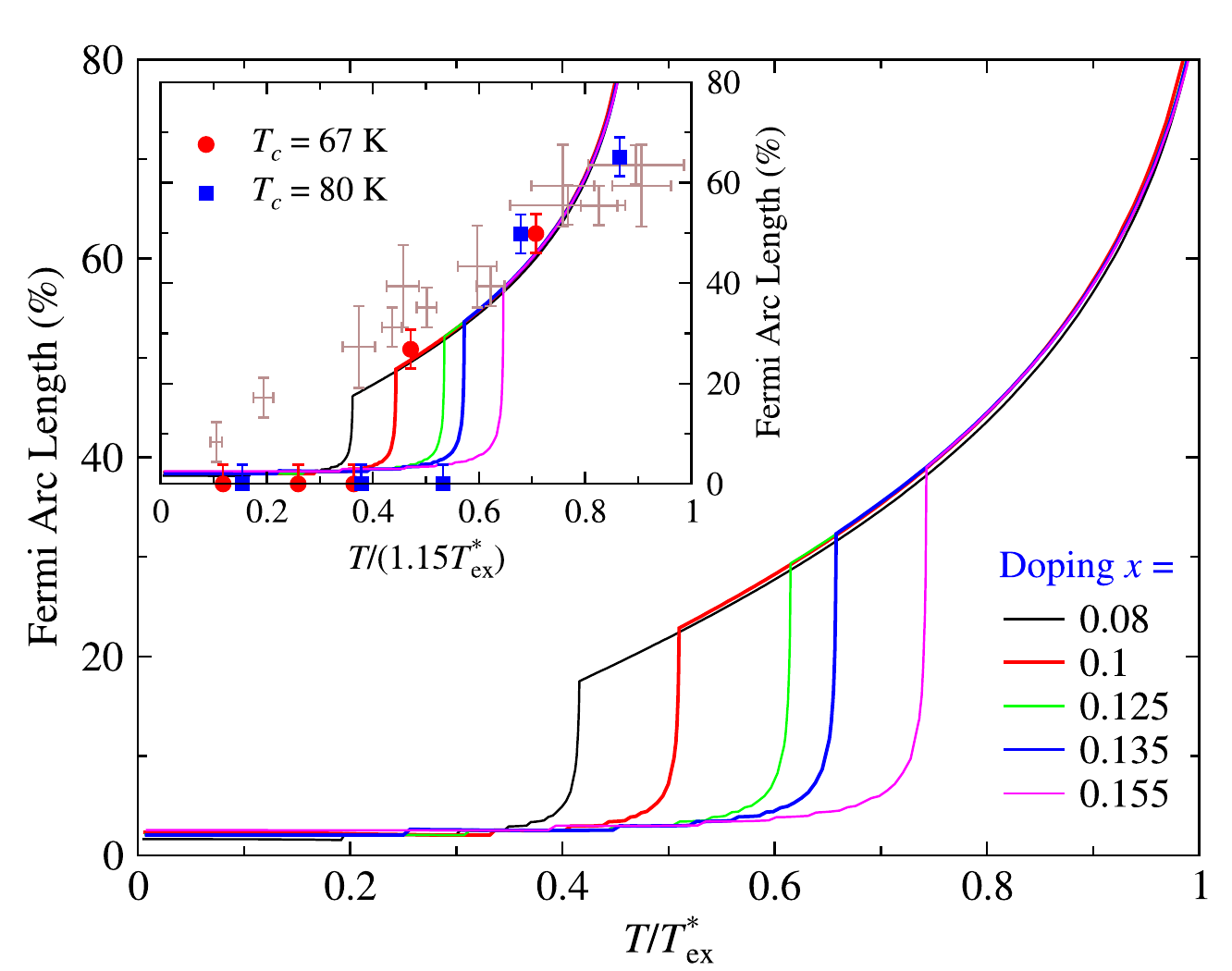}
\caption{ARPES comparisons in cuprates showing collapse of the Fermi arcs at the superconducting transition. The figure compares experimental data points~\cite{Kanigel2007} with theoretical curves~\cite{Chen2008}. Here $T^{*}_{\text{ex}}$ is the experimental $T^*$ determined by ARPES data. From \textcite{Chen2008}.}
\label{fig:26}
\end{figure}

Figure~\ref{fig:26} addresses the temperature dependence of the Fermi arcs and their sharp collapse~\cite{Chen2008} from above to below $T_\text{c}$. Note that here it is assumed (for simplicity) that the broadening parameter $\gamma$ is temperature independent, as the non-condensed pairs, which persist below $T_\text{c}$, continue to be distinguished from the condensate there. Plotted is the percentage of arc length as a function of $T/T^*$ and for different doping concentrations from the optimal to the underdoped regime. There is a clear universality seen in the normal state, in both theory and experiment~\cite{Kanigel2007} (shown in the inset).

\subsection{Behavior of the finite-\texorpdfstring{$\omega$}{} conductivity}

There is a substantial interest~\cite{Basov2005,Bilbro2011} in the
complex ac-conductivity
$\sigma(\omega) = \sigma_1(\omega) + i \sigma_2(\omega)$ in the
cuprates, notably both in the optical regime and at THz frequencies.
These experiments are particularly useful as they can reveal important
information about low-energy excitations and charge dynamics.  Both
gapped fermions and non-condensed Cooper pairs can contribute to
$\sigma(\omega)$.  In theory work summarized here, only the fermionic
contributions were considered and this might reasonably be viewed as a shortcoming.

\begin{figure*}
\centering
\includegraphics[width=6.in,clip] {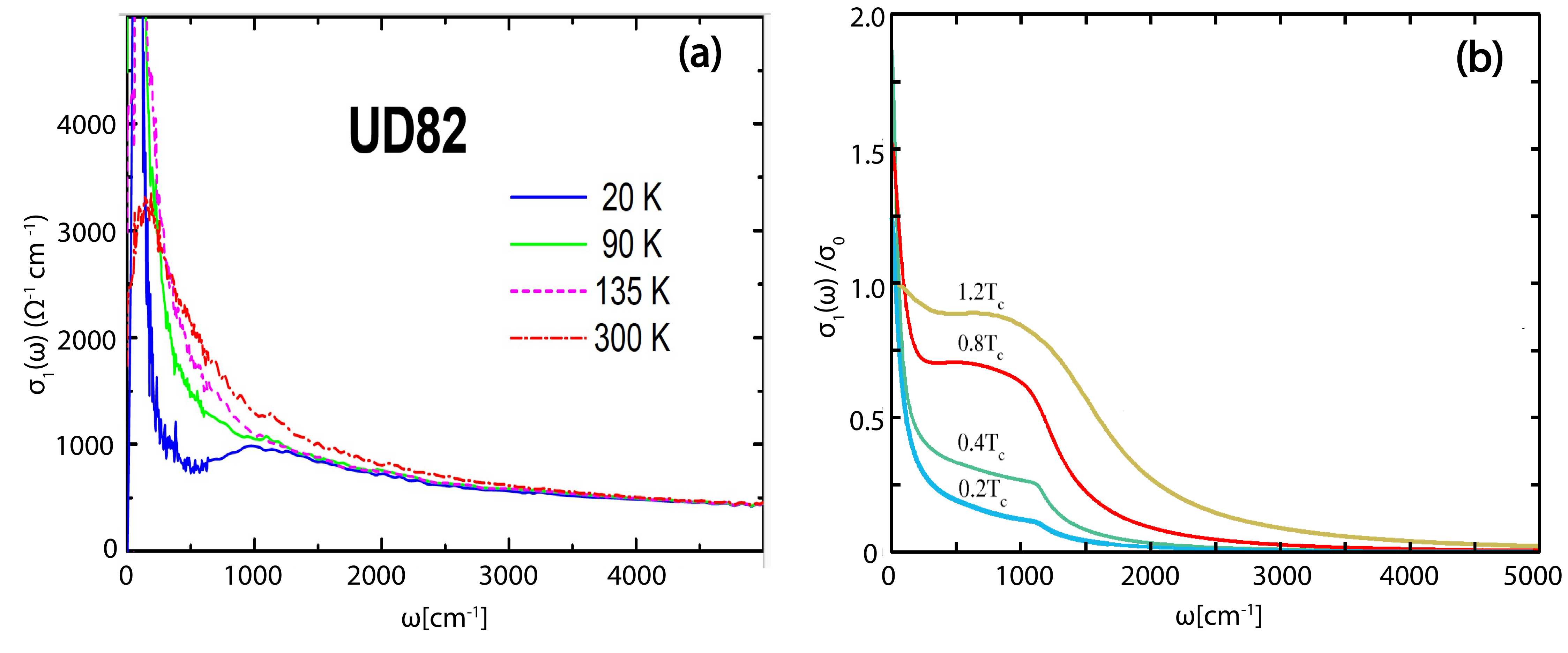}
\caption{Mid-infrared conductivity plots in cuprates showing that in the theory from \textcite{Wulin2012}, plotted on the right, and experiment from \textcite{Hwang2007}, shown on the left for an underdoped (UD) Bi2212 superconductor with $T_\text{c}=82$~K.
Both the theory and experimental figures show the real part of the frequency dependent conductivity $\sigma_1(\omega)$, 
at different indicated temperatures. The mid-infrared peak is presumed to be associated with the presence of a pseudogap.}
\label{fig:27}
\end{figure*}

A key feature of the in-plane $\sigma_1(\omega)$ is its two component
nature consisting of a ``coherent'' Drude-like low-$\omega$ feature
followed by an approximately $T$-independent mid-infrared (MIR)
peak~\cite{Basov2005, Lee2005, SantanderSyro2004}. This is illustrated
in Fig.~\ref{fig:27}. As stated in \textcite{Lee2005}: 
``\textit{The two component conductivity extends to the pseudogap boundary in the phase diagram... 
Moreover a softening of the mid-infrared band with doping resembles the decrease of the pseudogap temperature $T^*$.}''
Also of importance is the fact ~\cite{Kamaras1990} that 
``\textit{high $T_\text{c}$ materials are in the clean limit}'' and also that
``\dots\textit{the MIR feature is seen above and below $T_\text{c}$.}''
Thus, it appears that this MIR feature is not associated with
disordered superconductivity and related momentum non-conserving
processes, but rather it is due to the unconventional nature of the finite-frequency response~\cite{Basov2005}.

Within the crossover scenario, the presence of non-condensed pairs both above and below $T_\text{c}$ yields~\cite{Wulin2012} a mid-infrared peak. This peak occurs around the energy needed to break pairs and thereby create conducting fermions. Its position is doping dependent, and only weakly temperature dependent, following the weak $T$ dependence of the excitation gap $\Delta(T)$. As $T$ decreases below $T_\text{c}$, the relatively high frequency spectral weight from these pseudogap effects, present in the normal phase, is transferred to the condensate. This leads to a narrowing of the low-$\omega$ Drude feature, as can be seen in both plots in Fig.~\ref{fig:27}.

\begin{figure*}
\centering
\includegraphics[width=6.in]
{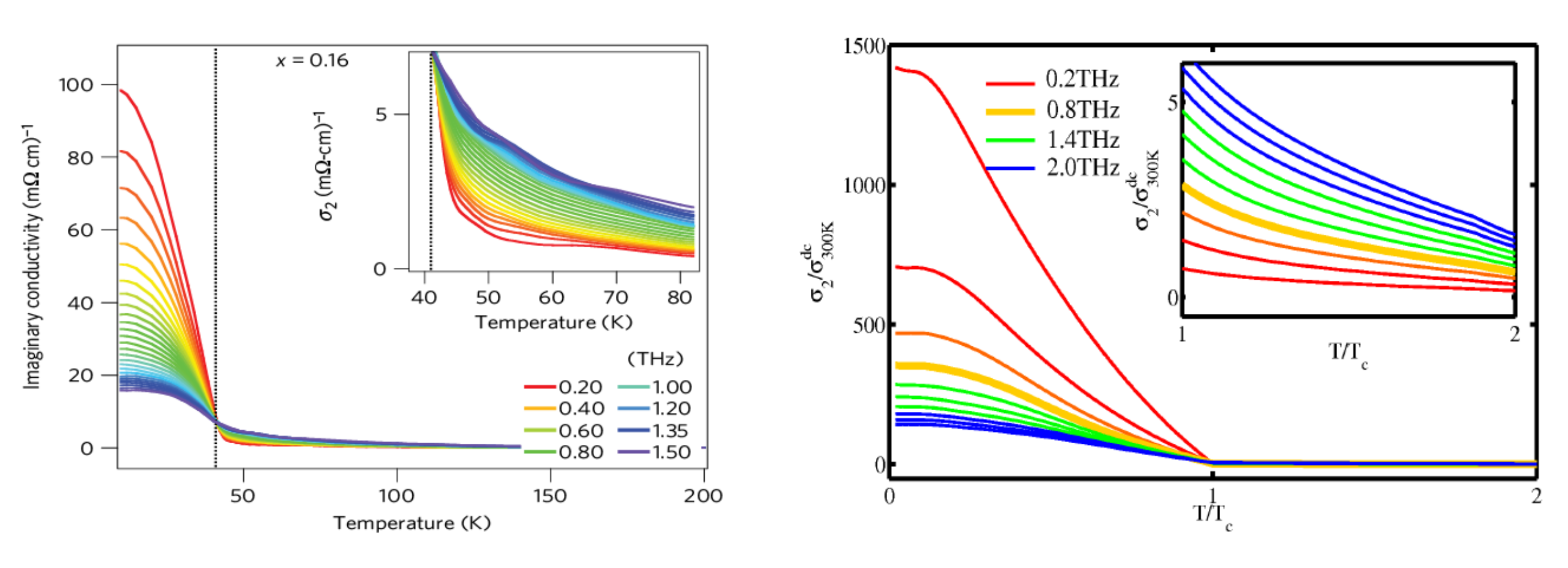}
\caption{Comparison of the behavior of the imaginary part of the THz
  conductivity, ($\sigma_2$), in cuprates, at different frequencies as a
  function of temperature.  Experimental data from \textcite{Bilbro2011} at optimal doping
  ($x=0.16$)  are plotted on the left and
  theory from \textcite{Wulin2012a} on the right.
  A moderately large normal-state $\sigma_2$ is thought to reflect the
  presence of a dynamical or fluctuating superfluid density. For this
  reason there is an enhanced plot of the normal-state region in the
  inset accompanying both plots.}
\label{fig:28}
\end{figure*}

Figure~\ref{fig:28} shows the theoretical prediction~\cite{Wulin2012a} and experimental behavior~\cite{Bilbro2011} found for the imaginary part of the THz conductivity, $\sigma_2(\omega)$, in the right and left panels, respectively. With decreasing temperature, at roughly $T_\text{c}$, $\sigma_2$ shows a sharp upturn at low $\omega$, of the form $\sigma_2 \propto n_s/\omega$, where $n_s$ is the superfluid density. The low-$\omega$ contribution above $T_\text{c}$ is of interest to the extent that it may reflect the presence of dynamical superfluid correlations. This is shown in the insets which present  an expanded view of the temperature dependencies near $T_\text{c}$. Both theory and experiment show that  the nesting of the $\sigma_2$ versus $T$ curves switches order above $T_\text{c}$. It should be emphasized that for this particular class of experiments the contribution from preformed pairs does not extend to very high temperatures. Indeed, here the effects are confined to temperatures in the vicinity of $T_\text{c}$, well below $T^*$. This is in contrast to other fluctuation experiments. It is notable, moreover, that the experimental data shows a more pronounced normal-state contribution than found in theory.

\begin{figure*}
\centering
\includegraphics[width=6.0in,clip] {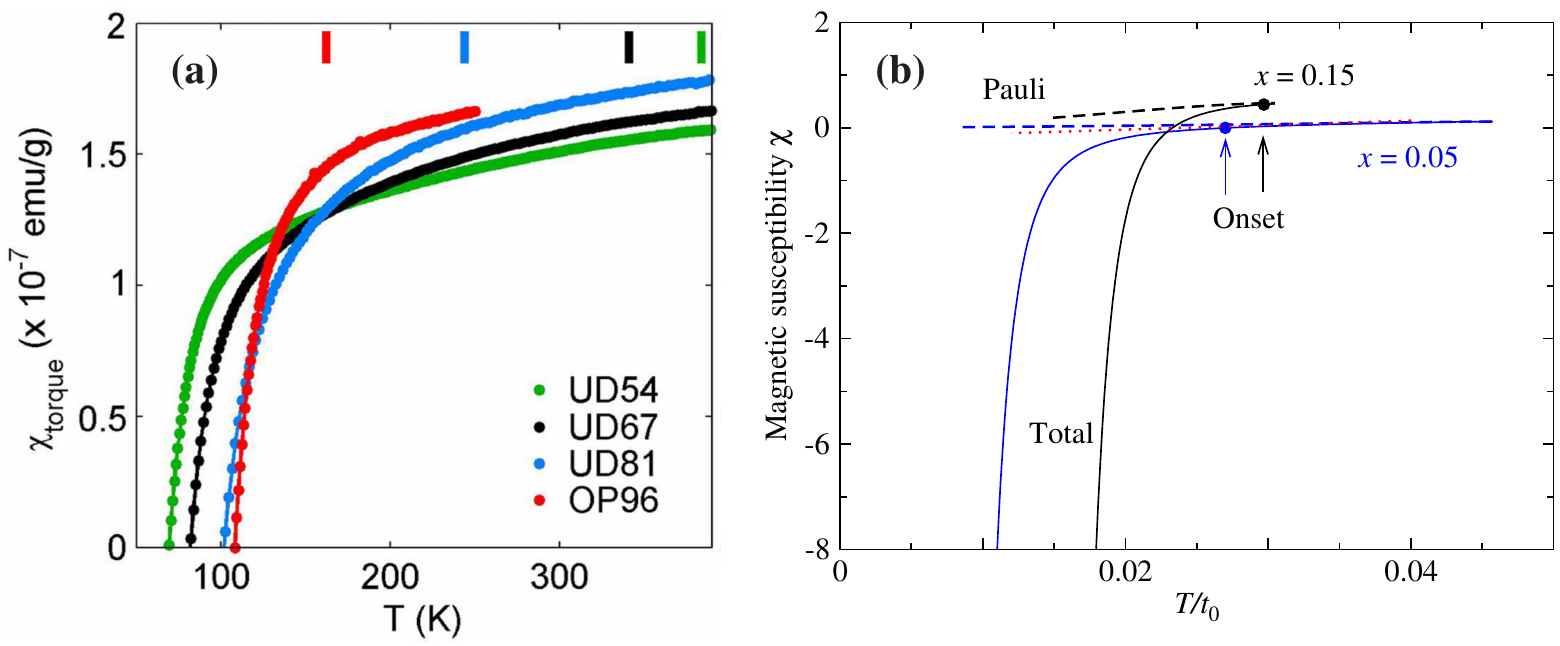}
\caption{Comparison of the behavior of the diamagnetic response above $T_\text{c}$ between (a) experiment taken from \textcite{Yu2019} and (b) theory taken from \textcite{Boyack2018}. In the theory plot, the (black) curve for optimal hole doping ($x=p=0.15$) and the (blue) curve for an underdoped system are labeled. The dashed lines are the Pauli paramagnetic susceptibility for each, while the solid lines are the sum of paramagnetic and diamagnetic contributions. The solid dots in (b) indicate the temperature where the onset of the diamagnetism occurs. For the underdoped case the red dotted lines are a linear fit to the high temperature data. }
\label{fig:29}
\end{figure*}

\subsection{Precursor diamagnetism}

The normal-state diamagnetic susceptibility in cuprates has also been widely discussed~\cite{Li2010}. Here, by contrast with the discussion surrounding $\sigma(\omega)$ above, the interest is focused on the bosonic contributions.  In conventional fluctuation theory~\cite{Larkin2009} the diamagnetic susceptibility, $\chi_{\dia}$, in the vicinity of $T \approx T_\text{c}$ can be relatively large as it scales (in 3D) as $1/\sqrt{T-T_{c}}$. What happens in BCS-BEC crossover theory as a consequence of the presence of a pseudogap? In a BCS-BEC crossover scenario $\chi_{\dia}$ now scales~\cite{Boyack2018} as $\sqrt{1/|\mu_{\pair}|}$ and, as can be seen from Eq.~\eqref{eq:28}, the principal effect is that the inverse pair chemical potential remains appreciable now for an extended range of temperatures well beyond the critical region around $T_\text{c}$, and strictly vanishing only at $T^*$.

This, in turn, suggests that there are fluctuation contributions to the diamagnetism at relatively higher temperatures than generally observed in conventional superconductors. It should be noted, however, that the visibility of fluctuation diamagnetism depends on other background, generally paramagnetic, contributions which are often difficult to quantify. A more detailed analysis leads to the results in Fig.~\ref{fig:29}, which shows a comparison between experiment~\cite{Yu2019} and theory~\cite{Boyack2018}.

\subsection{Other applications of BCS-BEC crossover: Features of the non-Fermi liquid}

By way of completeness, we end by including several other literature contributions which address BCS-BEC crossover theory in cuprates but for which there are no direct back-to-back experimental comparisons. These involve studies of how the non-Fermi liquid pseudogap state is reflected in quasi-particle-interference (QPI) experiments~\cite{Wulin2009} based on STM probes, and how it is reflected in quantum oscillations~\cite{Scherpelz2013a}.
In particular, it is found that the observation of a QPI pattern consistent with the so-called~\cite{Kohsaka2008} ``octet model'' is a direct signature of coherent superconducting order. 
It appears from theory that the QPI pattern in the pseudogap state~\cite{Wulin2009} is distinctly different from that in the superconducting phase.

\section{Conclusions}
\label{sec:Concl}

\subsection{Summary}

This Review article has been written in response to the large and relatively recent experimental literature on strongly correlated superconductors that are thought to exhibit BCS-BEC crossover phenomena. Many of these derive from artificial materials such as magic-angle twisted bilayer and trilayer graphene, quantum Hall bi-layers, or ionic-gate tuned semiconductors, as well as single unit cell and interfacial superconducting films. Also exciting are naturally grown superconductors, such as the Fe chalcogenides and the organic superconductor $\kappa$-(BEDT-TTF)$_4$Hg$_{2.89}$Br$_8$.

Because of the widespread interest, it is important to establish more precisely what BCS-BEC crossover theory is and what it is not. We have done so in this Review and in the process clarified distinctions between the Fermi gas and solid-state superconductors, between two  and three-dimensional materials, between $s$- and $d$-wave order parameter symmetries, and we have established distinguishing features of the BEC phase.

More generally, in this Review and in the context of different experiments, we addressed the three distinct ways of promoting a system out of the BCS  and into the crossover regime via either (i) small electronic energy scales, (ii) two dimensionality, or (iii) strong pairing ``glue''. We have emphasized that superconducting ``domes'' and ``pseudogaps'' are ubiquitous for crossover systems in periodic lattices. 

The narrative arc of this Review is encapsulated through the evolution
from Fig.~\ref{fig:1} to the next figure we now discuss,
Fig.~\ref{fig:30}.  Figure~\ref{fig:1} introduced the concept of
BCS-BEC crossover by raising the question of how to treat
superconductivity in the presence of progressively stronger attractive
interaction strengths. Notably, in contrast to the cold Fermi gases,
solid state experiments have little access to this interaction
strength parameter.

\begin{figure*}
\centering
\includegraphics[width=5.in,clip]{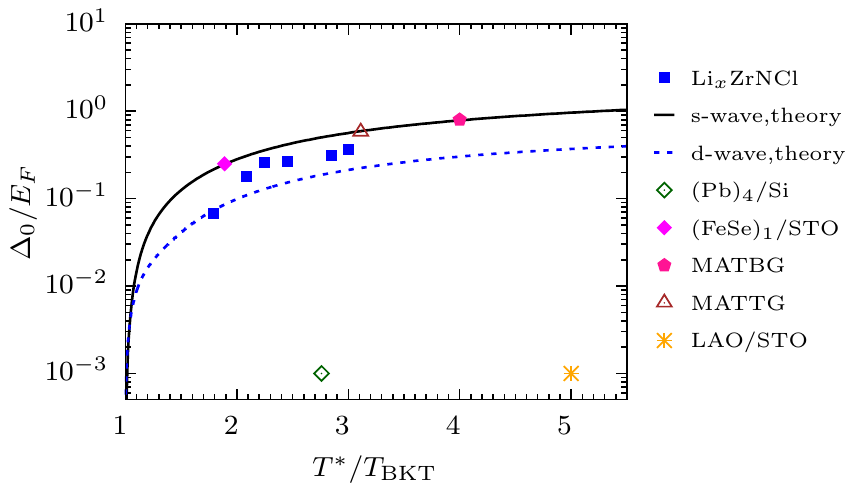}
\caption{
  Comparison between 2D BCS-BEC crossover theoretical predictions and experimental systems discussed in this paper. The two theory curves correspond to $s$- and $d$-wave pairing results obtained for a square lattice. In the vertical axis, the value of $\Delta_0$ is assumed to be
at $T=0$. The data points (see Appendix \ref{sec:App}) come from experiments on the lithium-intercalated nitride films \cite{Nakagawa2021},  one unit cell FeSe on strontium titanate~\cite{Faeth2021} and  magic-angle twisted bilayer and trilayer graphene \cite{Cao2018,Oh2021,Park2021,Kim2022}. Two additional data sets are associated with strongly disordered Pb films~\cite{Zhao2013} and from the interface superconductor
$\rm{LaAlO_3/SrTiO_3}$~\cite{Bozovic2020}. 
Among these the disordered Pb films are clearly
not related to BCS-BEC crossover nor are the $\rm{LaAlO_3/SrTiO_3}$ films, which appear also to be subject to disorder~\cite{Chen2018}. 
Importantly, this figure suggests a clear separation between superconductors which are compatible with
BCS-BEC crossover physics and those which are not.}
\label{fig:30}
\end{figure*}

Figure \ref{fig:30}, which represents a summary of many of the various 2D superconducting materials discussed in this Review, allows us to compare crossover theory and experiment. 
This is made possible by effectively representing the dimensionless interaction strength parameter in BCS-BEC crossover theory through dimensionless ratios of physically accessible parameters,
such as $T^*/T_\BKT$ and $\Delta_0/E_\text{F}$.
One could similarly consider $k_\text{F} \xi_0^{\coh}$ in counterpart plots.
All of these are strongly inter-connected and, importantly, the figure indicates that their 
inter-dependencies are generally robust to variations in the pairing symmetry (here from $s$-wave to $d$-wave).

Plotted on the vertical axis in a logarithmic scale is $\Delta_0/E_\text{F}$, where $\Delta_0$ is the zero-temperature excitation gap, while on the horizontal axis in a linear scale is $T^*/T_{\BKT}$ for two-dimensional superconductors. The upper (black) and lower (blue) theory curves are for $s$- and $d$-wave pairing symmetries, respectively. The data points come from the lithium-intercalated nitride films~\cite{Nakagawa2021}, from one unit cell FeSe on strontium titanate~\cite{Faeth2021} and from magic-angle twisted bilayer as well as trilayer graphene~\cite{Cao2018,Oh2021,Park2021,Kim2022}. 

Two additional data sets are associated with 
strongly disordered Pb films~\cite{Zhao2013} and from the interface superconductor
 $\rm{LaAlO_3/SrTiO_3}$~\cite{Bozovic2020}; the latter system does not fall into any simple category. In this plot, because of
their small $\Delta_0/E_\text{F}$ ratios, both are clearly distinct from BCS-BEC crossover candidate materials.
A comparison of theory and experiment in this replotting, thus, serves to highlight the distinction between strong pairing and strong disorder. In this way, the figure serves as a template for helping identify BCS-BEC crossover systems. The existence of a pseudogap (through the deviation of $T^*/T_\text{c}$ from unity), as well as observations of $2e$ pairing, appear insufficient. 

Additionally, we have addressed the question of under what circumstances should one expect to reach the BEC regime for a solid-state superconductor. In general, in this regime, rather than a very large transition temperature, one finds very small magnitudes of $T_\text{c}$ or $T_\BKT$. This point is often missed in the literature because the standard for the BCS-BEC crossover phase diagrams is based on Fermi-gas physics, 
where the BEC asymptote is large. This distinction is emphasized in Fig.~\ref{fig:1} of this Review.

In the BEC regime, all signs of a Fermi surface have disappeared.
Thus far, we are not able to report any unambiguous evidence that
candidate systems have reached the BEC regime. Some signatures of the
BEC we invoked earlier are that in this regime the character of the
states within vortex cores~\cite{Chien2006b} is distinctly different.
Similarly, in this regime, coherence peaks in the quasiparticle
tunnelling characteristics will be absent. Theoretical indications are
that a BEC superconductor can occur when either
$T^*/T_\BKT$ is much larger, say of the order of $10$, accompanied by
more conventional values of $\Delta_0/E_\text{F}$ or alternatively with
$\Delta_0/E_\text{F}$ of the order of $10$ or more, accompanied by more
conventional values of $T^*/T_\BKT$. The latter relates to the
interesting scenario in which superconductivity occurs in the presence
of very flat energy bands with nontrivial band topology and quantum geometry.

It is important to emphasize that, to establish a given superconductor as
belonging to the crossover regime can be done through a two-parameter analysis 
(both $\Delta_0/E_\text{F}$ and $T^*/T_\text{c}$
must be moderately large as in Fig.~\ref{fig:30}) or through a one-parameter analysis, 
by showing that $k_\text{F} \xi_0^{\coh}$ is
moderately small but in excess of the lower bounds set by
Eqs.~(\ref{eq:new1}) and (\ref{eq:new2}).  These bounds arise because
the dimensionless coherence length is readily quantified in terms of a
fundamental variable of crossover physics: the normalized pair density
$n_\pair/n$ at the transition temperature.  This necessarily varies
continuously from 0 in the strict BCS limit to exactly $1/2$
(discounting small thermal effects) in the BEC regime, where 
$k_\text{F} \xi_0^{\coh}$ saturates.  As discussed in this Review,
such a compact expression for the coherence length follows from the
Schafroth-like equation for $T_\text{c}$ in Eq.~\eqref{eq:2}.  We note
that $k_\text{F}$ here reflects the fixed density of electrons in the
superconductor and, thus, does not contain many-body effects or other
band-structure complexities.  Finally, it is most gratifying that
experiments studying superconductivity in the solid state (as distinct
from the cold gases) have access (albeit with some
uncertainty~\cite{Suzuki1991}) to this parameter, as outlined in
Sec.~\ref{sec:Application}.

\subsection{Outlook}

More generally in looking toward the future,
we are poised at the beginning of an extremely exciting era where the development of
synthetic superconductors seems limitless. 
Tunable 2D superconductors (such as MATBG~\cite{Cao2018, Oh2021}, MATTG~\cite{Park2021,
Kim2022}, Li$_{x}$ZrNCl~\cite{Nakagawa2021} etc.) are likely candidates  
for realizing superconductivity in the strong-coupling regime. The coupling strength and Fermi energy can be dramatically and precisely tuned by twisting, gating, and doping, which
provides the best platform to observe BCS-BEC crossover physics and to compare with theory.

Importantly, the present review can serve as a blueprint for future
experimental endeavors, as it establishes concrete, experimentally
falsifiable criteria to determine whether a given superconductor is in
the BCS-BEC crossover regime. A singular observation of only the
pseudogap phase or pairing above $T_{c}$ no longer suffices. Future
experimental studies will need to combine measurements of $\Delta$,
$E_{F}$, $T^*$, and $T_{c}$ or $T_{\BKT}$ to place candidate materials
on Fig.~\ref{fig:30}.  Critical tests will be to perform these
measurements with a continuous tuning parameter (gating, doping,
twisting, or isovalent substitution), to enable the comparison between
theory and experiment in an extended region of Fig.~\ref{fig:30}. An
example of such very complete studies is the work summarized here on
Li$_{x}$ZrNCl~\cite{Nakagawa2021}.

\begin{table*}
    \center
    \caption{Experimental data collected for Fig.~\ref{fig:30}. 
Here we identify the low temperature gap with $\Delta_0$.
For  Li$_x$ZrNCl  different rows  are for different carrier densities.
    }
\begin{ruledtabular}
    \begin{tabular}{ccccccc}
        ~~~~~~ Materials~~~~~~   & ~~~~~~$T_{\mathrm{BKT}}$ ~~~~~~ &  ~~~~~~$T^*$~~~~~~  &  ~~~~~~~$\Delta_0$~~~~~~~     &  ~~~~~~$E_\text{F}$~~~~~~~     &  ~~~~~~ $T^*/T_\text{c}$~~~~~~    &  ~~~~~~~$\Delta_0/E_\text{F}$~~~~~~~      \\
\colrule
          (FeSe)$_1$/STO        &  38 K   &   72 K   &    15 meV                &    60 meV  &  1.89   & 0.25   \\
          (Pb)$_4$/Si                                  &  2.4 K    &   6.9 K  &      0.35 meV             &   380 meV  &  2.9   &  0.001   \\
           (001) LAO/STO                           & 100 mK  &  500 mK    &      65 $\mu$eV               &  47 meV    & 5    &  0.001    \\
           Li$_x$ZrNCl    &   0.031 $T_\text{F}$   &    0.055 $T_\text{F}$     &   --    &   --   &   1.78    &  0.067 \\
                                          &    0.061     $T_\text{F}$    &        0.13           $T_\text{F}$            &     --      &  --     &  2.1     &  0.18  \\
                                          &  0.088    $T_\text{F}$    &         0.20              $T_\text{F}$             &       --    &    --  &  2.25   &   0.26   \\
                                          &   0.097     $T_\text{F}$     &         0.24             $T_\text{F}$                 &        --   &   --   &  2.45   &   0.27   \\
                                        &   0.10     $T_\text{F}$    &         0.30              $T_\text{F}$                 &      --    &     -- &  2.84   &   0.31   \\
                                        &   0.12     $T_\text{F}$   &        0.35                  $T_\text{F}$             &      --     &    --   &  3.0    &   0.36   \\
          MATBG              & 1.0 K   &   4 K  &      1.4 meV   & 20 K    &  4    &   0.8   \\
           MATTG                         &    2.25 K                 &       7 K               &    1.6 meV         &    32 K   &  3.1      &  0.58  \\
    \end{tabular}
\end{ruledtabular}
    \label{tab:Fig30Data}
\end{table*}

It should be noted that other tunable 2D superconductors such as
twisted transition metal dichalcogenides can also host flat
bands~\cite{Devakul2021, Li2021}, and should be viewed as future
candidates for superconductivity in the BCS-BEC crossover regime. It
has also been predicted that nonequilibrium optical driving on twisted
bilayer graphene can induce flat-band behavior associated with an
effective Floquet Hamiltonian~\cite{Assi2021}; this provides a
route towards the strong-coupling limit.  The
implications of the BCS-BEC crossover scenario in the general
nonequilibrium context will be important to address.  Ultrafast
spectroscopic experiments should more generally be explored to
characterize this band-structure engineering and its
potentially new forms of superconductivity.

Additionally, the study of high-$T_{c}$ Fe-based superconductors will
lead to new opportunities and challenges to explore the connection
between the BCS-BEC crossover physics, high-$T_{c}$ superconductivity,
and topological superconductivity. It is worth noting that the
disparity between the transport $T_{c}$ ($\sim$ 40~K) and the
spectroscopic $T^*$ ($\sim$ 70~K) has been a fundamental issue
undermining further progress on monolayer FeSe/SrTiO$_{3}$
systems~\cite{Faeth2021}. This Review can serve as the starting point
to systematically explore crossover physics for understanding this
remarkable 2D high-$T_{c}$ superconductor. A systematic tuning
experiment using gating, doping, or Se:Te substitution will need to be
performed. Importantly, with a specific Se:Te ratio $=x:1-x$ between
$x=0.45$ and $x=0.55$ the FeTe$_{1-x}$Se$_x$ bulk system exhibits a
nontrivial topology with a superconducting topological surface
state~\cite{Zhang2018}. It remains to understand what the role of this
topology is in the crossover physics.

Among new theoretical challenges, BCS-BEC crossover theories of
superconductivity will need to accommodate the effect of magnetic
fields, which will complete understanding of the canonical
superconducting phase diagrams. What is the nature of the
non-condensed pairs in the presence of a magnetic
field~\cite{Scherpelz2013}? How does condensation proceed when the
dimensions of the system are effectively reduced by the presence of
Landau levels~\cite{Lee1972,Schafroth1955} and how does one understand
the dynamics of vortices~\cite{Mozyrsky2019} from BCS to BEC?
Conceptually related is the central and difficult issue: how to
generalize the Bogoliubov de Gennes equations to the crossover
situation at finite temperature. This would enable other important
calculations, for example, describing Andreev tunneling, effects of
proximitization and addressing the vast number of situations that involve spatially dependent superconductivity. It is notably a
difficult problem as one needs to incorporate two distinct types of
(now spatially dependent) gaps, associated with condensed and
non-condensed pairs.

In a discipline where theory and experiment work hand-in-hand, it
should be clear that the multiple experimental platforms described in
this section collectively present enormous opportunities for future
theoretical developments.  In the process they enhance our
understanding of this generalized BCS theory in a deeper and much
broader sense.

\section{Acknowledgments}

We would like to thank our various collaborators who contributed to this work and to our understanding over the years: B. Jank\'o, I. Kosztin, J. Maly, J. Stajic, A. Iyengar, S. Tan, C.-C. Chien, Y. He, H. Guo, D. Wulin, P. Scherpelz, B. Anderson, C.-T. Wu, X. Wang, K. M. Shen, and D. G. Schlom. This work was partially (K. L., Z. W.) supported by Department of Energy (DE-SC0019216). Q. C. was supported by the Innovation Program for Quantum Science and Technology (Grant No. 2021ZD0301904). R. B. was supported by the Department of Physics and Astronomy, Dartmouth College. S.-L. Y. acknowledges the support by the U.S. National Science Foundation through Grant 2145373.


\appendix

\begin{appendices}

\begin{figure*}
 \centering
\includegraphics[width=0.8\linewidth]{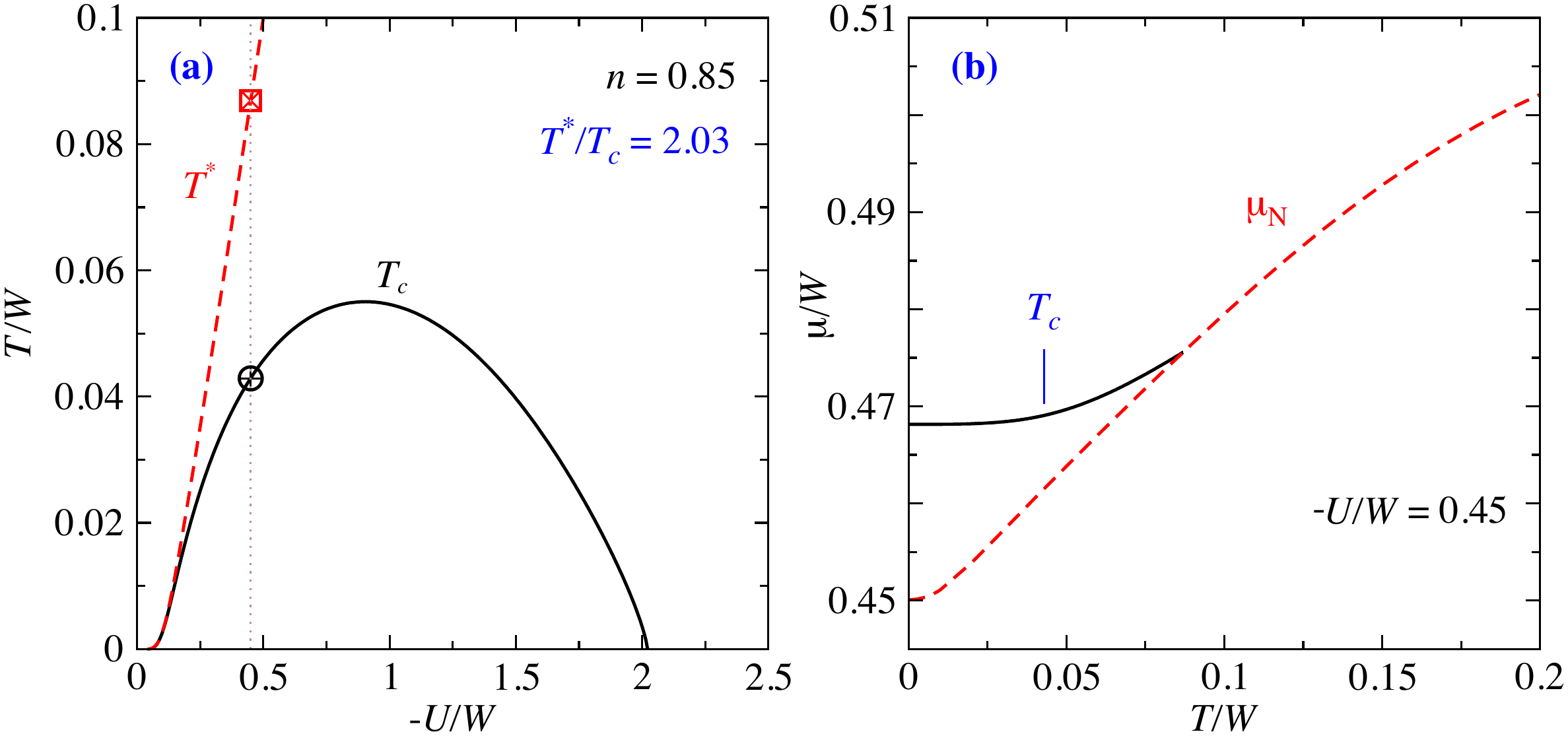}
\caption{(a) $T_\text{c}$ -- $U$ phase diagram for a $d$-wave
  superconductor with electron density $n=0.85$ on a quasi-2D square
  lattice. The energy dispersion is
  $\epsilon_{\vect k} = (4t+ 4 t^\prime +2 t_z) - 2t (\cos k_x +\cos
  k_y) - 4 t^\prime \cos k_x \cos k_y - 2 t_z \cos k_z$ with
  $t'=-0.3t$ and $t_z/t = 0.01$. All energies are normalized by
  $W=4t$. The pairing gap is $\Delta_\mb{k} =\Delta\varphi^{}_\mb{k}$
  with $\varphi^{}_\mb{k} =\cos k_x-\cos k_y$. (b) Temperature
  dependencies of the chemical potential $\mu$ and the extrapolated
  normal-state $\mu_{\text{N}}$, for interaction strength
  $U/W = -0.45$, corresponding to the vertical dotted line in (a).
  Emphasizing the small variations in $\mu$, here $\mu$ changes by
  $-0.5\%$ from $T=0$ to the pairing onset $T^*$, and
  $(\mu-\mu_{\text{N}})/\mu_{\text{N}}$ is found to be 3.8\% at $T=0$.
}
\label{fig:31}
\end{figure*}

\section{Experimental Data for 2D Superconductors}
\label{sec:App}

In this Appendix, we present in Table \ref{tab:Fig30Data} the data
collected for Fig.~\ref{fig:30} from various sources.  In this table,
if $T_{\mathrm{BKT}}$ is not available, we use the corresponding
$T_\text{c}$. The abbreviations are: (FeSe)$_1$/STO = monolayer FeSe
grown on the \STO~ substrate, (Pb)$_4$/Si = 4-monolayer Pb film grown
on the Si substrate, (001) LAO/STO = (001)-oriented
LaAlO$_3$/SrTiO$_3$ interface, MATBG = magic-angle twisted bilayer
graphene, and MATTG = magic-angle twisted trilayer graphene.
   
The sources of the data are as follows: for (FeSe)$_1$/STO,
$\{ T_{\mathrm{BKT}}, T^*\}$ are taken from \textcite{Faeth2021}, and
$\{\Delta_0, E_\text{F}\}$ from \textcite{Liu2012}.  For (Pb)$_4$/Si
the data for $\{ T_{\mathrm{BKT}}, T^*\}$ are from
\textcite{Zhao2013}. To estimate $\Delta_0/E_\text{F}$ we use
\textcite{Zhang2010}, where the sample used is actually a monolayer
Pb film on Si substrate, (Pb)$_1$/Si. We do not expect
$\Delta_0/E_\text{F}$ to differ much between (Pb)$_4$/Si and
(Pb)$_1$/Si.
 
The data for Li$_x$ZrNCl are taken from \textcite{Nakagawa2021}.  For
(001)LAO/STO we use \textcite{Pai2018} for $T_{\mathrm{BKT}}$,
\textcite{Richter2013} for $\{T^*, \Delta_0 \}$, and
\textcite{Sulpizio2014,Pai2018} for $E_\text{F}$.  In this system we
have used the $d_{xy}$ orbital band to arrive at $E_\text{F}$, and the
data collected all roughly correspond to the same gating voltage
$V_g\approx -100$~V.

The values of $\{T_{\mathrm{BKT}}, T^*, E_\text{F}\}$ for MATBG are
taken from \textcite{Cao2018} for a twist angle
$\theta\approx 1.05^\circ$. Here, $T^*$ is estimated from the Ohmic
recovery point from the $V$-$I$ characteristic measurement. $\Delta_0$
is obtained from \textcite{Oh2021}, which is appropriate to a very
close but slightly different twist angle $\theta\approx 1.01^\circ$
system.
   
For MATTG we use \textcite{Park2021} for $T_{\mathrm{BKT}}$ and \textcite{Kim2022} for $\{T^*,\Delta_0 \}$. The value of $E_\text{F}$ is estimated by Stevan Nadj-Perge and provided through a private communication.


\section{General BCS-BEC Crossover Theory for D-wave Case Near Half Filling}
\label{sec:AppB}

In this appendix, we present additional details about BCS-BEC crossover
theory in the $d$-wave case, focusing on the region around half filling
in the electron band. The results here are presumed to be generally
appropriate to nodal superconductors in this half-filled regime where
(as discussed in the text) a BEC is not accessible. 
In Appendix~\ref{sec:AppC}, we make contact with some aspects of cuprate
experiments, but it is important not to confuse the phenomenological
appendix with the more precise predictions we present here.

For definiteness, we look at a typical band structure that happens to
be used for cuprates (but otherwise is of no particular
consequence). We take
$\epsilon_{\vect k} = (4t+4 t^\prime +2 t_z) - 2t (\cos k_x +\cos k_y)
- 4 t^\prime \cos k_x \cos k_y - 2 t_z \cos k_z$ with
$t^\prime/t= -0.3$.  This band structure is more complicated than that
used in the main text of the paper (for both $s$- and $d$-wave
systems), as it has a van Hove singularity which is prominent for the
band fillings we address.  This is found to affect some properties of
the crossover.

The goal of this appendix is to present the general behavior of the
$T^*$ and $ T_\text{c}$ phase diagrams, and the associated properties
of the chemical potential. The latter is useful to establish because
it can, in principle, be measured.  Moreover, the size of the fermionic
chemical potential is often viewed as a measure of where a given
system is in the crossover spectrum.  By contrast, we emphasize here,
unlike in the Fermi gases, how improbable it is to find a solid state
superconductor anywhere in proximity to a BEC. As discussed in the
main text there are better indicators of crossover physics than found
in $\mu$, through the behavior, for example, of $T^*/T_\text{c}$ and
the coherence length.

Fig.~\ref{fig:31}(a) plots a $d$-wave phase diagram at a hole
concentration $p= 1-n = 0.15$ as a function of attractive coupling
constant. Indicated are representative values of $T^*$ and
$T_\text{c}$.  In the next figure, the solid line in
Fig.~\ref{fig:31}(b) serves to characterize the behavior of the
self-consistently determined fermionic chemical potential $\mu(T)$ for
this particular interaction strength, as a function of temperature
$T$.  The dashed line indicates the counterpart value in the
extrapolated normal state, $\mu_{\text{N}}(T)$, obtained by turning
off the attraction.  A crucial point follows by comparing
Figs.~\ref{fig:31}(a) and \ref{fig:31}(b), where we see that, although
there is an appreciable separation between $T^*$ and $T_\text{c}$, the
chemical potential differs only slightly from its normal-state value.

\begin{figure}[b]
\centering
\includegraphics[width=0.85\linewidth]
{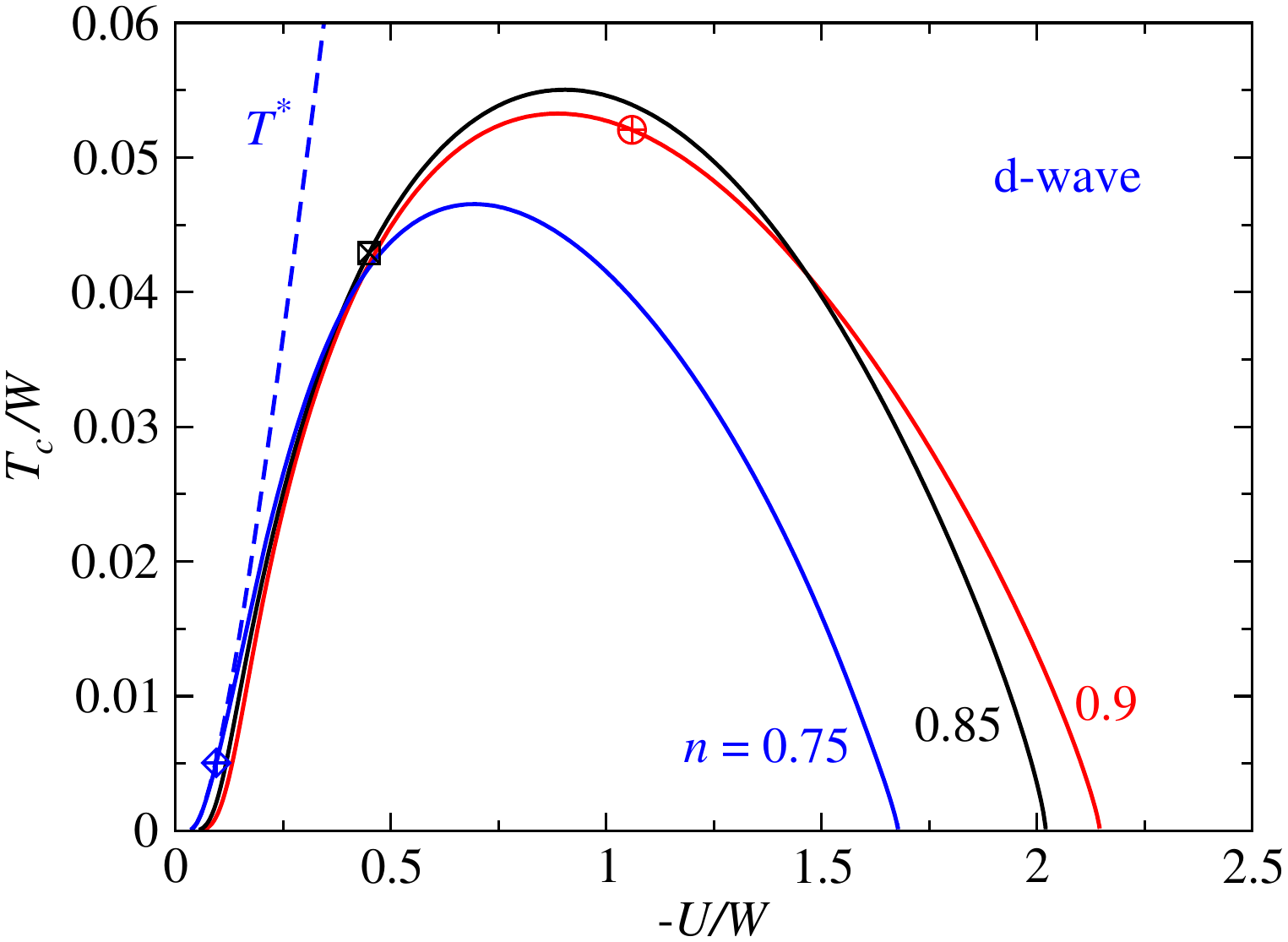}
\caption{$T_\text{c}$ -- $U$ phase diagrams for quasi-2D d-wave superconductors with the same energy dispersion as in Fig.~\ref{fig:31}, computed for different electron densities $n=1-p$, as labeled, where $p$ is the hole doping. The symbols indicate where a given system (represented by the $n$ value and $T^*/T_\text{c}$) is located in the corresponding experimental phase diagram~\cite{Hashimoto2014}. For clarity, here we show the $T^*$ line for $n=0.75$ only. }
\label{fig:32}
\end{figure}

Figure \ref{fig:32} presents results for a range of hole
concentrations, near half filling. For reasons that will become clear
later, we choose $T^*/T_\text{c}$ to be $4.7$ to illustrate the
behavior for a slightly lower hole doping $p=0.1$, while
$T^*/T_\text{c} = 1.05$ for a system with higher doping corresponding
to $p =0.25$. These two cases respectively show the effects of
increasing and decreasing the size of the pseudogap.

Table \ref{tab:Fig32Data} summarizes some central findings. Here we
tabulate results for all three hole doping levels,
$p=\{0.1,0.15,0.25\}$, including the behavior of the chemical
potentials. This table presents the ratios of the zero-temperature
chemical potential $\mu$ to their normal-state counterparts. The
difference from unity is small and in the most extreme case, still
less than 10\%.  From this comparison, one might view these systems as
conventional BCS superconductors, but it should be emphasized that they
all belong to the BCS-BEC crossover regime as $T_\text{c}$ and $T^*$
are quite distinct.

\begin{table}
    \center
    \caption{Table showing changes in chemical potential associated with different values of $T^*/T_\text{c}$. Here $W=4t$.} 
    \begin{ruledtabular}
    \begin{tabular}{cccc}
        ~~hole doping ($p$)~~                            &     ~~$T^*/T_\text{c}$ ~~                 &  ~~$|U|/W$~~~                &    ~~$\mu(T=0)/\mu_{\text{N}}(T=0)$~~~  \\
        \colrule
               $p=0.10$                                         &          4.73                          &  1.06      &  1.09   \\
                $p=0.15$                                        &         2.03                          &  0. 45     &   1.04  \\
                 $p=0.25$                                     &              1.05                  &  0.095      &   1.003   \\
    \end{tabular}
    \end{ruledtabular}
    \label{tab:Fig32Data}
\end{table}

\begin{figure}[htp]
\centering
\includegraphics[width=0.8\linewidth]
{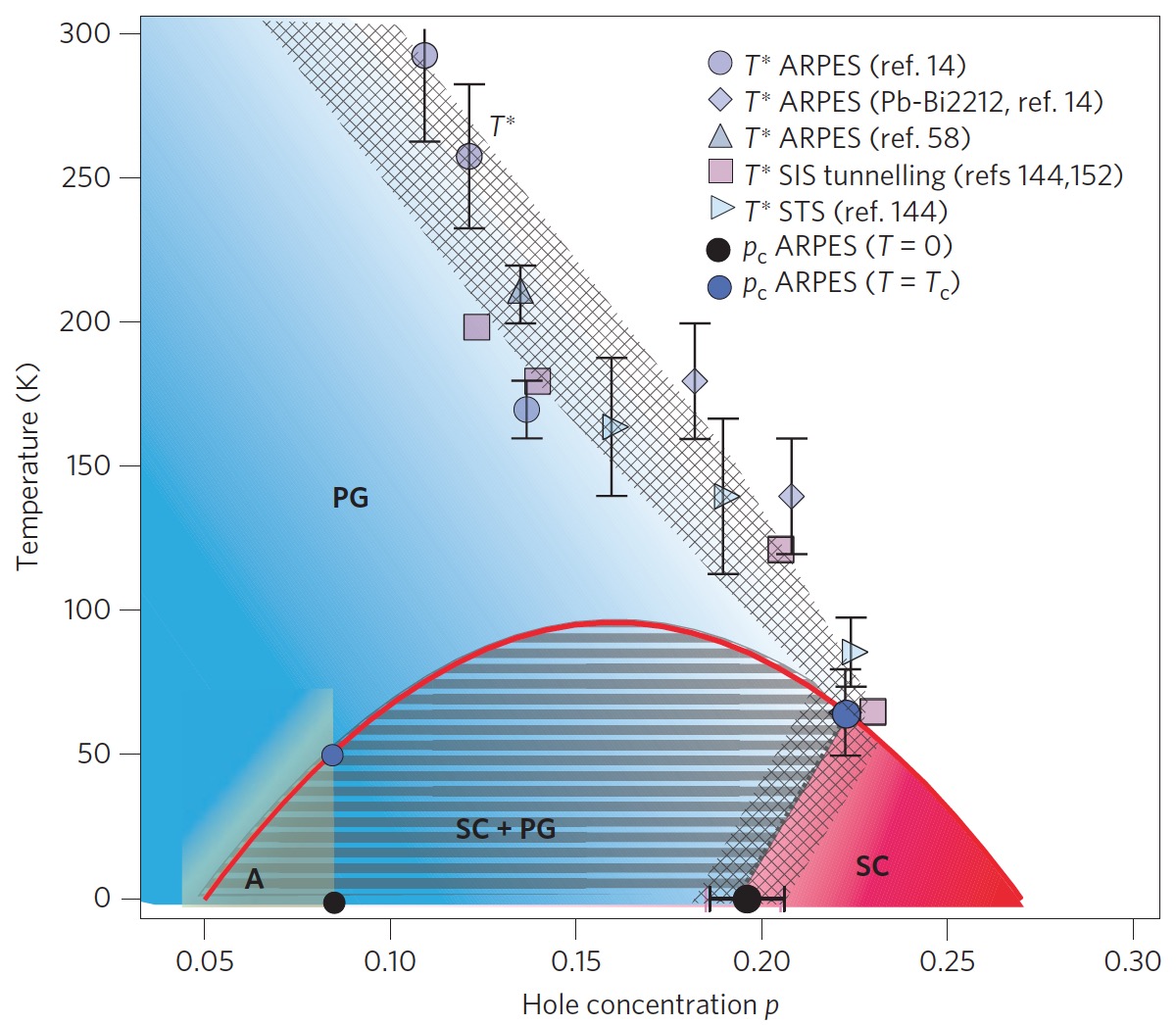}
\caption{ 
Experimental cuprate phase diagram, taken from \textcite{Hashimoto2014}. }
\label{fig:33}
\end{figure}

\section{Implications of the Cuprate Phase Diagram and Relation to Twisted Graphene
Family}
\label{sec:AppC}

\begin{table*}[tp]
\begin{minipage}[c]{\textwidth}
    \center
    \caption{Key parameters for hole-doped cuprates. In some sense these are near weak coupling which reflects the fact that the
cuprate $T^*/T_\text{c}$ are not very large except at extreme underdoping. Here $\Delta_0= 2 \Delta$, which is the zero temperature spectral gap $|\Delta_{\vect k}|=|\Delta (\cos k_x -\cos k_y)|$ at $(k_x,k_y)=(\pi,0)$ as measured in ARPES. }
       \begin{ruledtabular}
    \begin{tabular}{ccccccccc}
        ~~hole doping ($p$)~~                            &     ~~$T^*$ (K) ~~                 &  ~~$T_\text{c}$ (K)~~~                &   ~~ $T^*/T_\text{c}$~~~     &  $t$ (meV) ~~&~~$2\Delta_0 / k_\text{B}T_\text{c}$       & ~~$T_\text{F}$ (K)~~ &$T_\text{c}/T_\text{F}$  & $|U|$ (meV)\\
        \colrule
               $p=0.10$                                        &     260               & 55                          & 4.73                 &  22.7         &~~25.9~~          &~~502~~~&0.11 &  96.4  \\
               $p=0.15$                                       &    190               & 93                           &2.03               &   46.6         &~~9.85~~            &~975~~&0.095  & 84.0  \\
                $p=0.25$                                          &     32               & 30.6                         & 1.05                &  130   &~~4.28~~             &~2466~~~&0.012  & 49.3 \\
    \end{tabular}
    \end{ruledtabular}
    \label{tab:CuprateData}
  \end{minipage}
\end{table*}


Whether any of the above discussion is relevant to the cuprates cannot
be unequivocally established. But it is useful to explore what the
consequences are if we assume the values of $n$ and $T^*/T_\text{c}$ chosen
above and then establish the implications of this $d$-wave BCS-BEC
crossover.  Indeed, the correspondence between both of these
parameters can be seen to be reasonably compatible with the cuprate
phase diagram, which is shown in Fig.~\ref{fig:33}
\cite{Hashimoto2014}.  This compatibility of the parameter set, of
course, depends on assuming that the measured $T^*$ is related to pairing.

We emphasize that there are complexities concerning this phase diagram
which are still not fully settled. Among these is the observation of a
second characteristic temperature~\cite{Vishik2018}, which
is not shown in the plot.  This temperature
is typically about $20\%$ above $T_\text{c}$, although significantly below
$T^*$ for heavily underdoped cuprates; one might speculate that this
is associated with the onset of a more extended fluctuation regime
where bosonic transport, derived from quasi-stable pre-formed pairs
near condensation, is significant.  Here we focus only on the presumed
gap opening temperature $T^*$ plotted above.  We emphasize that there is
no unanimity about whether one should associate the experimental $T^*$
with pairing or an alternative energy scale, for example, deriving
from possible ordering (e.g., $d$-density wave~\cite{Chakravarty2001})
or fluctuations in the particle-hole channel.

We view the ratio $T^*/T_\text{c}$ and corresponding density as input
parameters. However, one test of the applicability of this theory
comes from establishing the corresponding size of the electronic
energy scales needed to match the size of the measured $T_\text{c}$ and
$T^*$, say in Kelvin. At issue is the hopping matrix elements $t$,
which determine the bandwidth and Fermi energy for each cuprate with a
different hole concentration. 

One might estimate that $T_\text{c}/T_\text{F}$ is around $0.1$ in the
underdoped cuprates, as is confirmed in Table~\ref{tab:CuprateData},
where we present a more precise analysis.  It should be emphasized here
that in the literature the observation that
$T_\text{c}/T_\text{F} \approx 0.1$ is often misinterpreted as
representing the BEC limit of a Fermi gas. By contrast, the analysis
here shows that this characteristic number is associated with a solid-state superconductor that is very far from the BEC regime.

More specific cuprate parameters are presented in
Table~\ref{tab:CuprateData} which indicates the (only)
adjustable parameter, $t$, in the fifth column of the table. It should
be noted that this fitting suggests that the effective bandwidths will
have to decrease as the system becomes more underdoped.
Moreover, the attractive interaction $U$ appears to become stronger
as the insulator is approached. This should have some consequences
for the origin of the pairing ``glue".

We note that the values
for $T_\text{F}$ shown appear to be slightly smaller, but not by orders of
magnitude, than those presented by Uemura ~\cite{Uemura1997}.
As yet this remains an unsettled issue.

We take note of recent work applying BCS-BEC
crossover theory to the cuprates~ \cite{Harrison2022}. Here it was
suggested that the cuprates with a ``magic" ratio of
$2\Delta_0/T_\text{c} = 6.5$ can be identified with the unitary point in a
three dimensional cold Fermi gas.  This unitary point relates to the
location of an isolated two-body bound state.  However, as emphasized
in this review, the superconducting phase diagrams of solid-state
superconductors and Fermi gases are quite different, making such an
identification difficult to support.
In particular, from Table~\ref{tab:CuprateData} it follows that even
at optimal doping $p=0.15$, we have $2\Delta_0/T_\text{c} =9.85$,
which is, indeed, also consistent with numbers obtained from
photoemission experiments~\cite{He2018}. This value is larger than
$6.5$ and it follows that, on the basis of the analysis of the
chemical potential (Table~\ref{tab:Fig32Data}), such systems are far
from the BEC as well.

\begin{figure}
\centering
\includegraphics[width=0.9\linewidth]{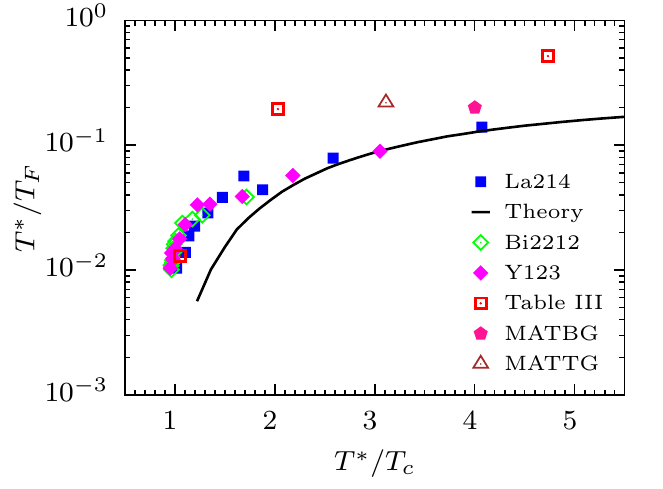}
\caption{ This figure provides some evidence that cuprates may belong
  to the BCS-BEC crossover family and that cuprates and both the
  twisted graphene superconducting families, MATBG and MATTG, seem to
  be rather similar.  The cuprate data of La214, Bi2212, and Y123 are
  the same as in Fig.~\ref{fig:8}.  In the legend, ``Table III"
  represents the additional two cuprate data points from
  Table~\ref{tab:CuprateData} for hole doping $p=0.1$ and $p=0.15$.
  The solid line is the predicted behavior for a $d$-wave crossover
  superconductor.  }
\label{fig:34}
\end{figure}

We end this Review with a figure (Fig. \ref{fig:34}) consolidating 
the results in the above table with those in Figs. \ref{fig:8} and \ref{fig:30}.
This presents a combination of the key parameters associated with both
MATBG and MATTG and a collection of counterpart data on
the hole-doped cuprates.
Indeed, one can see that the two graphene points are sandwiched between the
two most underdoped cuprates ($p=0.10$ and $p=0.15$). While it has been conjectured
~\cite{Oh2021} that MATBG bears a striking similarity to
the cuprates, the figure has presented some quantitative evidence
in support of this point.

\section{Convention and Notations}
\label{sec:Convention}

\subsection{Notations}

We follow standard notations as much as possible. They are summarized below.\\

$\hbar$ --- (reduced) Planck constant

$k_\text{B}$ --- Boltzmann constant

$c$ --- Speed of light

$e$ --- Electron charge

$E_\text{F}$ --- Fermi energy

$k_\text{F}$ --- Fermi momentum

$T_\text{c}$ --- Critical temperature for (superfluid/superconducting) phase
transition

$T_\text{BKT}$, $T_\varphi$ --- BKT transition temperature for (quasi-)2D superfluids.

$T^*$, $T_\Delta, T_{\text{pair}}$ --- Pair formation or pseudogap onset temperature.

$T$ --- Temperature

$\mu$, $\mu_{\pair}$, $\mu_{\text{B}}$ --- Fermionic, pair and bosonic chemical
potential, respectively.

$\mu_\text{N}$ --- Normal-state fermion chemical potential (which could be extrapolated down to $T=0$.

$\Delta$ --- Fermionic excitation gap

$\Delta_{\sc}$ --- Superconducting/superfluid order parameter

$\Delta_{\pg}$ --- Pseudogap 

$\Delta_\text{BCS}$ --- Mean-field gap obtained from BCS theory.

$\Delta_{0}\equiv \Delta(T=0)$ --- zero-temperature gap 

Four-vector $k \equiv (i\omega_n, \mathbf{k})$, $\sum_k \equiv T\sum_n
\sum_{\bf k} $, where $\omega_n = (2n+1)\pi k_\text{B} T/\hbar$ is the odd (fermionic)
Matsubara frequency, with $n \in \mathbb{Z}$.

Four-vector $q\equiv (i\Omega_l, \mathbf{q})$, $\sum_q \equiv T\sum_l
\sum_{\bf q} $, where $\Omega_l = 2l\pi k_\text{B}T/\hbar$ is the even (bosonic)
Matsubara frequency, with $l \in \mathbb{Z}$.

$f(x) = 1/(e^{x/k_\text{B}T}+1)$ --- Fermi-Dirac distribution function

$b(x) = 1(e^{x/k_\text{B}T}-1)$ --- Bose-Einstein distribution function

$G(k)$, $G_0(k)$ --- Full and bare Green's functions for fermions

$\Sigma(k)$ --- Self energy of fermions

$\Sigma_\text{sc}(k)$ --- Superconducting self energy of fermions

$\Sigma_\text{pg}(k)$ --- Pseudogap self energy of fermions

$\chi(q)$ --- Pair susceptibility

$t(q)$ --- $t$-matrix

$U<0$ --- Strength of the attractive interaction between fermions.

$U_\text{c}$ --- Critical interaction strength at which the two-body scattering
length diverges in free space, or more generally, the strength at which a bound state starts to emerge.

$V_{\mathbf{k,k'}} = U\varphi_{\mathbf{k}}\varphi_{\mathbf{k}'}$ --- Separable pairing interaction, with strength $U < 0$ and the symmetry factor $\phik$. For a contact potential or the attractive Hubbard model,  $\varphi_{\mathbf{k}} = 1$;  for the cuprates, $\varphi_{\mathbf{k}} = \cos k_x - \cos k_y$.

$E_\text{kin}$ --- Characteristic kinetic-energy scale, which can be taken to be half of the band width at moderate density or $E_\text{F}$ at low density.

$\ek = \mathbf{k}^2/2m$ --- Bare fermion dispersion in free space, with $\hbar = 1$. 

 $\ek = 2t(2-\cos k_x - \cos k_y) + 4t^\prime(1-\cos k_x\cos k_y) + 2t_z(1-\cos k_z)$ --- Bare fermion dispersion in a quasi-2D square lattice, where $t$ and $t^\prime$ are the nearest- and next-nearest-neighbor in-plane hopping integral, respectively, and $t_z$ is the out-of-plane hopping integral. Here the lattice constants have been set to unity, $a=b=c=1$.

$\xi_{\bf k} =\ek -\mu$ --- Bare fermion dispersion measured from
the chemical potential.

$\Ek$ --- Bogoliubov quasiparticle dispersion

$\uk^2 = \frac{1}{2}(1+\xi_{\bf k}/\Ek)$, $\vk^2= \frac{1}{2}(1-\xi_{\bf
  k}/\Ek)$ 
--- Coherence factors as given in BCS theory.

$\Psi^\text{BCS}$ --- Ground-state BCS wave function

$n$ --- Fermion number density

$p = (1-n)$ or $x=(1-n)$ --- Hole-doping concentration (in the cuprates)

$n_{\text{B}} \equiv n_\text{pair}$ --- Fermion pair or boson number density

$M_{\text{B}} \equiv M_\text{pair}$ --- Effective mass of fermion pairs or bosons

$N_\mathbf{q}/N$ --- Quasi-condensate fraction (in 2D Fermi gas experiment)

$\rho_s$ --- Superfluid phase stiffness, having dimension $[n]/[m]$.

$a_s$ --- $s$-wave inter-fermion scattering length

$a_\text{2D}$ --- 2D $s$-wave inter-fermion scattering length

$d$ --- Interparticle distance $\equiv d_\text{particle} $ (in MATBG and MATTG) and inter-layer distance in the double-layer quantum Hall context. 

$\xi_0^\text{coh}$ --- GL coherence length

$\xi_0$ --- Pair size

$H_{c2}$ --- Upper critical field

$B$ --- Magnetic field strength

$\ell_B=\sqrt{\hbar/|eB|}$ --- Magnetic length

$\Phi_0$ --- Flux quantum

$\rho_{xx}$ --- Longitudinal resistivity

$\rho_{xy}$ --- Transverse resistivity

$R_{\text{H}}, R_{xy}$  --- Hall, transverse resistance

$R_{xx}^\text{counter}$ --- Longitudinal counter flow resistance measured in the double-layer quantum Hall systems

$R_{xy}^\text{drag}$ --- Hall drag resistance

$\sigma_1, \sigma_2$ --- Real and imaginary parts of the conductivity $\sigma(\omega)$

$\chi_\text{dia}$ --- Diamagnetic susceptibility

$M_\text{dia}$ --- Diamagnetic response in magnetization

${\cal D}^\text{crit}_\text{pair}$ --- Critical value associated with the phase space density of pairs for TBK transition.

$1/T_1$ --- Nuclear spin-lattice relaxation rate

$V_g , V_\text{gate}$--- Gating voltage

$\nu$ --- Electronic band filling factor (in MATBG and MATTG)

$\theta$ --- Twist angle (in MATBG and MATTG)\\

We always refer to the absolute value when we refer to the interaction
parameter $U$ as increasing or decreasing.

\subsection{Convention for units}

Throughout this Review, we use the convention for units where it is not
explicitly spelled out:\\

$\hbar = k_\text{B} = c = 1$.

In numerics, we set the volume to unity, and $E_\text{F} = T_\text{F} = k_\text{F} = 2m = 1$ for the free space cases, which leads to $n = 1/(3\pi^2)$ in 3D.

For the lattice cases, we take the half bandwidth $W = zt = 1$ and lattice constants $a=b=c=1$. In a simple (quasi-)2D square or 3D cubic lattice, $n=1$ at half filling. 

Our fermionic chemical potential $\mu$ is measured with respect to the
bottom of the non-interacting energy band, such that $\epsilon_{\mathbf{k}=0}^{} = 0$. This leads to (i) $\mu = E_\text{F}$ in the
non-interacting limit at $T=0$, and (ii) $\mu$ changes sign when the
system crosses the boundary between fermionic and bosonic regimes.

\subsection{Abbreviations}

3D --- Three dimensions

2D --- Two dimensions

1UC --- One unit cell (thickness)

AL --- Aslamazov-Larkin (theory)

AFM --- Antiferromagnet  (or atomic force microscope)

ARPES --- Angle-resolved photoemission spectroscopy

BCS --- Bardeen-Cooper-Schrieffer (theory)

BEC --- Bose-Einstein condensation

BKT --- Berezinskii-Kosterlitz-Thouless (transition)

BSCCO, Bi2212 --- Bi$_2$Sr$_2$CaCu$_2$O$_{8+ \delta}$

CF --- Composite fermion

DC --- Direct current

DMFT --- Dynamical mean field theory

GL --- Ginzburg-Landau (theory)

GP --- Gross-Pitaevskii (equation)

LAO/STO --- LaAlO$_3$/SrTiO$_3$ (interface)

LSCO, La214 --- La$_{1-x}$Sr$_{x}$CuO$_4$

LL --- Landau levels

MIR --- Mid-infrared (conductivity)

MATBG (MATTG) --- Magic angle twisted bilayer (trilayer) graphene

meV --- milli-electron volts

NMR --- Nuclear magnetic resonance

NSR --- Nozi\'eres and Schmitt-Rink

OD --- Overdoped (cuprates)

PG --- Pseudogap

QMC --- Quantum Monte Carlo (simulations)

QPI --- Quasi-particle interference

RF -- Radio frequency (spectroscopy)

RPA --- Random phase approximation 

TDGL --- Time-dependent Ginzburg-Landau (theory)

TMA --- $t$-matrix approximation

SC --- Superconductor

SCTA --- Self-consistent $t$-matrix approximation

SI --- Superconductor-insulator (transition)

SIN --- Superconductor-insulator-normal metal (tunneling junction)

STM/STS --- Scan tunneling microscopy/spectroscopy

UD --- Underdoped (cuprates)

YBCO --- YBa$_2$Cu$_3$O$_{7-\delta}$

Y123 --- Y$_{0.8}$Ca$_{0.2}$Ba$_2$Cu$_3$O$_{7-\delta}$

YRZ --- Yang, Rice, Zhang (theory)

$\mu$SR --- Muon spin resonance/rotation/relaxation

\end{appendices}


\bibliographystyle{apsrev4-2}
\bibliography{ReferencesAll,magneton,ReferencesNewest,Suzuki}

\end{document}